\documentclass[conference,onecolumn]{IEEEtran}
\usepackage{amsmath,amssymb,amsfonts}
\usepackage{algorithmic}
\usepackage{graphicx}
\usepackage{textcomp}
\usepackage[dvipsnames,table]{xcolor}
\usepackage{dirtytalk}
\usepackage{circuitikz}
\usepackage{pgfplots}
\usepgfplotslibrary{fillbetween}
\pgfplotsset{width=10cm,compat=1.9}
\usepackage{adjustbox}
\usepackage{siunitx}
\usepackage{fontawesome}
\usepackage{listings}
\usepackage{tabularx}
\usepackage{tabulary}
\usepackage{tabu}
\usepackage{longtable}
\usepackage{tikz}
\usetikzlibrary{shapes.geometric, arrows}
\usepackage{circuitikz}
\usepackage[acronym]{glossaries}

\usepackage[backend=biber, style=ieee, url=false]{biblatex}

\newacronym{ABV}{ABV}{Assertion Based Verification}
\newacronym{ADAS}{ADAS}{Advanced Driver Assistance Systems}
\newacronym{ADC}{ADC}{Analog to Digital Converter}
\newacronym{API}{API}{Application Program Interface}
\newacronym{ASIL}{ASIL}{Automotive Safety Integrity Level}
\newacronym{ATPG}{ATPG}{Automated Test Pattern Generation}
\newacronym{BDD}{BDD}{Binary Decision Diagram}
\newacronym{BMC}{BMC}{Bounded Model Checking}
\newacronym{CPU}{CPU}{Central Processing Unit}
\newacronym{CAD}{CAD}{Computer Aided Design}
\newacronym{CEX}{CEX}{Counter Example}
\newacronym{CNF}{CNF}{Conjunctive Normal Form}
\newacronym{COI}{COI}{Cone of Influence}
\newacronym{ConfigVerMet}{ConfigVerMet}{Configuration Verification Methodology}
\newacronym{CBAW}{CBAW}{Configure-By-Air-Wires}
\newacronym{COV}{COV}{Coverage}
\newacronym{CDV}{CDV}{Coverage Driven Verification}
\newacronym{CSR}{CSR}{Control \& Status Register}
\newacronym{CSV}{CSV}{Comma Separated Values}
\newacronym{CRV}{CRV}{Constrained Random Verification}
\newacronym{CTL}{CTL}{Computational Tree Logic}
\newacronym{DAC}{DAC}{Digital to Analog Converter}
\newacronym{DB}{DB}{Database}
\newacronym{DD}{DD}{Decision Diagram}
\newacronym{DUT}{DUT}{Design Under Test}
\newacronym{DUV}{DUV}{Design Under Verification}
\newacronym{EDA}{EDA}{Electronic Design Automation}
\newacronym{FEC}{FEC}{Formal Equivalence Checking}
\newacronym{FEV}{FEV}{Formal Equivalence Verification}
\newacronym{FPV}{FPV}{Formal Property Verification}
\newacronym{FPGA}{FPGA}{Field Programmable Gate Array}
\newacronym{FSM}{FSM}{Finite State Machine}
\newacronym{FuSa}{FuSa}{Functional Safety}
\newacronym{FV}{FV}{Formal Verification}
\newacronym{GUI}{GUI}{Graphical User Interface}
\newacronym{HDL}{HDL}{Hardware Description Language}
\newacronym{HVL}{HVL}{Hardware Verification Language}
\newacronym{IC}{IC}{Integrated Circuit}
\newacronym{IP}{IP}{Intellectual Property}
\newacronym{I2C}{I2C}{Inter-Integrated Circuit}
\newacronym{IEEE}{IEEE}{Institute of Electrical and Electronics Engineers}
\newacronym{ISO}{ISO}{International Organization for Standardization}
\newacronym{ITRS}{ITRS}{International Technology Roadmap for Semiconductors}
\newacronym{JG}{JG}{JasperGold}
\newacronym{LTL}{LTL}{Linear Temporal Logic}
\newacronym{MDV}{MDV}{Metric Driven Verification}
\newacronym{OBDD}{OBDD}{Ordered Binary Decision Diagram}
\newacronym{OVM}{OVM}{Open Verification Methodology}
\newacronym{OVL}{OVL}{Open Verification Library}
\newacronym{PICT}{PICT}{Pairwise Independent Combinatorial Testing}
\newacronym{PSL}{PSL}{Property Specification Language}
\newacronym{ROBDD}{ROBDD}{Reduced Ordered Binary Decision Diagram}
\newacronym{RTL}{RTL}{Register Transfer Level}
\newacronym{SAT}{SAT}{Satisfiability}
\newacronym{SECDED}{SECDED}{Single Error Correction and Double Error Detection}
\newacronym{SEooC}{SEooC}{Safety Element out of Context}
\newacronym{SoC}{SoC}{System-on-Chip}
\newacronym{SVA}{SVA}{SystemVerilog Assertions}
\newacronym{UVM}{UVM}{Universal Verification Methodology}
\newacronym{UVC}{UVC}{UVM Verification Component}
\newacronym{VHSIC}{VHSIC}{Very High Speed Integrated Circuit}
\newacronym{VHDL}{VHDL}{VHSIC Hardware Description Language}
\newacronym{vPlan}{vPlan}{Verification Plan}
\newacronym{VSIF}{VSIF}{Verification Session Input File}
\newacronym{XML}{XML}{Extensible Markup Language}
\begin{document}

\lstdefinelanguage{Verilog}{morekeywords={accept_on,alias,always,always_comb,always_ff,always_latch,and,assert,assign,assume,automatic,before,begin,bind,bins,binsof,bit,break,buf,bufif0,bufif1,byte,case,casex,casez,cell,chandle,checker,class,clocking,cmos,config,const,constraint,context,continue,cover,covergroup,coverpoint,cross,deassign,default,defparam,design,disable,dist,do,edge,else,end,endcase,endchecker,endclass,endclocking,endconfig,endfunction,endgenerate,endgroup,endinterface,endmodule,endpackage,endprimitive,endprogram,endproperty,endspecify,endsequence,endtable,endtask,enum,event,eventually,expect,export,extends,extern,final,first_match,for,force,foreach,forever,fork,forkjoin,function,generate,genvar,global,highz0,highz1,if,iff,ifnone,ignore_bins,illegal_bins,implements,implies,import,incdir,include,initial,inout,input,inside,instance,int,integer,interconnect,interface,intersect,join,join_any,join_none,large,let,liblist,library,local,localparam,logic,longint,macromodule,matches,medium,modport,module,nand,negedge,nettype,new,nexttime,nmos,nor,noshowcancelled,not,notif0,notif1,null,or,output,package,packed,parameter,pmos,posedge,primitive,priority,program,property,protected,pull0,pull1,pulldown,pullup,pulsestyle_ondetect,pulsestyle_onevent,pure,rand,randc,randcase,randsequence,rcmos,real,realtime,ref,reg,reject_on,release,repeat,restrict,return,rnmos,rpmos,rtran,rtranif0,rtranif1,s_always,s_eventually,s_nexttime,s_until,s_until_with,scalared,sequence,shortint,shortreal,showcancelled,signed,small,soft,solve,specify,specparam,static,string,strong,strong0,strong1,struct,super,supply0,supply1,sync_accept_on,sync_reject_on,table,tagged,task,this,throughout,time,timeprecision,timeunit,tran,tranif0,tranif1,tri,tri0,tri1,triand,trior,trireg,type,typedef,union,unique,unique0,unsigned,until,until_with,untyped,use,uwire,var,vectored,virtual,void,wait,wait_order,wand,weak,weak0,weak1,while,wildcard,wire,with,within,wor,xnor,xor,`uvm_create, `uvm_rand_send_with},morecomment=[l]{//}}

\title{A Semi-Formal Verification Methodology for Efficient Configuration Coverage of Highly Configurable Digital Designs}

\author{\IEEEauthorblockN{Aman Kumar}
\IEEEauthorblockA{Infineon Technologies \\
Dresden, Germany \\
Aman.Kumar@infineon.com}
\and
\IEEEauthorblockN{Sebastian Simon}
\IEEEauthorblockA{Infineon Technologies \\
Dresden, Germany \\
Sebastian.Simon@infineon.com}
}

\maketitle

\begin{abstract}
Nowadays, a majority of System-on-Chips (SoCs) make use of \acrfull{IP} in order to shorten development cycles. When such \acrshort{IP}s are developed, one of the main focuses lies in the high configurability of the design. This flexibility on the design side introduces the challenge of covering a huge state space of \acrshort{IP} configurations on the verification side to ensure the functional correctness under every possible parameter setting. The vast number of possibilities does not allow a brute-force approach, and therefore, only a selected number of settings based on typical and extreme assumptions are usually verified. Especially in automotive applications, which need to follow the ISO 26262 functional safety standard, the requirement of covering all significant variants needs to be fulfilled in any case. State-of-the-Art existing verification techniques such as simulation-based verification and formal verification have challenges such as time-space explosion and state-space explosion respectively and therefore, lack behind in verifying highly configurable digital designs efficiently. This paper is focused on a semi-formal verification methodology for efficient configuration coverage of highly configurable digital designs. The methodology focuses on reduced runtime based on simulative and formal methods that allow high configuration coverage. The paper also presents the results when the developed methodology was applied on a highly configurable microprocessor \acrshort{IP} and discusses the gained benefits.
\end{abstract}

\begin{IEEEkeywords}
Formal Verification, \acrfull{ConfigVerMet}, Pairwise Verification, Equivalence Class Verification, Formal Structural Checking, \acrfull{FuSa}, ISO 26262
\end{IEEEkeywords}

\section{Introduction}
Modern \acrshort{SoC} designs are becoming more and more complex due to factors such as technology scaling\footnote{Technology scaling is the decrease in chip dimensions and increase in processor speed.}, mixed-signal designs, safety and security-critical devices, more demand by customers from a single chip and many more. In an observation presented in \cite{Moore}, Gordon E.~Moore estimated that \say{the number of transistors on an \acrfull{IC}, doubles every 18-24 months}. This empirical observation came to be known as Moore's law. The number of transistors on a single chip is increasing with better scaling, and Moore's law continues to be relevant until the present day.

With the increase in the scalability of the \acrshort{IC}, the possibility of adding more and more features to a single chip also increased. Modern chips are comparable to a Swiss-knife that can perform multiple functionalities using State-of-the-Art design methodologies. A study from \cite{Itrs} depicts that the complexity of chips will continue to grow further. An increase in the number of computation logics and CPUs to cope up with the advancement in technology is predicted in the forecast. This indicates that the issues faced in product development due to the rising complexity need to be addressed using new methods.

A common concept of generic Intellectual Properties (IPs) came into the picture by the chip developers, which focuses on generically reusing the existing components by adopting well-defined methodologies and development processes. This process is becoming more tedious since multiple factors and assumptions need to be kept in mind while developing a generic \acrshort{IP}. One of the biggest challenges of rising complexity is to ensure the functional correctness of the designs. Nevertheless, as the design process developed with time to inculcate complex functionalities, the verification process also matured by introducing State-of-the-Art verification methodologies like \acrfull{OVM} and \acrfull{UVM}. However, these methodologies do not address the problem of verifying a highly configurable digital design within reasonable time and effort in order to prove the functional correctness completely.

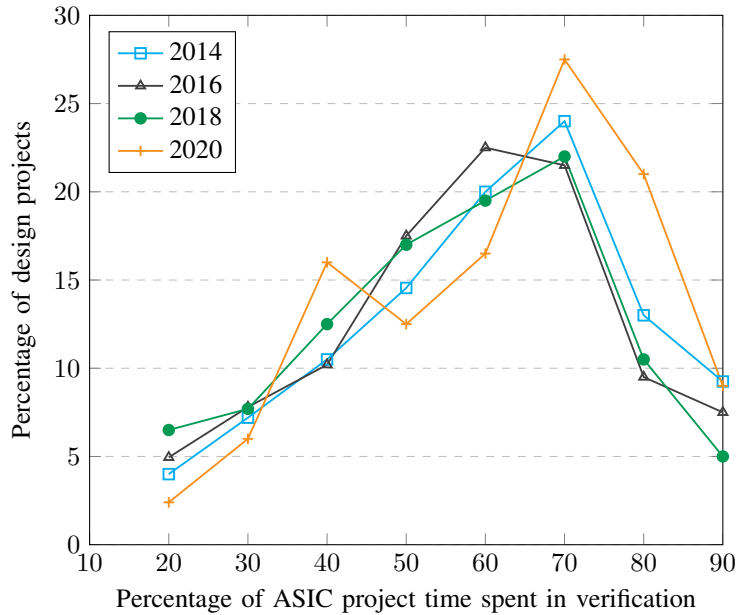
\begin{figure}[h]
\centering
\begin{tikzpicture}
\begin{axis}[
    xlabel={Percentage of ASIC project time spent in verification},
    ylabel={Percentage of design projects},
    xmin=10, xmax=90,
    ymin=0, ymax=30,
    xtick={10,20,30,40,50,60,70,80,90},
    ytick={0,5,10,15,20,25,30},
    legend pos=north west,
    ymajorgrids=true,
    grid style=dashed,
]

\addplot[
    color=cyan,
    mark=square,
    line width=0.25mm,
    ]
    coordinates {
    (20,4)(30,7.2)(40,10.5)(50,14.55)(60,20)(70,24)(80,13)(90,9.25)
    };
    \addlegendentry{2014}

\addplot[
    color=darkgray,
    mark=triangle,
    line width=0.25mm,
    ]
    coordinates {
    (20,4.95)(30,7.8)(40,10.2)(50,17.5)(60,22.5)(70,21.5)(80,9.5)(90,7.5)
    };
    \addlegendentry{2016}

\addplot[
    color=ForestGreen,
    mark=*,
    line width=0.25mm,
    ]
    coordinates {
    (20,6.5)(30,7.7)(40,12.5)(50,17)(60,19.5)(70,22)(80,10.5)(90,5)
    };
    \addlegendentry{2018}

\addplot[
    color=BurntOrange,
    mark=+,
    line width=0.25mm,
    ]
    coordinates {
    (20,2.4)(30,6)(40,16)(50,12.5)(60,16.5)(70,27.5)(80,21)(90,9)
    };
    \addlegendentry{2020}

\end{axis}
\end{tikzpicture}
\caption{Verification efforts required in overall product development \cite{VerStudy}}
\label{asic_ver_time}
\end{figure}

A recent study in Fig.~\ref{asic_ver_time} shows that verification consumes approximately \SI{60}{\percent} of the overall project time. It can be depicted from the figure that the percentage of projects spending more than \SI{70}{\percent} of their time in verification increased in 2020 compared to 2018. One of the major reasons for such a drastic increase is the rise in complexity of the designs. Nevertheless, to guarantee a completely bug-free design is also difficult to realize given the fact that some corner cases may miss out even after careful consideration. To address the complexity of verification of digital designs, several State-of-the-Art methods and approaches such as \acrfull{CRV}, directed testing, \acrfull{ABV} and \acrfull{FV} are used in the industry. Verification methodologies such as \acrshort{OVM} and \acrshort{UVM} have also been developed in the past years that capture best practices for verification and have become the de-facto standard concerning the simulation-based verification methods. It is also important to define methodologies since it enables reusability and drives standardization.

Nowadays, a majority of \acrshort{SoC}s make use of \acrshort{IP} blocks in order to shorten development cycles with respect to digital design, functional verification and physical implementation. When such \acrshort{IP}s are developed, one of the main focuses lies in the high configurability of the design in order to meet a high range of top-level requirements from different systems. This configurability in the design is introduced by using parameters.

The practice of using highly configurable design has its advantages in terms of flexibility and integrity. However, this flexibility on design side introduces the challenge of covering a huge state space of configurations on the verification side to ensure the functional correctness under every possible parameter setting. The vast number of possibilities does not allow a brute-force approach, and therefore, only a selected number of settings based on typical and extreme assumptions are usually verified. Especially in automotive applications, which needs to follow the ISO 26262 automotive functional safety standard, the requirement of covering all significant variants needs to be fulfilled in every case \cite{iso}. In addition to this, \acrfull{SEooC} designs that use configurations in order to enable reusability also needs to be verified exhaustively for all possible combinations of its configurations.

In order to overcome the explained shortcomings, a more sophisticated approach is required, which efficiently reduces the state-space of settings without neglecting any relevant configuration. This is accomplished by the semi-formal verification methodology explained in this paper that allow the closure of targeted configuration coverage. The new approach was applied in an exemplary way to a microprocessor \acrshort{IP} to demonstrate its added value.

\section{Background}

\subsection{Highly Configurable Digital Design}
A configurable digital design is made in the \acrshort{RTL} description by using \say{generic} in VHDL or \say{parameter} in Verilog/SystemVerilog. It can also be referred to as a parameterized design. A highly configurable digital design can have multiple parameters or parameters that have multiple possible values (also referred to as the cardinality of the parameter) or both.

\subsection{Safety Element out of Context}
The automotive industry develops generic elements for different applications and different customers. These generic elements can be developed independently by different organizations. In such cases, assumptions are made about the requirements and the design; including the safety requirements that are allocated to the element by higher design levels and on the design external to the element \cite{iso}.

A \acrfull{SEooC} is one such aspect of generic elements that are defined in the safety standards for ISO 26262. A \acrshort{SEooC} is a safety-related element which is not developed for a specific item. This means, it is not developed in the context of a particular system or vehicle \cite{iso}. In other words, a \acrshort{SEooC} is developed for an assumed context and not for a specific automotive vehicle. A \acrshort{SEooC} can be used as a plug and play device. To realize this, the design can be made configurable in order to make it more generic.

\tikzset{every picture/.style={line width=1pt}} %set default line width to 0.75pt        
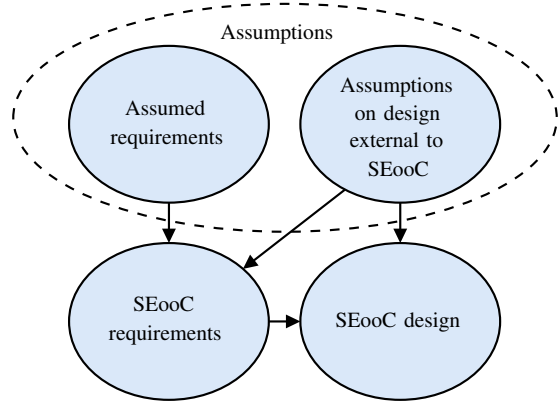
\begin{figure}[h]
\centering
\resizebox{0.4\textwidth}{!}{%
%\begin{adjustbox}{max width=8cm}
\begin{tikzpicture}[x=0.75pt,y=0.75pt,yscale=-1,xscale=1]
%uncomment if require: \path (0,536); %set diagram left start at 0, and has height of 536

%Shape: Ellipse [id:dp9694052040423473] 
\draw   [fill={rgb, 255:red, 74; green, 144; blue, 226 }  ,fill opacity=0.2 ] (306,170.5) .. controls (306,143.71) and (333.65,122) .. (367.75,122) .. controls (401.85,122) and (429.5,143.71) .. (429.5,170.5) .. controls (429.5,197.29) and (401.85,219) .. (367.75,219) .. controls (333.65,219) and (306,197.29) .. (306,170.5) -- cycle ;
%Shape: Ellipse [id:dp4290153616911201] 
\draw   [fill={rgb, 255:red, 74; green, 144; blue, 226 }  ,fill opacity=0.2 ] (163,170.5) .. controls (163,143.71) and (190.65,122) .. (224.75,122) .. controls (258.85,122) and (286.5,143.71) .. (286.5,170.5) .. controls (286.5,197.29) and (258.85,219) .. (224.75,219) .. controls (190.65,219) and (163,197.29) .. (163,170.5) -- cycle ;
%Shape: Ellipse [id:dp48500031312172265] 
\draw   [fill={rgb, 255:red, 74; green, 144; blue, 226 }  ,fill opacity=0.2 ] (163,292.5) .. controls (163,265.71) and (190.65,244) .. (224.75,244) .. controls (258.85,244) and (286.5,265.71) .. (286.5,292.5) .. controls (286.5,319.29) and (258.85,341) .. (224.75,341) .. controls (190.65,341) and (163,319.29) .. (163,292.5) -- cycle ;
%Shape: Ellipse [id:dp25419037169391934] 
\draw   [fill={rgb, 255:red, 74; green, 144; blue, 226 }  ,fill opacity=0.2 ] (306,292.5) .. controls (306,265.71) and (333.65,244) .. (367.75,244) .. controls (401.85,244) and (429.5,265.71) .. (429.5,292.5) .. controls (429.5,319.29) and (401.85,341) .. (367.75,341) .. controls (333.65,341) and (306,319.29) .. (306,292.5) -- cycle ;
%Shape: Ellipse [id:dp018284306411413542] 
\draw  [dash pattern={on 4.5pt off 4.5pt}] (128.5,166.5) .. controls (128.5,127.56) and (203.72,96) .. (296.5,96) .. controls (389.28,96) and (464.5,127.56) .. (464.5,166.5) .. controls (464.5,205.44) and (389.28,237) .. (296.5,237) .. controls (203.72,237) and (128.5,205.44) .. (128.5,166.5) -- cycle ;
%Straight Lines [id:da034899027577526054] 
\draw    (224.75,219) -- (224.75,241) ;
\draw [shift={(224.75,244)}, rotate = 270] [fill={rgb, 255:red, 0; green, 0; blue, 0 }  ][line width=0.08]  [draw opacity=0] (8.93,-4.29) -- (0,0) -- (8.93,4.29) -- cycle    ;
%Straight Lines [id:da520971740918774] 
\draw    (333.5,211) -- (272.87,258.16) ;
\draw [shift={(270.5,260)}, rotate = 322.13] [fill={rgb, 255:red, 0; green, 0; blue, 0 }  ][line width=0.08]  [draw opacity=0] (8.93,-4.29) -- (0,0) -- (8.93,4.29) -- cycle    ;
%Straight Lines [id:da5463395150012806] 
\draw    (367.75,219) -- (367.75,241) ;
\draw [shift={(367.75,244)}, rotate = 270] [fill={rgb, 255:red, 0; green, 0; blue, 0 }  ][line width=0.08]  [draw opacity=0] (8.93,-4.29) -- (0,0) -- (8.93,4.29) -- cycle    ;
%Straight Lines [id:da14170930211069388] 
\draw    (286.5,292.5) -- (303,292.5) ;
\draw [shift={(306,292.5)}, rotate = 180] [fill={rgb, 255:red, 0; green, 0; blue, 0 }  ][line width=0.08]  [draw opacity=0] (8.93,-4.29) -- (0,0) -- (8.93,4.29) -- cycle    ;

% Text Node
\draw (329,140) node [anchor=north west][inner sep=0.75pt]   [align=left] {Assumptions};
% Text Node
\draw (255,107) node [anchor=north west][inner sep=0.75pt]   [align=left] {Assumptions};
% Text Node
\draw (338,157) node [anchor=north west][inner sep=0.75pt]   [align=left] {on design};
% Text Node
\draw (335,174) node [anchor=north west][inner sep=0.75pt]   [align=left] {external to};
% Text Node
\draw (346,192) node [anchor=north west][inner sep=0.75pt]   [align=left] {SEooC};
% Text Node
\draw (196,154) node [anchor=north west][inner sep=0.75pt]   [align=left] {Assumed};
% Text Node
\draw (187,172) node [anchor=north west][inner sep=0.75pt]   [align=left] {requirements};
% Text Node
\draw (201,275) node [anchor=north west][inner sep=0.75pt]   [align=left] {SEooC};
% Text Node
\draw (186,293) node [anchor=north west][inner sep=0.75pt]   [align=left] {requirements};
% Text Node
\draw (326,285) node [anchor=north west][inner sep=0.75pt]   [align=left] {SEooC design};

\end{tikzpicture}}
\caption{Relationship between assumptions and \acrshort{SEooC} development \cite{iso}}
\label{seooc}
%\end{adjustbox}
\end{figure}

Fig.~\ref{seooc} shows the relationship between assumptions and \acrshort{SEooC} development. The development of a \acrshort{SEooC} can start at a certain hierarchical-level of requirements and design. Each individual requirement or design prerequisite is pre-determined in the status \say{assumed} \cite{iso}. On the other hand, verification activities should demonstrate that the developed \acrshort{SEooC}, at any level, is consistent with the requirements in the context where it is used \cite{iso}. This means a \acrshort{SEooC} is verified once for all the configurations possible and then used in any context. The important point for verification is that a \acrshort{SEooC} should be exhaustively verified for all possible configurations of the parameters and if not, arguments must be given in support of it \cite{iso}.

\subsection{Simulation-Based Verification}
Simulation-based verification uses input stimuli to drive the inputs of the \acrfull{DUV} and check the output to verify the functional correctness of the design. Simulation techniques such as \acrshort{UVM} are the de-facto standard being used in the industry. They make use of methods such as \acrfull{CRV} and \acrfull{CDV} to verify designs. For parameterized designs, the testbenches are made robust using techniques such as \acrshort{UVM} harness \cite{uvmharness} but do not address the problem of verifying huge combinations of all the parameters. In general, only extreme values in addition with some random values from the parameter range is usually verified. Simulation-based verification also suffers time-space explosion and therefore, is not optimum for verifying highly configurable digital designs.

\subsection{Formal Verification}
Formal verification is the use of tools that mathematically analyze the space of possible behaviours of a design, rather than computing results for particular values \cite{fvbook}. It is an exhaustive verification technique that uses mathematical proof methods to verify if the design implementation matches design specifications.

\tikzset{every picture/.style={line width=1pt}} %set default line width to 0.75pt        
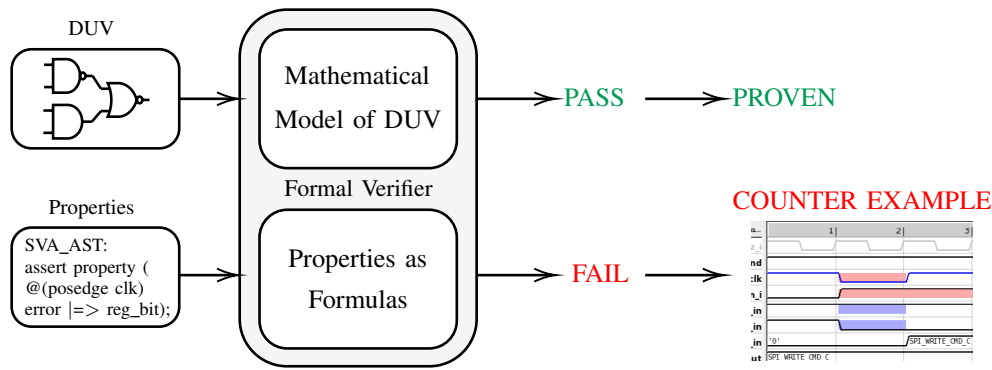
\begin{figure}[h]
\centering
\begin{tikzpicture}[x=0.75pt,y=0.75pt,yscale=-1,xscale=1]
%uncomment if require: \path (0,487); %set diagram left start at 0, and has height of 487

%Rounded Rect [id:dp4165598725149069] 
\draw [fill={rgb, 255:red, 155; green, 155; blue, 155 }  ,fill opacity=0.1 ]  (271,133.9) .. controls (271,120.7) and (281.7,110) .. (294.9,110) -- (366.6,110) .. controls (379.8,110) and (390.5,120.7) .. (390.5,133.9) -- (390.5,266.1) .. controls (390.5,279.3) and (379.8,290) .. (366.6,290) -- (294.9,290) .. controls (281.7,290) and (271,279.3) .. (271,266.1) -- cycle ;
%Rounded Rect [id:dp9339450857220801] 
\draw [fill={rgb, 255:red, 255; green, 255; blue, 255 }  ,fill opacity=1 ] (281,134) .. controls (281,126.27) and (287.27,120) .. (295,120) -- (366.5,120) .. controls (374.23,120) and (380.5,126.27) .. (380.5,134) -- (380.5,176) .. controls (380.5,183.73) and (374.23,190) .. (366.5,190) -- (295,190) .. controls (287.27,190) and (281,183.73) .. (281,176) -- cycle ;
%Rounded Rect [id:dp15725489998606723] 
\draw [fill={rgb, 255:red, 255; green, 255; blue, 255 }  ,fill opacity=1 ] (281,225) .. controls (281,217.27) and (287.27,211) .. (295,211) -- (366.5,211) .. controls (374.23,211) and (380.5,217.27) .. (380.5,225) -- (380.5,267) .. controls (380.5,274.73) and (374.23,281) .. (366.5,281) -- (295,281) .. controls (287.27,281) and (281,274.73) .. (281,267) -- cycle ;
%Rounded Rect [id:dp23064821910884348] 
\draw   (156,140) .. controls (156,134.48) and (160.48,130) .. (166,130) -- (230.5,130) .. controls (236.02,130) and (240.5,134.48) .. (240.5,140) -- (240.5,170) .. controls (240.5,175.52) and (236.02,180) .. (230.5,180) -- (166,180) .. controls (160.48,180) and (156,175.52) .. (156,170) -- cycle ;
%Rounded Rect [id:dp08722625659251793] 
\draw   (156,230) .. controls (156,224.48) and (160.48,220) .. (166,220) -- (230.5,220) .. controls (236.02,220) and (240.5,224.48) .. (240.5,230) -- (240.5,260) .. controls (240.5,265.52) and (236.02,270) .. (230.5,270) -- (166,270) .. controls (160.48,270) and (156,265.52) .. (156,260) -- cycle ;
%Straight Lines [id:da5451179594593565] 
\draw    (240.5,155) -- (268.5,155) ;
\draw [shift={(270.5,155)}, rotate = 180] [color={rgb, 255:red, 0; green, 0; blue, 0 }  ][line width=0.75]    (10.93,-3.29) .. controls (6.95,-1.4) and (3.31,-0.3) .. (0,0) .. controls (3.31,0.3) and (6.95,1.4) .. (10.93,3.29)   ;
%Straight Lines [id:da932307846658921] 
\draw    (240.5,244) -- (268.5,244) ;
\draw [shift={(270.5,244)}, rotate = 180] [color={rgb, 255:red, 0; green, 0; blue, 0 }  ][line width=0.75]    (10.93,-3.29) .. controls (6.95,-1.4) and (3.31,-0.3) .. (0,0) .. controls (3.31,0.3) and (6.95,1.4) .. (10.93,3.29)   ;
%Shape: Nand Gate [id:dp4973421698975471] 
\draw   (176.89,135) -- (183.69,135) .. controls (187.43,135) and (190.48,138.79) .. (190.48,143.45) .. controls (190.48,148.11) and (187.43,151.9) .. (183.69,151.9) -- (176.89,151.9) -- (176.89,135) -- cycle (172.36,137.82) -- (176.89,137.82) (172.36,149.08) -- (176.89,149.08) (193.2,143.45) -- (196.82,143.45) (190.48,143.45) .. controls (190.48,142.52) and (191.09,141.76) .. (191.84,141.76) .. controls (192.59,141.76) and (193.2,142.52) .. (193.2,143.45) .. controls (193.2,144.38) and (192.59,145.14) .. (191.84,145.14) .. controls (191.09,145.14) and (190.48,144.38) .. (190.48,143.45) -- cycle ;
%Shape: Nor Gate [id:dp3384573831898732] 
\draw   (204.44,147.11) -- (210.1,147.11) .. controls (214.05,147.26) and (217.58,150.56) .. (219.16,155.56) .. controls (217.58,160.57) and (214.05,163.86) .. (210.1,164.01) -- (204.44,164.01) .. controls (206.87,158.78) and (206.87,152.34) .. (204.44,147.11) -- cycle (201.04,149.93) -- (205.57,149.93) (201.04,161.2) -- (205.57,161.2) (221.88,155.56) -- (225.5,155.56) (219.16,155.56) .. controls (219.16,154.63) and (219.77,153.87) .. (220.52,153.87) .. controls (221.27,153.87) and (221.88,154.63) .. (221.88,155.56) .. controls (221.88,156.5) and (221.27,157.25) .. (220.52,157.25) .. controls (219.77,157.25) and (219.16,156.5) .. (219.16,155.56) -- cycle ;
%Shape: And Gate [id:dp5535652210284228] 
\draw   (176.89,158.1) -- (184.23,158.1) .. controls (188.28,158.1) and (191.57,161.89) .. (191.57,166.55) .. controls (191.57,171.21) and (188.28,175) .. (184.23,175) -- (176.89,175) -- (176.89,158.1) -- cycle (172,160.92) -- (176.89,160.92) (172,172.18) -- (176.89,172.18) (191.57,166.55) -- (196.46,166.55) ;
%Straight Lines [id:da6137311540297312] 
\draw    (201.04,161.2) -- (196.46,166.55) ;
%Straight Lines [id:da9346897980882642] 
\draw    (201.04,149.93) -- (196.82,143.45) ;

%Straight Lines [id:da23863669298323864] 
\draw    (390.5,155) -- (428.5,155) ;
\draw [shift={(430.5,155)}, rotate = 180] [color={rgb, 255:red, 0; green, 0; blue, 0 }  ][line width=0.75]    (10.93,-3.29) .. controls (6.95,-1.4) and (3.31,-0.3) .. (0,0) .. controls (3.31,0.3) and (6.95,1.4) .. (10.93,3.29)   ;
%Straight Lines [id:da2720188774020682] 
\draw    (390.5,244) -- (428.5,244) ;
\draw [shift={(430.5,244)}, rotate = 180] [color={rgb, 255:red, 0; green, 0; blue, 0 }  ][line width=0.75]    (10.93,-3.29) .. controls (6.95,-1.4) and (3.31,-0.3) .. (0,0) .. controls (3.31,0.3) and (6.95,1.4) .. (10.93,3.29)   ;
%Straight Lines [id:da0006533934546508746] 
\draw    (475.5,155) -- (513.5,155) ;
\draw [shift={(515.5,155)}, rotate = 180] [color={rgb, 255:red, 0; green, 0; blue, 0 }  ][line width=0.75]    (10.93,-3.29) .. controls (6.95,-1.4) and (3.31,-0.3) .. (0,0) .. controls (3.31,0.3) and (6.95,1.4) .. (10.93,3.29)   ;
%Straight Lines [id:da9360060120526879] 
\draw    (475.5,244) -- (513.5,244) ;
\draw [shift={(515.5,244)}, rotate = 180] [color={rgb, 255:red, 0; green, 0; blue, 0 }  ][line width=0.75]    (10.93,-3.29) .. controls (6.95,-1.4) and (3.31,-0.3) .. (0,0) .. controls (3.31,0.3) and (6.95,1.4) .. (10.93,3.29)   ;
%Image [id:dp3393365110156814] 
\draw (585,252) node  {\includegraphics[width=83.63pt,height=52.5pt]{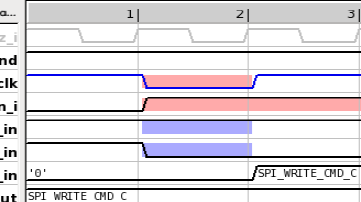}};

% Text Node
\draw (292,194) node [anchor=north west][inner sep=0.75pt]  [font=\small] [align=left] {Formal Verifier};
% Text Node
\draw (292,137) node [anchor=north west][inner sep=0.75pt]   [align=left] {Mathematical};
% Text Node
\draw (287,159) node [anchor=north west][inner sep=0.75pt]   [align=left] {Model of DUV};
% Text Node
\draw (294,229) node [anchor=north west][inner sep=0.75pt]   [align=left] {Properties as};
% Text Node
\draw (304,250) node [anchor=north west][inner sep=0.75pt]   [align=left] {Formulas};
% Text Node
\draw (183,114) node [anchor=north west][inner sep=0.75pt]  [font=\footnotesize] [align=left] {DUV};
% Text Node
\draw (161,224) node [anchor=north west][inner sep=0.75pt]  [font=\scriptsize] [align=left] {SVA\_AST:\\assert property (\\@(posedge clk)\\error $|$=$>$ reg\_bit);};
% Text Node
\draw (173,204) node [anchor=north west][inner sep=0.75pt]  [font=\footnotesize] [align=left] {Properties};
% Text Node
\draw (433,148) node [anchor=north west][inner sep=0.75pt]   [align=left] {\textcolor{ForestGreen}{PASS}};
% Text Node
\draw (437,237) node [anchor=north west][inner sep=0.75pt]   [align=left] {\textcolor{red}{FAIL}};
% Text Node
\draw (518,148) node [anchor=north west][inner sep=0.75pt]   [align=left] {\textcolor{ForestGreen}{PROVEN}};
% Text Node
\draw (518,200) node [anchor=north west][inner sep=0.75pt]   [align=left] {\textcolor{red}{COUNTER EXAMPLE}};

\end{tikzpicture}
\caption{Formal verifier}
\label{formal_verifier}
\end{figure}

Fig.~\ref{formal_verifier} shows the working of a formal verifier. There are two inputs to the formal verifier tool. On the one hand, the \acrfull{DUV} is fed into the tool which is converted into a mathematical model. On the other hand, properties, written in \acrfull{SVA} that capture the intent of the design is fed into the tool. The tool then converts these properties into mathematical formulas. In the next step, the tool tries to prove these mathematical formulas on the mathematical model of the \acrshort{DUV}. If the properties do not hold, it is said to be failed, and a \acrfull{CEX} is generated by the tool to debug further. In general, the absence of a \acrshort{CEX} is nothing but a pass or proven result.

Formal verification uses clever algorithms to verify the functional correctness of the design exhaustively \cite{fvbook}. However, for larger designs, it suffers state-space explosion and often takes longer than reasonable evaluation times for a conclusive proof result \cite{fvbook}. Although solutions such as \acrfull{CBAW} technique in formal verification have been proposed in \cite{seq_arm} and \cite{param_ver_cadence} which suggests to convert parameters into wires and make use of pseudo-constants, nevertheless, this technique can only be applied to designs having fewer parameters and cardinality. The technique does not hold for bus width parameterizations as well. In addition to this, not every design \acrshort{IP} can be fully verified using formal verification due to known limitations and challenges involved in it.

\section{Semi-Formal Configuration Verification Methodology}

\subsection{Overview of the Methodology}
\tikzset{every picture/.style={line width=1pt}} %set default line width to 0.75pt        
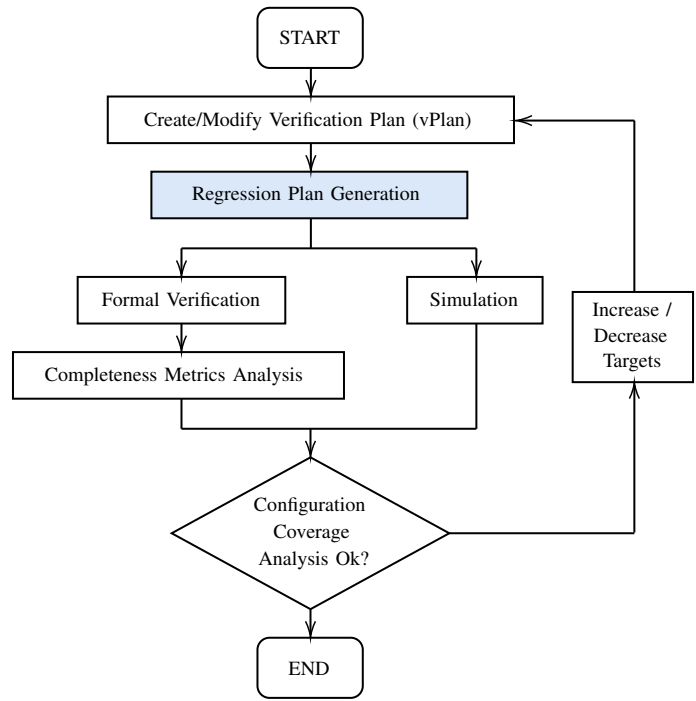
\begin{figure}[h]
\centering
\resizebox{0.5\textwidth}{!}{%
\begin{tikzpicture}[x=0.75pt,y=0.75pt,yscale=-1,xscale=1]
%uncomment if require: \path (0,726); %set diagram left start at 0, and has height of 726

%Rounded Rect [id:dp8718323734264437] 
\draw   (261,79) .. controls (261,74.58) and (264.58,71) .. (269,71) -- (323,71) .. controls (327.42,71) and (331,74.58) .. (331,79) -- (331,103) .. controls (331,107.42) and (327.42,111) .. (323,111) -- (269,111) .. controls (264.58,111) and (261,107.42) .. (261,103) -- cycle ;
%Shape: Rectangle [id:dp8245693128897191] 
\draw   (161.5,130) -- (431.5,130) -- (431.5,161) -- (161.5,161) -- cycle ;
%Shape: Rectangle [id:dp4942607292441048] 
\draw  [fill={rgb, 255:red, 74; green, 144; blue, 226 }  ,fill opacity=0.2 ] (190.5,180) -- (400.5,180) -- (400.5,211) -- (190.5,211) -- cycle ;
%Shape: Rectangle [id:dp937490425712483] 
\draw   (141.5,250) -- (279.5,250) -- (279.5,281) -- (141.5,281) -- cycle ;
%Shape: Rectangle [id:dp9954862992199565] 
\draw   (360.5,250) -- (450.5,250) -- (450.5,281) -- (360.5,281) -- cycle ;
%Shape: Rectangle [id:dp40122474242182427] 
\draw   (98.5,300) -- (318.5,300) -- (318.5,331) -- (98.5,331) -- cycle ;
%Flowchart: Decision [id:dp3579645395475737] 
\draw   (296.25,370) -- (388.5,420.5) -- (296.25,471) -- (204,420.5) -- cycle ;
%Rounded Rect [id:dp46441469151785597] 
\draw   (261,498) .. controls (261,493.58) and (264.58,490) .. (269,490) -- (323,490) .. controls (327.42,490) and (331,493.58) .. (331,498) -- (331,522) .. controls (331,526.42) and (327.42,530) .. (323,530) -- (269,530) .. controls (264.58,530) and (261,526.42) .. (261,522) -- cycle ;
%Straight Lines [id:da9794044299695686] 
\draw    (296.5,111) -- (296.5,128) ;
\draw [shift={(296.5,130)}, rotate = 270] [color={rgb, 255:red, 0; green, 0; blue, 0 }  ][line width=0.75]    (10.93,-3.29) .. controls (6.95,-1.4) and (3.31,-0.3) .. (0,0) .. controls (3.31,0.3) and (6.95,1.4) .. (10.93,3.29)   ;
%Straight Lines [id:da45565954505139494] 
\draw    (296.5,161) -- (296.5,178) ;
\draw [shift={(296.5,180)}, rotate = 270] [color={rgb, 255:red, 0; green, 0; blue, 0 }  ][line width=0.75]    (10.93,-3.29) .. controls (6.95,-1.4) and (3.31,-0.3) .. (0,0) .. controls (3.31,0.3) and (6.95,1.4) .. (10.93,3.29)   ;
%Straight Lines [id:da47995231294653884] 
\draw    (210.5,231) -- (210.5,248) ;
\draw [shift={(210.5,250)}, rotate = 270] [color={rgb, 255:red, 0; green, 0; blue, 0 }  ][line width=0.75]    (10.93,-3.29) .. controls (6.95,-1.4) and (3.31,-0.3) .. (0,0) .. controls (3.31,0.3) and (6.95,1.4) .. (10.93,3.29)   ;
%Straight Lines [id:da9837037852532813] 
\draw    (406.5,231) -- (406.5,248) ;
\draw [shift={(406.5,250)}, rotate = 270] [color={rgb, 255:red, 0; green, 0; blue, 0 }  ][line width=0.75]    (10.93,-3.29) .. controls (6.95,-1.4) and (3.31,-0.3) .. (0,0) .. controls (3.31,0.3) and (6.95,1.4) .. (10.93,3.29)   ;
%Straight Lines [id:da11966463814864658] 
\draw    (210.5,231) -- (406.5,231) ;
%Straight Lines [id:da45503815978336704] 
\draw    (296.5,211) -- (296.5,231) ;
%Straight Lines [id:da657536617300073] 
\draw    (210.5,281) -- (210.5,298) ;
\draw [shift={(210.5,300)}, rotate = 270] [color={rgb, 255:red, 0; green, 0; blue, 0 }  ][line width=0.75]    (10.93,-3.29) .. controls (6.95,-1.4) and (3.31,-0.3) .. (0,0) .. controls (3.31,0.3) and (6.95,1.4) .. (10.93,3.29)   ;
%Straight Lines [id:da007274758792701119] 
\draw    (296.25,351) -- (296.25,368) ;
\draw [shift={(296.25,370)}, rotate = 270] [color={rgb, 255:red, 0; green, 0; blue, 0 }  ][line width=0.75]    (10.93,-3.29) .. controls (6.95,-1.4) and (3.31,-0.3) .. (0,0) .. controls (3.31,0.3) and (6.95,1.4) .. (10.93,3.29)   ;
%Straight Lines [id:da1848221524936866] 
\draw    (210.5,331) -- (210.5,351) ;
%Straight Lines [id:da1874467603235319] 
\draw    (406.5,281) -- (406.5,351) ;
%Straight Lines [id:da9500977110044615] 
\draw    (210.5,351) -- (406.5,351) ;
%Straight Lines [id:da4072950647081939] 
\draw    (296.25,471) -- (296.25,488) ;
\draw [shift={(296.25,490)}, rotate = 270] [color={rgb, 255:red, 0; green, 0; blue, 0 }  ][line width=0.75]    (10.93,-3.29) .. controls (6.95,-1.4) and (3.31,-0.3) .. (0,0) .. controls (3.31,0.3) and (6.95,1.4) .. (10.93,3.29)   ;
%Shape: Rectangle [id:dp8542446981787548] 
\draw   (470.5,260) -- (550.5,260) -- (550.5,320) -- (470.5,320) -- cycle ;
%Straight Lines [id:da6741523144172463] 
\draw    (511.5,420) -- (511.5,322) ;
\draw [shift={(511.5,320)}, rotate = 450] [color={rgb, 255:red, 0; green, 0; blue, 0 }  ][line width=0.75]    (10.93,-3.29) .. controls (6.95,-1.4) and (3.31,-0.3) .. (0,0) .. controls (3.31,0.3) and (6.95,1.4) .. (10.93,3.29)   ;
%Straight Lines [id:da5277429823662507] 
\draw    (388.5,420.5) -- (511.5,420.5) ;
%Straight Lines [id:da7169896078462137] 
\draw    (511.5,146) -- (511.5,260) ;
%Straight Lines [id:da5047819233810578] 
\draw    (511.5,146) -- (434.5,146) ;
\draw [shift={(432.5,146)}, rotate = 360] [color={rgb, 255:red, 0; green, 0; blue, 0 }  ][line width=0.75]    (10.93,-3.29) .. controls (6.95,-1.4) and (3.31,-0.3) .. (0,0) .. controls (3.31,0.3) and (6.95,1.4) .. (10.93,3.29)   ;

% Text Node
\draw (274,84.5) node [anchor=north west][inner sep=0.75pt]   [align=left] {START};
% Text Node
\draw (184,139) node [anchor=north west][inner sep=0.75pt]   [align=left] {Create/Modify Verification Plan (vPlan)};
% Text Node
\draw (216,188) node [anchor=north west][inner sep=0.75pt]   [align=left] {Regression Plan Generation};
% Text Node
\draw (156,258.5) node [anchor=north west][inner sep=0.75pt]   [align=left] {Formal Verification};
% Text Node
\draw (374,258.5) node [anchor=north west][inner sep=0.75pt]   [align=left] {Simulation};
% Text Node
\draw (118,308.5) node [anchor=north west][inner sep=0.75pt]   [align=left] {Completeness Metrics Analysis};
% Text Node
\draw (257,395) node [anchor=north west][inner sep=0.75pt]   [align=left] {Configuration};
% Text Node
\draw (270,413) node [anchor=north west][inner sep=0.75pt]   [align=left] {Coverage};
% Text Node
\draw (260,431) node [anchor=north west][inner sep=0.75pt]   [align=left] {Analysis Ok?};
% Text Node
\draw (280,504) node [anchor=north west][inner sep=0.75pt]   [align=left] {END};
% Text Node
\draw (482,265.5) node [anchor=north west][inner sep=0.75pt]   [align=left] {Increase /};
% Text Node
\draw (483,282.5) node [anchor=north west][inner sep=0.75pt]   [align=left] {Decrease};
% Text Node
\draw (489,299.5) node [anchor=north west][inner sep=0.75pt]   [align=left] {Targets};

\end{tikzpicture}}
\caption{Overview of the methodology}
\label{methodology_flow}
\end{figure}

The verification of digital designs starts with creating a \acrfull{vPlan} that is derived from the specifications. It contains all the verification items as requirements that need to be covered during the course of verification. Once the \acrshort{vPlan} is well defined, the next step is the regression plan generation where relevant configurations for the design that needs to be verified are identified. These configurations are generated using several regression reduction techniques discussed in the coming sections and aim to unveil configuration-dependent deficiencies in the design. After the regression plan generation, based on the results from the generation, formal and simulation-based verification is carried out. For formal verification, completeness metrics analysis is carried out which essentially checks the structural coverage for the \acrshort{DUV}. After performing formal and simulation runs, a configuration coverage analysis is done where it is checked if all the relevant configurations identified in the regression plan generation are covered. If this is true, the verification is complete. If not, then the \acrshort{vPlan} needs to be modified, or the probable reasons for not achieving the complete configuration coverage needs to be investigated until coverage closure is met.

The main focus of the methodology remains in the regression plan generation, as highlighted in Fig.~\ref{methodology_flow}. The subsequent sections will discuss the generation flow in details.

\subsection{Regression Plan Generation}
The main task of the proposed methodology lies in the regression plan generation. The flow should select the number of configurations and the corresponding regressions in such a way that the total number of regressions are significantly less. At the same time, the number of configuration-dependent deficiencies in the design unveiled is as high as possible. To find the optimum number of regressions required, we make use of an automated regression plan generation flow.

\begin{figure}[h]
%\begin{adjustbox}{max width=\textwidth}
\centering
\tikzset{every picture/.style={line width=1pt}} %set default line width to 0.75pt        

\begin{tikzpicture}[x=0.75pt,y=0.75pt,yscale=-1,xscale=1]
%uncomment if require: \path (0,680); %set diagram left start at 0, and has height of 680

%Rounded Rect [id:dp810641014648342] 
\draw  [fill={rgb, 255:red, 74; green, 144; blue, 226 }  ,fill opacity=0.15 ] (248.5,489.2) .. controls (248.5,476.94) and (258.44,467) .. (270.7,467) -- (354.3,467) .. controls (366.56,467) and (376.5,476.94) .. (376.5,489.2) -- (376.5,555.8) .. controls (376.5,568.06) and (366.56,578) .. (354.3,578) -- (270.7,578) .. controls (258.44,578) and (248.5,568.06) .. (248.5,555.8) -- cycle ;
%Rounded Rect [id:dp17656261237024662] 
\draw  [fill={rgb, 255:red, 74; green, 144; blue, 226 }  ,fill opacity=0.15 ] (177.5,186.4) .. controls (177.5,165.74) and (194.24,149) .. (214.9,149) -- (410.1,149) .. controls (430.76,149) and (447.5,165.74) .. (447.5,186.4) -- (447.5,298.6) .. controls (447.5,319.26) and (430.76,336) .. (410.1,336) -- (214.9,336) .. controls (194.24,336) and (177.5,319.26) .. (177.5,298.6) -- cycle ;
%Flowchart: Document [id:dp1674753469733321] 
\draw  [fill={rgb, 255:red, 173; green, 212; blue, 131 }  ,fill opacity=0.52 ] (5,368) -- (112.5,368) -- (112.5,426.58) .. controls (45.31,426.58) and (58.75,447.7) .. (5,434.03) -- cycle ;
%Image [id:dp8166276678691289] 
%\draw (15.75,427.03) node  {\includegraphics[width=16.13pt,height=15pt]{Images/Automation.png}};
\draw (15.75,427.03) node  {\faGears};
%Flowchart: Document [id:dp8457442973551343] 
\draw  [fill={rgb, 255:red, 173; green, 212; blue, 131 }  ,fill opacity=0.52 ] (133.5,368) -- (241,368) -- (241,426.58) .. controls (173.81,426.58) and (187.25,447.7) .. (133.5,434.03) -- cycle ;
%Straight Lines [id:da13910085524862903] 
\draw    (112.5,403) -- (130.5,403) ;
\draw [shift={(133.5,403)}, rotate = 180] [fill={rgb, 255:red, 0; green, 0; blue, 0 }  ][line width=0.08]  [draw opacity=0] (8.93,-4.29) -- (0,0) -- (8.93,4.29) -- cycle    ;
%Image [id:dp5134467771319375] 
%\draw (144.25,427.03) node  {\includegraphics[width=16.13pt,height=15pt]{Images/Automation.png}};
\draw (144.25,427.03) node  {\faGears};
%Flowchart: Punched Tape [id:dp13012705025734972] 
\draw  [fill={rgb, 255:red, 155; green, 155; blue, 155 }  ,fill opacity=0.47 ] (269,381.9) .. controls (269,384.61) and (278.68,386.8) .. (290.63,386.8) .. controls (302.57,386.8) and (312.25,384.61) .. (312.25,381.9) .. controls (312.25,379.19) and (321.93,377) .. (333.88,377) .. controls (345.82,377) and (355.5,379.19) .. (355.5,381.9) -- (355.5,421.1) .. controls (355.5,418.39) and (345.82,416.2) .. (333.88,416.2) .. controls (321.93,416.2) and (312.25,418.39) .. (312.25,421.1) .. controls (312.25,423.81) and (302.57,426) .. (290.63,426) .. controls (278.68,426) and (269,423.81) .. (269,421.1) -- cycle ;
%Right Arrow [id:dp24354998015103346] 
\draw   (241,397.75) -- (257.5,397.75) -- (257.5,394) -- (268.5,401.5) -- (257.5,409) -- (257.5,405.25) -- (241,405.25) -- cycle ;
%Flowchart: Document [id:dp39150712301779245] 
\draw  [fill={rgb, 255:red, 173; green, 212; blue, 131 }  ,fill opacity=0.48 ] (383,368) -- (490.5,368) -- (490.5,426.58) .. controls (423.31,426.58) and (436.75,447.7) .. (383,434.03) -- cycle ;
%Right Arrow [id:dp2773350485960413] 
\draw   (355.5,397.75) -- (372,397.75) -- (372,394) -- (383,401.5) -- (372,409) -- (372,405.25) -- (355.5,405.25) -- cycle ;
%Image [id:dp7553948309607266] 
%\draw (393.75,427.03) node  {\includegraphics[width=16.13pt,height=15pt]{Images/Automation.png}};
\draw (393.75,427.03) node  {\faGears};
%\draw (393.75,427.03) node  {\includegraphics[width=16.13pt,height=15pt]{Images/Automation.png}};
%\draw (408,427.03) node  {\faCheckCircle};
%Flowchart: Document [id:dp0852098502292915] 
\draw  [fill={rgb, 255:red, 173; green, 212; blue, 131 }  ,fill opacity=0.52 ] (533,312) -- (640.5,312) -- (640.5,370.58) .. controls (573.31,370.58) and (586.75,391.7) .. (533,378.03) -- cycle ;
%Image [id:dp3688684156954172] 
%\draw (543.75,371.03) node  {\includegraphics[width=16.13pt,height=15pt]{Images/Automation.png}};
\draw (543.75,371.03) node  {\faGears};
%Flowchart: Document [id:dp49667355322921325] 
\draw  [fill={rgb, 255:red, 173; green, 212; blue, 131 }  ,fill opacity=0.52 ] (533,418.58) -- (640.5,418.58) -- (640.5,477.15) .. controls (573.31,477.15) and (586.75,498.27) .. (533,484.6) -- cycle ;
%Image [id:dp4631638952057353] 
%\draw (543.75,477.6) node  {\includegraphics[width=16.13pt,height=15pt]{Images/Automation.png}};
\draw (543.75,477.6) node  {\faGears};
%Straight Lines [id:da2139800734870363] 
\draw    (511.5,348) -- (529.5,348) ;
\draw [shift={(532.5,348)}, rotate = 180] [fill={rgb, 255:red, 0; green, 0; blue, 0 }  ][line width=0.08]  [draw opacity=0] (8.93,-4.29) -- (0,0) -- (8.93,4.29) -- cycle    ;
%Straight Lines [id:da9352222567556892] 
\draw    (511.5,454) -- (529.5,454) ;
\draw [shift={(532.5,454)}, rotate = 180] [fill={rgb, 255:red, 0; green, 0; blue, 0 }  ][line width=0.08]  [draw opacity=0] (8.93,-4.29) -- (0,0) -- (8.93,4.29) -- cycle    ;
%Straight Lines [id:da21252642257747212] 
\draw    (511.5,348) -- (511.5,454) ;
%Straight Lines [id:da7222250897552467] 
\draw    (490.5,403) -- (511.5,403) ;
%Flowchart: Document [id:dp14772175402151966] 
\draw  [fill={rgb, 255:red, 173; green, 212; blue, 131 }  ,fill opacity=0.52 ] (190,253) -- (297.5,253) -- (297.5,311.58) .. controls (230.31,311.58) and (243.75,332.7) .. (190,319.03) -- cycle ;
%Flowchart: Document [id:dp3560608925406006] 
\draw  [fill={rgb, 255:red, 173; green, 212; blue, 131 }  ,fill opacity=0.52 ] (327.5,253) -- (435,253) -- (435,311.58) .. controls (367.81,311.58) and (381.25,332.7) .. (327.5,319.03) -- cycle ;
%Flowchart: Punched Tape [id:dp1435414397467456] 
\draw  [fill={rgb, 255:red, 245; green, 166; blue, 35 }  ,fill opacity=0.48 ] (258.5,166.28) .. controls (258.5,170) and (270.92,173.02) .. (286.25,173.02) .. controls (301.58,173.02) and (314,170) .. (314,166.28) .. controls (314,162.55) and (326.42,159.53) .. (341.75,159.53) .. controls (357.08,159.53) and (369.5,162.55) .. (369.5,166.28) -- (369.5,220.25) .. controls (369.5,216.53) and (357.08,213.51) .. (341.75,213.51) .. controls (326.42,213.51) and (314,216.53) .. (314,220.25) .. controls (314,223.98) and (301.58,227) .. (286.25,227) .. controls (270.92,227) and (258.5,223.98) .. (258.5,220.25) -- cycle ;
%Right Arrow [id:dp7365325146991464] 
\draw   (308.22,466.93) -- (308.39,439.42) -- (304.63,439.39) -- (312.27,421.1) -- (319.67,439.49) -- (315.91,439.46) -- (315.74,466.98) -- cycle ;
%Right Arrow [id:dp044919554986647015] 
\draw   (316.5,336.09) -- (316.2,363.6) -- (319.96,363.64) -- (312.25,381.9) -- (304.93,363.48) -- (308.69,363.52) -- (308.98,336.01) -- cycle ;
%Flowchart: Document [id:dp09632796520733278] 
\draw  [fill={rgb, 255:red, 173; green, 212; blue, 131 }  ,fill opacity=0.52 ] (256,492) -- (368.5,492) -- (368.5,558) .. controls (298.19,558) and (312.25,581.8) .. (256,566.4) -- cycle ;
%Image [id:dp18456550765399982] 
%\draw (266.75,559.4) node  {\includegraphics[width=16.13pt,height=15pt]{Images/Automation.png}};
\draw (266.75,559.4) node  {\faGears};
%Image [id:dp1695483796632986] 
%\draw (200.75,312.03) node  {\includegraphics[width=16.13pt,height=15pt]{Images/Automation.png}};
\draw (200.75,312.03) node  {\faGears};
%Image [id:dp881471039339399] 
%\draw (338.25,312.03) node  {\includegraphics[width=16.13pt,height=15pt]{Images/Automation.png}};
\draw (338.25,312.03) node  {\faGears};
%Straight Lines [id:da7828343063016263] 
\draw    (241.5,234) -- (241.5,250) ;
\draw [shift={(241.5,253)}, rotate = 270] [fill={rgb, 255:red, 0; green, 0; blue, 0 }  ][line width=0.08]  [draw opacity=0] (8.93,-4.29) -- (0,0) -- (8.93,4.29) -- cycle    ;
%Straight Lines [id:da3326086937929835] 
\draw    (382.5,234) -- (382.5,250) ;
\draw [shift={(382.5,253)}, rotate = 270] [fill={rgb, 255:red, 0; green, 0; blue, 0 }  ][line width=0.08]  [draw opacity=0] (8.93,-4.29) -- (0,0) -- (8.93,4.29) -- cycle    ;
%Straight Lines [id:da11834267025350709] 
\draw    (241.5,234) -- (382.5,234) ;
%Straight Lines [id:da8843675279772605] 
\draw    (314,220.25) -- (314,234) ;
%Image [id:dp525180565806106] 
%\draw (509.05,568) node  {\includegraphics[width=16.12pt,height=15pt]{Images/Automation.png}};
\draw (509.05,568) node  {\faGears};
%\draw (509.05,585) node  {\faCheckCircle};
%\draw [color={rgb, 255:red, 245; green, 166; blue, 35 }  ,draw opacity=0.48 ][fill={rgb, 255:red, 245; green, 166; blue, 35 }  ,fill opacity=0.48 ] (503,580) -- (516,580) -- (516,590) -- (503,590) -- cycle ;
%\draw [color={rgb, 255:red, 173; green, 212; blue, 131 }  ,draw opacity=0.52 ][fill={rgb, 255:red, 173; green, 212; blue, 131 }  ,fill opacity=0.52 ] (503,597) -- (516,597) -- (516,607) -- (503,607) -- cycle ;
%Pentagon Arrow [id:dp631230068435078] 
\draw  [fill={rgb, 255:red, 155; green, 155; blue, 155 }  ,fill opacity=0.14 ] (4.88,99.98) -- (156.5,99.98) -- (177.5,114.98) -- (156.5,129.98) -- (4.88,129.98) -- cycle ;
%Chevron Arrow [id:dp27517884337096765] 
\draw  [fill={rgb, 255:red, 155; green, 155; blue, 155 }  ,fill opacity=0.14 ] (188.5,99.98) -- (417.32,99.98) -- (439.5,114.98) -- (417.32,129.98) -- (188.5,129.98) -- (210.68,114.98) -- cycle ;
%Chevron Arrow [id:dp030533550138117427] 
\draw  [fill={rgb, 255:red, 155; green, 155; blue, 155 }  ,fill opacity=0.14 ] (452.5,99.98) -- (617.8,99.98) -- (639.8,114.98) -- (617.8,129.98) -- (452.5,129.98) -- (474.5,114.98) -- cycle ;

% Text Node
\draw (22,372) node [anchor=north west][inner sep=0.75pt]   [align=left] {\textbf{{\footnotesize Configuration}}};
% Text Node
\draw (27,386) node [anchor=north west][inner sep=0.75pt]   [align=left] {{\footnotesize \textbf{Information}}};
% Text Node
\draw (17,401) node [anchor=north west][inner sep=0.75pt]   [align=left] {{\scriptsize Parameters extracted}};
% Text Node
\draw (25,412) node [anchor=north west][inner sep=0.75pt]   [align=left] {{\scriptsize from the design}};
% Text Node
\draw (142,372) node [anchor=north west][inner sep=0.75pt]   [align=left] {\textbf{{\footnotesize{Pairwise Template}}}};
% Text Node
\draw (142,386) node [anchor=north west][inner sep=0.75pt]   [align=left] {{\footnotesize Input file generated}};
% Text Node
\draw (141,397) node [anchor=north west][inner sep=0.75pt]   [align=left] {{\footnotesize with parameters for}};
% Text Node
\draw (141,408) node [anchor=north west][inner sep=0.75pt]   [align=left] {{\footnotesize Pairwise generation}};
% Text Node
\draw (277,396) node [anchor=north west][inner sep=0.75pt]   [align=left] {\textbf{{\footnotesize \acrshort{ConfigVerMet}}}};
% Text Node
\draw (394,371) node [anchor=north west][inner sep=0.75pt]   [align=left] {\textbf{{\scriptsize{Optimized Pairwise}}}};
% Text Node
\draw (411,382) node [anchor=north west][inner sep=0.75pt]   [align=left] {\textbf{{\scriptsize{Regressions}}}};
% Text Node
\draw (396,395) node [anchor=north west][inner sep=0.75pt]   [align=left] {{\scriptsize Optimized Pairwise}};
% Text Node
\draw (394,406) node [anchor=north west][inner sep=0.75pt]   [align=left] {{\scriptsize generation using MS}};
% Text Node
\draw (415,417) node [anchor=north west][inner sep=0.75pt]   [align=left] {{\scriptsize PICT tool}};
% Text Node
\draw (543,315) node [anchor=north west][inner sep=0.75pt]   [align=left] {\textbf{{\footnotesize{SV Configuration}}}};
% Text Node
\draw (552,328) node [anchor=north west][inner sep=0.75pt]   [align=left] {\textbf{{\footnotesize{Coverage File}}}};
% Text Node
\draw (560,343) node [anchor=north west][inner sep=0.75pt]   [align=left] {{\scriptsize Generate SV}};
% Text Node
\draw (554,353) node [anchor=north west][inner sep=0.75pt]   [align=left] {{\scriptsize config. coverage}};
% Text Node
\draw (577,364) node [anchor=north west][inner sep=0.75pt]   [align=left] {{\scriptsize file}};
% Text Node
\draw (572.5,422) node [anchor=north west][inner sep=0.75pt]   [align=left] {\textbf{{\footnotesize{VSIF}}}};
% Text Node
\draw (548,437) node [anchor=north west][inner sep=0.75pt]   [align=left] {{\scriptsize Generate VSIF for}};
% Text Node
\draw (540,449) node [anchor=north west][inner sep=0.75pt]   [align=left] {{\scriptsize regression (Simulation}};
% Text Node
\draw (564,461) node [anchor=north west][inner sep=0.75pt]   [align=left] {{\scriptsize + Formal)}};
% Text Node
\draw (260,475) node [anchor=north west][inner sep=0.75pt]   [align=left] {\textbf{{\footnotesize Bottom-Up Approach}}};
% Text Node
%\draw (287.25,174.02) node [anchor=north west][inner sep=0.75pt]   [align=left] {{\footnotesize \textbf{\underline{REQ Tool}}}};
\draw (268,175.5) node [anchor=north west][inner sep=0.75pt]   [align=left] {{\footnotesize \textbf{Requirements Tool}}};
% Text Node
\draw (282,190) node [anchor=north west][inner sep=0.75pt]   [align=left] {{\scriptsize Constraints and}};
% Text Node
\draw (274,201) node [anchor=north west][inner sep=0.75pt]   [align=left] {{\scriptsize Equivalence Classes}};
% Text Node
\draw (289,546) node [anchor=north west][inner sep=0.75pt]   [align=left] {{\scriptsize verification}};
% Text Node
\draw (215,256) node [anchor=north west][inner sep=0.75pt]   [align=left] {{\footnotesize \textbf{Constraints}}};
% Text Node
\draw (205,270) node [anchor=north west][inner sep=0.75pt]   [align=left] {{\scriptsize Extract constraints}};
% Text Node
\draw (195,283) node [anchor=north west][inner sep=0.75pt]   [align=left] {{\scriptsize from XML for Pairwise}};
% Text Node
\draw (220,296) node [anchor=north west][inner sep=0.75pt]   [align=left] {{\scriptsize verification}};
% Text Node
\draw (337,256) node [anchor=north west][inner sep=0.75pt]   [align=left] {{\footnotesize \textbf{Equivalence Class}}};
% Text Node
\draw (340,269.5) node [anchor=north west][inner sep=0.75pt]   [align=left] {{\scriptsize Extract Equivalence}};
% Text Node
\draw (335,281.5) node [anchor=north west][inner sep=0.75pt]   [align=left] {{\scriptsize Classes from XML for}};
% Text Node
\draw (339,293.5) node [anchor=north west][inner sep=0.75pt]   [align=left] {{\scriptsize Pairwise verification}};
% Text Node
\draw (271,524) node [anchor=north west][inner sep=0.75pt]   [align=left] {{\scriptsize Generate candidates}};
% Text Node
\draw (266,495) node [anchor=north west][inner sep=0.75pt]   [align=left] {{\footnotesize \textbf{Formal Structural}}};
% Text Node
\draw (288,508) node [anchor=north west][inner sep=0.75pt]   [align=left] {{\footnotesize \textbf{Checking}}};
% Text Node
\draw (280,535) node [anchor=north west][inner sep=0.75pt]   [align=left] {{\scriptsize for Block-Level}};
% Text Node
\draw (377,161) node [anchor=north west][inner sep=0.75pt]   [align=left] {\textbf{{\footnotesize Top-Down}}};
% Text Node
\draw (377,176) node [anchor=north west][inner sep=0.75pt]   [align=left] {\textbf{{\footnotesize Approach}}};
% Text Node
\draw (522,561) node [anchor=north west][inner sep=0.75pt]   [align=left] {{\scriptsize Automated using Python}};
% Text Node
\draw (53,107) node [anchor=north west][inner sep=0.75pt]   [align=left] {\textbf{Prepare}};
% Text Node
\draw (284.5,107) node [anchor=north west][inner sep=0.75pt]   [align=left] {\textbf{Extract}};
% Text Node
\draw (516.5,107) node [anchor=north west][inner sep=0.75pt]   [align=left] {\textbf{Execute}};

\end{tikzpicture}
\caption{Regression plan generation flow}
\label{regplanflow}
%\end{adjustbox}
\end{figure}
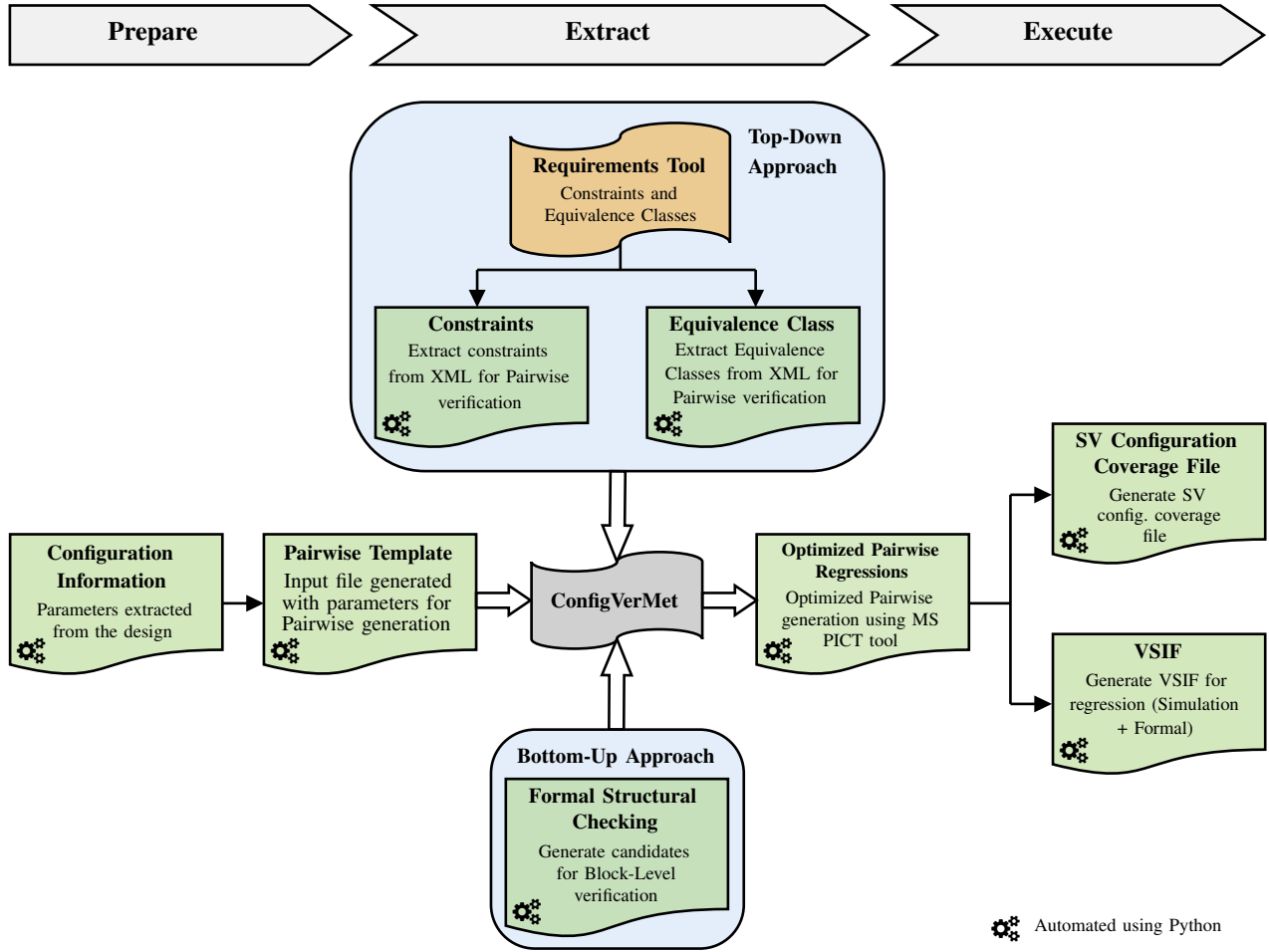

The regression plan generation flow is an automated generation flow implemented using Python scripts, as shown in Fig.~\ref{regplanflow}. The flow is essentially divided into three steps: prepare, extract and execute. These steps can be invoked via Makefile\footnote{A makefile is a file (by default named \say{Makefile}) containing a set of directives used by a make build automation tool to generate a target/goal.}. Additionally, we also name the methodology as \acrfull{ConfigVerMet}.

Each of the steps shown in Fig.~\ref{regplanflow} are discussed in details in the subsequent sections.

\section{ConfigVerMet: Prepare}
In the prepare step, the configuration information is gathered, meaning the parameters and their possible values are extracted from the design. This is done using automated scripts in Python and any formal verification tool (formal tools have dedicated switches to show parameters used in the \acrshort{DUV}). Once the configuration information is collected, a Pairwise template is generated automatically. This template contains all the parameters and their values written in a syntax supported by the Pairwise generation tool. This template is the input file for the Pairwise generation tool that generates the Pairwise regressions. The Pairwise generation tool used in our methodology is the open-source Microsoft \acrfull{PICT} tool\footnote{Access online: https://github.com/microsoft/pict} which is used predominantly in the software industry \cite{pairwise_study} \cite{pict_use}. It is also important at this stage to introduce the Pairwise verification technique.

\subsection{Pairwise Verification}
Pairwise verification is an effective test case generation technique that is based on the observation that faults are most commonly triggered by one or interaction of fewer than three parameters. This observation, in fact, is based on multiple findings published by various authors and Fig.~\ref{pairwise_graph} from \cite{pairwise} shows a plot between the degree of interaction in the parameters and the cumulative percentage of failures triggered by these interactions. The data used to plot these curves were accumulated for 15 years \cite{pairwise}.

\tikzset{every picture/.style={line width=0.75pt}} %set default line width to 0.75pt
\begin{figure}[h]
\centering
    \begin{tikzpicture}
    \begin{axis}[
        xlabel={Interaction},
        ylabel={Cummulative Percentage},
        xmin=1, xmax=6,
        ymin=0, ymax=100,
        xtick={1,2,3,4,5,6},
        ytick={0,25,50,75,100},
        legend pos=south east,
        ymajorgrids=true,
        xmajorgrids=true,
        grid style=dashed,
    ]
    
    \addplot[
        color=cyan,
        mark=square,
        line width=0.25mm,
        ]
        coordinates {
        (1,68)(2,98)(3,99)(4,100)
        };
        \addlegendentry{Medical devices}
    
    \addplot[
        color=darkgray,
        mark=triangle,
        line width=0.25mm,
        ]
        coordinates {
        (1,29)(2,77)(3,95)(4,97)(5,97.5)(6,100)
        };
        \addlegendentry{Browser}
    
    \addplot[
        color=ForestGreen,
        mark=*,
        line width=0.25mm,
        ]
        coordinates {
        (1,42)(2,70.5)(3,89)(4,96)(5,96.5)(6,100)
        };
        \addlegendentry{Server}
    
    \addplot[
        color=BurntOrange,
        mark=+,
        line width=0.25mm,
        ]
        coordinates {
        (1,69)(2,94)(3,99)(4,100)
        };
        \addlegendentry{NASA distributed database}
    
    \end{axis}
    \end{tikzpicture}
\caption{Error-detection rates for four-to six-way interactions between parameters \cite{pairwise}}
\label{pairwise_graph}
\end{figure}
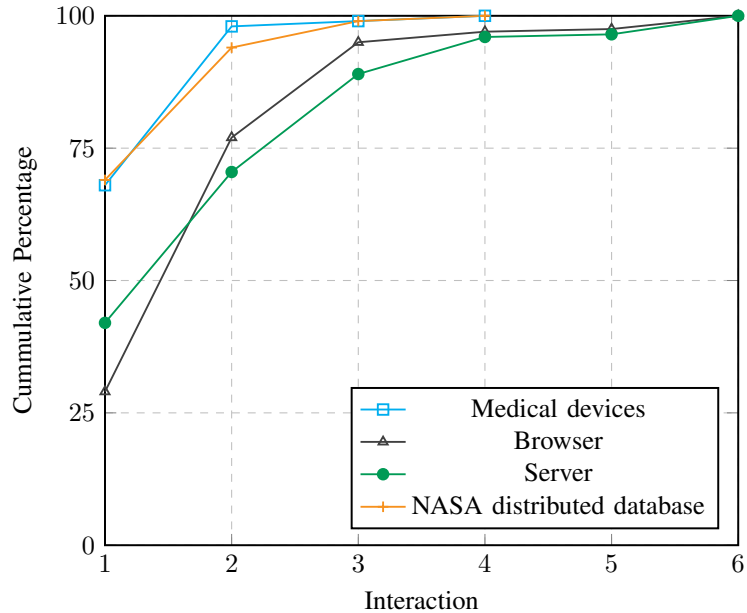

From Fig.~\ref{pairwise_graph}, it can be concluded that on average, \SI{67}{\percent} of failures are triggered by only a single parameter value, \SI{93}{\percent} or fewer failures are caused for 2-way pairs or pairwise interactions and \SI{98}{\percent} or fewer by 3-way combinations \cite{pairwise}. A fault triggered by a seven-way interaction has not yet appeared \cite{pairwise}. Pairwise is especially very effective when there are many parameters having a high cardinality \textit{C} (values that a parameter can take). For example, a manufacturing automation system that has 20 controls, each with 10 possible settings - a total of 10\textsuperscript{20} configurations are possible. In order to cover all such combinations, 10\textsuperscript{20} regressions with different configuration values need to be run. This is not realistic given the fact that modern simulators cannot handle such huge regressions for a \acrshort{DUV} and will take years to complete each one of them \cite{fvbook}. On the other hand, we can check all pairs of these values with only 180 regressions if they are carefully constructed \cite{pairwise}. To generalize, the number of regressions required in order to perform pairwise verification for \textit{\textbf{n}}-way combinatorial verification of \textit{\textbf{N}} parameters with \textit{\textbf{C}} values (cardinality) is proportional to $\boldsymbol{C}^{\boldsymbol{n}} \boldsymbol{\log} \boldsymbol{N}$ \cite{pairwise}.

\tikzset{every picture/.style={line width=0.75pt}} %set default line width to 0.75pt
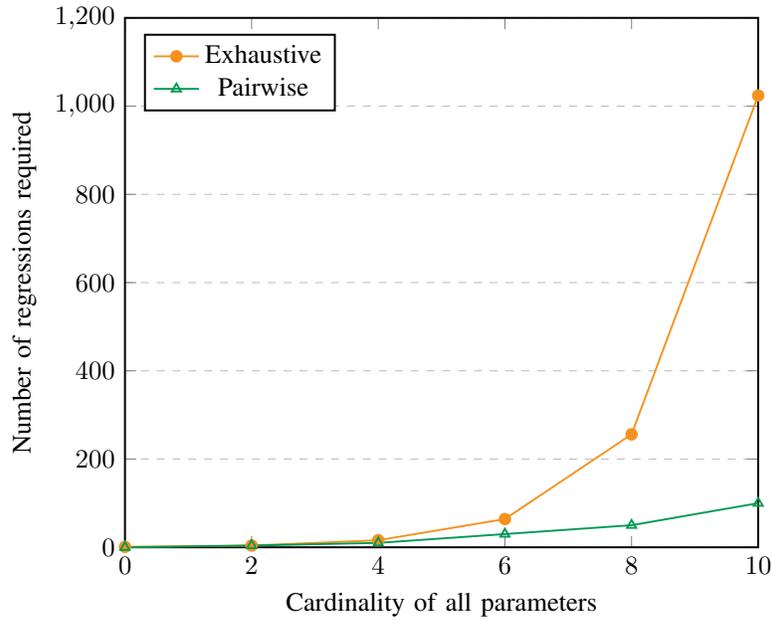
\begin{figure}[h]
\centering
\begin{tikzpicture}
\begin{axis}[
    xlabel={Cardinality of all parameters},
    ylabel={Number of regressions required},
    xmin=0, xmax=10,
    ymin=0, ymax=1200,
    xtick={0,2,4,6,8,10},
    ytick={0,200,400,600,800,1000,1200},
    legend pos=north west,
    ymajorgrids=true,
    grid style=dashed,
]
\addplot[
    color=BurntOrange,
    line width=0.25mm,
    mark=*,name path=A
    ]
    coordinates {
    (0,1)(2,4)(4,16)(6,64)(8,256)(10,1024)
    };
    \addlegendentry{Exhaustive}
\addplot[
    color=ForestGreen,
    line width=0.25mm,
    mark=triangle,name path=B
    ]
    coordinates {
    (0,0)(2,4)(4,10)(6,30)(8,50)(10,100)
    };
    \addlegendentry{Pairwise}
\end{axis}
\end{tikzpicture}
\caption{Increase in the number of exhaustive and Pairwise regressions with the cardinality of parameters \cite{pict}}
\label{exhaustive_pairwise}
\end{figure}

Fig.~\ref{exhaustive_pairwise} shows the comparison of the increase in the number of exhaustive and Pairwise regressions with the cardinality of parameters. In contrast to many combinatorial strategies that help testers maximize the probability of detecting defects such as random testing, n-wise testing and others, Pairwise proves to be the most prominent among these \cite{pict}.

Pairwise verification is argued to be the most effective method for uncovering potential bugs, and as we increase the \textit{n}-value, there are diminished benefits in terms of finding faults. Nevertheless, the cost of covering those is disproportionately high and often unlucky to find bugs in these \cite{pairwise_study}. Having said that, it should also be kept in mind that for a particular design, one might already achieve a \SI{100}{\percent} bug finding using Pairwise and performing a higher-order \textit{n}-value test would eventually end up being redundant and time-consuming. Nevertheless, as soon as we increase the value of \textit{n}, the number of regressions required in order to perform \textit{n}-wise verification also increases which is contradictory to the original idea that we want to decrease the number of tests or regressions required in order to unveil bugs without compromising with the verification quality. Therefore, a thoroughness of degree 2 (true Pairwise) is considered to be an acceptable trade-off and balance between the thoroughness and the effort involved in finding the potential bug in a \acrshort{DUV} \cite{pairwise_study}.

The fundamental principle behind Pairwise generation is: \say{For every pair of parameters, verify every combination of that pair} \cite{nwise_pkg}. A pilot design, as shown in Fig.~\ref{pilot_design} is used to explain this in details.

\tikzset{every picture/.style={line width=1pt}} %set default line width to 0.75pt        
\begin{figure}[h]
\centering
\begin{tikzpicture}[x=0.75pt,y=0.75pt,yscale=-1,xscale=1]
%uncomment if require: \path (0,434); %set diagram left start at 0, and has height of 434

%Rounded Rect [id:dp25858848817775804] 
\draw   (361,94.8) .. controls (361,76.13) and (376.13,61) .. (394.8,61) -- (606.7,61) .. controls (625.37,61) and (640.5,76.13) .. (640.5,94.8) -- (640.5,196.2) .. controls (640.5,214.87) and (625.37,230) .. (606.7,230) -- (394.8,230) .. controls (376.13,230) and (361,214.87) .. (361,196.2) -- cycle ;
%Rounded Rect [id:dp3823283320865618] 
\draw   (381,99.9) .. controls (381,90.01) and (389.01,82) .. (398.9,82) -- (452.6,82) .. controls (462.49,82) and (470.5,90.01) .. (470.5,99.9) -- (470.5,182.1) .. controls (470.5,191.99) and (462.49,200) .. (452.6,200) -- (398.9,200) .. controls (389.01,200) and (381,191.99) .. (381,182.1) -- cycle ;
%Rounded Rect [id:dp43556170762137825] 
\draw   (480,91.8) .. controls (480,86.39) and (484.39,82) .. (489.8,82) -- (531.7,82) .. controls (537.11,82) and (541.5,86.39) .. (541.5,91.8) -- (541.5,121.2) .. controls (541.5,126.61) and (537.11,131) .. (531.7,131) -- (489.8,131) .. controls (484.39,131) and (480,126.61) .. (480,121.2) -- cycle ;
%Rounded Rect [id:dp9954447415810141] 
\draw   (558.5,91.8) .. controls (558.5,86.39) and (562.89,82) .. (568.3,82) -- (610.2,82) .. controls (615.61,82) and (620,86.39) .. (620,91.8) -- (620,121.2) .. controls (620,126.61) and (615.61,131) .. (610.2,131) -- (568.3,131) .. controls (562.89,131) and (558.5,126.61) .. (558.5,121.2) -- cycle ;
%Rounded Rect [id:dp9465009854833324] 
\draw   (480,160.8) .. controls (480,155.39) and (484.39,151) .. (489.8,151) -- (531.7,151) .. controls (537.11,151) and (541.5,155.39) .. (541.5,160.8) -- (541.5,190.2) .. controls (541.5,195.61) and (537.11,200) .. (531.7,200) -- (489.8,200) .. controls (484.39,200) and (480,195.61) .. (480,190.2) -- cycle ;
%Rounded Rect [id:dp6487166611421722] 
\draw   (558.5,160.8) .. controls (558.5,155.39) and (562.89,151) .. (568.3,151) -- (610.2,151) .. controls (615.61,151) and (620,155.39) .. (620,160.8) -- (620,190.2) .. controls (620,195.61) and (615.61,200) .. (610.2,200) -- (568.3,200) .. controls (562.89,200) and (558.5,195.61) .. (558.5,190.2) -- cycle ;
%Right Arrow [id:dp8005775577517891] 
\draw   (320.5,191.95) -- (344.8,191.95) -- (344.8,187.7) -- (361,196.2) -- (344.8,204.7) -- (344.8,200.45) -- (320.5,200.45) -- cycle ;
%Right Arrow [id:dp28414202117664034] 
\draw   (320.5,90.55) -- (344.8,90.55) -- (344.8,86.3) -- (361,94.8) -- (344.8,103.3) -- (344.8,99.05) -- (320.5,99.05) -- cycle ;
%Right Arrow [id:dp05249516431521961] 
\draw   (320.5,115.55) -- (344.8,115.55) -- (344.8,111.3) -- (361,119.8) -- (344.8,128.3) -- (344.8,124.05) -- (320.5,124.05) -- cycle ;
%Right Arrow [id:dp6957159287591883] 
\draw   (320.5,140.55) -- (344.8,140.55) -- (344.8,136.3) -- (361,144.8) -- (344.8,153.3) -- (344.8,149.05) -- (320.5,149.05) -- cycle ;
%Right Arrow [id:dp7296812365441967] 
\draw   (320.5,165.8) -- (344.8,165.8) -- (344.8,161.55) -- (361,170.05) -- (344.8,178.55) -- (344.8,174.3) -- (320.5,174.3) -- cycle ;
%Shape: Brace [id:dp6476445940124969] 
\draw   (190.5,81) .. controls (185.83,81) and (183.5,83.33) .. (183.5,88) -- (183.5,135.5) .. controls (183.5,142.17) and (181.17,145.5) .. (176.5,145.5) .. controls (181.17,145.5) and (183.5,148.83) .. (183.5,155.5)(183.5,152.5) -- (183.5,203) .. controls (183.5,207.67) and (185.83,210) .. (190.5,210) ;

% Text Node
\draw (408,137) node [anchor=north west][inner sep=0.75pt]   [align=left] {FIFO};
% Text Node
\draw (498,90) node [anchor=north west][inner sep=0.75pt]   [align=left] {I2C};
% Text Node
\draw (489,108) node [anchor=north west][inner sep=0.75pt]   [align=left] {Master};
% Text Node
\draw (577,90) node [anchor=north west][inner sep=0.75pt]   [align=left] {I2C};
% Text Node
\draw (498,158) node [anchor=north west][inner sep=0.75pt]   [align=left] {I2C};
% Text Node
\draw (577,158) node [anchor=north west][inner sep=0.75pt]   [align=left] {I2C};
% Text Node
\draw (489,178) node [anchor=north west][inner sep=0.75pt]   [align=left] {Master};
% Text Node
\draw (572,109) node [anchor=north west][inner sep=0.75pt]   [align=left] {Slave};
% Text Node
\draw (572,178) node [anchor=north west][inner sep=0.75pt]   [align=left] {Slave};
% Text Node
\draw (381,207) node [anchor=north west][inner sep=0.75pt]   [align=left] {Top-Level};
% Text Node
\draw (191,85) node [anchor=north west][inner sep=0.75pt]   [align=left] {Num\_Master [1:0]};
% Text Node
\draw (199,109) node [anchor=north west][inner sep=0.75pt]   [align=left] {Num\_Slave [1:0]};
% Text Node
\draw (200,135) node [anchor=north west][inner sep=0.75pt]   [align=left] {Fifo\_Depth [1:0]};
% Text Node
\draw (229,161) node [anchor=north west][inner sep=0.75pt]   [align=left] {Fifo\_Width};
% Text Node
\draw (245,187) node [anchor=north west][inner sep=0.75pt]   [align=left] {Clk\_Div};
% Text Node
\draw (76,136) node [anchor=north west][inner sep=0.75pt]   [align=left] {Configurations};

\end{tikzpicture}
\caption{Pilot configurable design for Pairwise generation}
\label{pilot_design}
\end{figure}
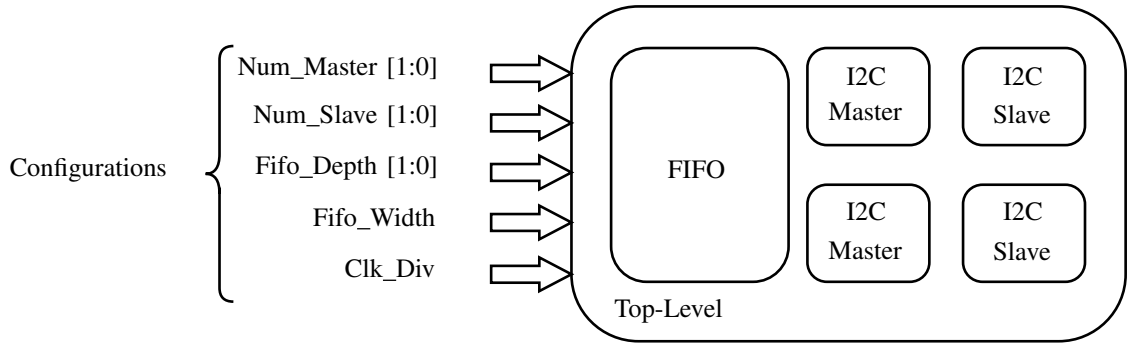

The pilot design has 6 parameters:
\begin{itemize}
    \item \textbf{P1}: Num\_Master (2 bits, 4 values)
    \item \textbf{P2}: Num\_Slave (2 bits, 4 values)
    \item \textbf{P3}: Fifo\_Depth (2 bits, 4 values)
    \item \textbf{P4}: Fifo\_Width (1 bit, 2 values)
    \item \textbf{P5}: Clk\_Div (1 bit, 2 values)
\end{itemize}

Depending upon the parameters mentioned above, the design can be configured into several states. In principle, there are a total of 256 combinations possible for the parameters. This means 256 times the simulation tests need to be run, or in other words, 256 regressions are required for the exhaustive verification. The first step for Pairwise generation is to identify the value-pairs $\boldsymbol{V}$ which are basically the possible values for the combination of two parameters. Table \ref{value_pairs} depicts the the value-pairs for the pilot design. To shorten the table, three constraints are considered for the parameters:
\begin{itemize}
    \item Num\_Master must be equal to Num\_Slave
    \item If Num\_Master is 3 then Clk\_Div should be 0
    \item If Num\_Master is 1 then Clk\_Div should be 1
\end{itemize}

\begin{table}[h]
\caption{Value-pairs for the pilot design}
\begin{center}
%\resizebox{\textwidth}{!}{%
\setlength{\arrayrulewidth}{1pt}
%\begin{adjustbox}{max width=\textwidth}
\begin{tabular}{|p{0.15\textwidth}|p{0.15\textwidth}|p{0.15\textwidth}|p{0.15\textwidth}|p{0.15\textwidth}|}
\hline 
 \cellcolor{black!20}{\textbf{\textcolor{white}{P1xP1}}} & \cellcolor{black!20}P2xP1 & \cellcolor{black!20}P3xP1 & \cellcolor{black!20}P4xP1 & \cellcolor{black!20}P5xP1 \\
\hline 
 P1xP2 \hfill \textbf{4} & \cellcolor{black!20}{\textbf{\textcolor{white}{P2xP2}}} & \cellcolor{black!20}P3xP2 & \cellcolor{black!20}P4xP2 & \cellcolor{black!20}P5xP2 \\
\hline 
 P1xP3 \hfill \textbf{16} & P2xP3 \hfill \textbf{16} & \cellcolor{black!20}{\textbf{\textcolor{white}{P3xP3}}} & \cellcolor{black!20}P4xP3 & \cellcolor{black!20}P5xP3 \\
\hline 
 P1xP4 \hfill \textbf{8} & P2xP4 \hfill \textbf{8} & P3xP4 \hfill \textbf{8} & \cellcolor{black!20}{\textbf{\textcolor{white}{P4xP4}}} & \cellcolor{black!20}P5xP4 \\
\hline 
 P1xP5 \hfill \textbf{6} & P2xP5 \hfill \textbf{6} & P3xP5 \hfill \textbf{8} & P4xP5 \hfill \textbf{4} & \cellcolor{black!20}{\textbf{\textcolor{white}{P5xP5}}} \\
\hline
\end{tabular}
\label{value_pairs}
\end{center}
\end{table}

Table \ref{value_pairs} shows in white cells that there are a total of 86 value-pairs that needs to be verified. As an example, P2 and P3 have 16 combinations possible. Similarly, P4 and P5 have 4 combinations possible. They all add up to the value of 86. The next step is to generate the minimum number of regressions that constitute all the value-pairs efficiently. For the pilot design, the minimum number of regressions to cover all value-pairs is 16, as shown in Table \ref{regression_table}.

\renewcommand{\arraystretch}{1.5}
\begin{table}[h]
\setlength\arrayrulewidth{0.75pt}
\caption{Regression generated covering all value-pairs for the pilot design}
\centering
\setlength{\tabcolsep}{10pt}
 \begin{tabular}{c c c c c}
 \hline
 \textbf{P1} & \textbf{P2} & \textbf{P3} & \textbf{P4} & \textbf{P5} \\
 \hline\hline
 0 & 0 & 0 & \cellcolor{ForestGreen}{\textbf{\textcolor{white}{0}}} & \cellcolor{ForestGreen}{\textbf{\textcolor{white}{0}}} \\
 \hline
 1 & 1 & 0 & \cellcolor{ForestGreen}{\textbf{\textcolor{white}{1}}} &\cellcolor{ForestGreen}{\textbf{\textcolor{white}{1}}} \\
 \hline
 2 & 2 & 0 & \cellcolor{ForestGreen}{\textbf{\textcolor{white}{1}}} & \cellcolor{ForestGreen}{\textbf{\textcolor{white}{0}}} \\
 \hline
 3 & 3 & 0 & 1 & 0 \\
 \hline
 0 & 0 & 1 & 1 & 1 \\
 \hline
 1 & 1 & 1 & \cellcolor{ForestGreen}{\textbf{\textcolor{white}{0}}} & \cellcolor{ForestGreen}{\textbf{\textcolor{white}{1}}} \\
 \hline
 \cellcolor{BurntOrange}{\textbf{\textcolor{white}{2}}} & 2 & 1 & 0 & \cellcolor{BurntOrange}{\textbf{\textcolor{white}{1}}} \\
 \hline
 3 & 3 & 1 & 0 & 0 \\
 \hline
 \cellcolor{BurntOrange}{\textbf{\textcolor{white}{0}}} & 0 & 2 & 0 & \cellcolor{BurntOrange}{\textbf{\textcolor{white}{1}}} \\
 \hline
 \cellcolor{BurntOrange}{\textbf{\textcolor{white}{1}}} & 1 & 2 & 1 & \cellcolor{BurntOrange}{\textbf{\textcolor{white}{1}}} \\
 \hline
 \cellcolor{BurntOrange}{\textbf{\textcolor{white}{2}}} & 2 & 2 & 0 & \cellcolor{BurntOrange}{\textbf{\textcolor{white}{0}}} \\
 \hline
 3 & 3 & 2 & 1 & 0 \\
 \hline
 \cellcolor{BurntOrange}{\textbf{\textcolor{white}{0}}} & 0 & 3 & 1 & \cellcolor{BurntOrange}{\textbf{\textcolor{white}{0}}} \\
 \hline
 1 & 1 & 3 & 0 & 1 \\
 \hline
 2 & 2 & 3 & 0 & 0 \\
 \hline
 \cellcolor{BurntOrange}{\textbf{\textcolor{white}{3}}} & 3 & 3 & 0 & \cellcolor{BurntOrange}{\textbf{\textcolor{white}{0}}} \\
 \hline
\end{tabular}
\label{regression_table}
\end{table}

It can be depicted from Table \ref{regression_table} that all the 86 value-pairs are covered in just 16 regressions. For example, there are four value-pairs for P4xP5, and all are covered as highlighted in \textcolor{ForestGreen}{green} in the table. Similarly, all 6 value-pairs for P1xP5 are covered as highlighted in \textcolor{BurntOrange}{orange}. Consequently, other value-pairs are also covered in just 16 regressions.

As mentioned above, \acrshort{PICT} uses an input file, also referred to as the template file in the methodology. The template file needs to follow the syntax of the tool for defining parameters and the constraints, if necessary. Listing \ref{pairwise_template} shows an example of the template that is generated automatically after extracting the configuration information.
%\vspace{0.25cm}
%\FloatBarrier
\vspace{0.25cm}
\begin{lstlisting}[language=tcl, caption=Example of generated Pairwise template (in \acrshort{PICT} syntax) used as an input to the \acrshort{PICT} tool, numbers=left, label={pairwise_template}]
##=======================================
## Parameters and their possible values
##=======================================
Num_Master: 0,1,2,3
Num_Slave:  0,1,2,3
Fifo_Depth: 0,1,2,3
Fifo_Width: 0,1
Clk_Div:    0,1

##=======================================
## Constraints for the parameters
##=======================================
[Num_Master] = [Num_Slave];
IF [Num_Master] = 3 THEN [Clk_Div] = 0;
IF [Num_Master] = 1 THEN [Clk_Div] = 1;
\end{lstlisting}

\section{ConfigVerMet: Extract}
The prepare step introduced Pairwise verification, which is used to generate the regressions. However, for highly configurable digital designs, Pairwise technique alone does not necessarily give a realistic number of regressions for complete verification. For example, a design having 6 parameters with 15 values (cardinality) would result in more than 300 Pairwise regressions. For this purpose, some more regression reduction techniques are introduced in the extract step. The extract step is divided into two parts: the top-down approach and the bottom-up approach.

The top-down approach is primarily influenced by the design specifications. These design specifications are generally coming from the \acrshort{IP} requirements or concept level at the beginning of \acrshort{SoC} development cycle. A State-of-the-Art requirements tool is used to specify constraints and Equivalence classes.

\subsection{Constraints}
Constraints are the conditions that apply to the parameters and help in limiting the Pairwise set. An example of such constraint is mentioned below:

\begin{equation}
\textrm{Address\_Width\_2} < \textrm{Address\_Width\_1}
\end{equation}

The constraint mentioned above depicts that the value of parameter Address\_Width\_2 must always be less than the value of the parameter Address\_Width\_1. These constraints are mentioned in the design specifications and are extracted automatically from the \acrfull{XML} files. The \acrshort{XML} file is located in the requirements tool. The extracted constraints are then added to the Pairwise template for the Pairwise generation.

Another regression reduction technique, known as Equivalence class verification, is also implemented in the top-down approach. The details of this approach are discussed in the subsequent section. Once the Equivalence classes are identified and specified in the requirements tool, they are automatically extracted from the \acrshort{XML} file and added to the Pairwise template.

\subsection{Equivalence Class Verification}
Equivalence class verification is a regression reduction technique, often used in software testing to identify the most efficient combinations of parameters that expose potential bugs. The fundamental principle behind this method is the formation of Equivalence classes. Input or output data is grouped or partitioned into sets of data that we expect to behave the same during functional verification using an Equivalence relation \cite{eq_class}. An Equivalence relation is a relation between the data of a set that make them behave similarly. An example of an Equivalence relation is illustrated below:

\begin{equation}
\textrm{Memory Access} = \textrm{Direct } \forall \textrm{ P1} \in \textrm{\{}1,2,...,16\textrm{\}}
\end{equation}

The Equivalence relation mentioned above depicts that the memory access is always direct for all values of the parameter P1 that belongs to the values 1 to 16.

\tikzset{every picture/.style={line width=1pt}} %set default line width to 0.75pt        
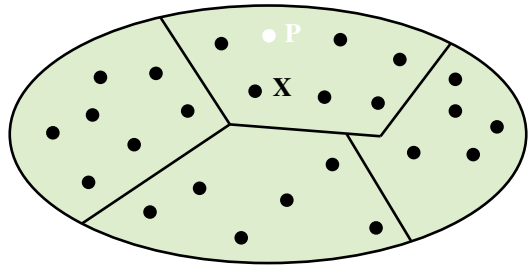
\begin{figure}[h]
\centering
\begin{tikzpicture}[x=0.75pt,y=0.75pt,yscale=-1,xscale=1]
%uncomment if require: \path (0,488); %set diagram left start at 0, and has height of 488

%Shape: Ellipse [id:dp20243212606513827] 
\draw  [fill={rgb, 255:red, 173; green, 212; blue, 131 }  ,fill opacity=0.4 ] (100,175) .. controls (100,139.1) and (158.31,110) .. (230.25,110) .. controls (302.19,110) and (360.5,139.1) .. (360.5,175) .. controls (360.5,210.9) and (302.19,240) .. (230.25,240) .. controls (158.31,240) and (100,210.9) .. (100,175) -- cycle ;
%Straight Lines [id:da5823944296618098] 
\draw    (176,116) -- (211,170) ;
%Straight Lines [id:da041013338437227986] 
\draw    (211,170) -- (287,176) ;
%Straight Lines [id:da8914398568311872] 
\draw    (287,176) -- (322.5,129) ;
%Straight Lines [id:da9556971787254493] 
\draw    (270,175) -- (302.5,229) ;
%Straight Lines [id:da28237732571441376] 
\draw    (211,170) -- (136.5,220) ;
%Shape: Circle [id:dp17592098183198668] 
\draw  [color={rgb, 255:red, 255; green, 255; blue, 255 }  ,draw opacity=1 ][fill={rgb, 255:red, 255; green, 255; blue, 255 }  ,fill opacity=1 ] (233.5,125.25) .. controls (233.5,123.73) and (232.27,122.5) .. (230.75,122.5) .. controls (229.23,122.5) and (228,123.73) .. (228,125.25) .. controls (228,126.77) and (229.23,128) .. (230.75,128) .. controls (232.27,128) and (233.5,126.77) .. (233.5,125.25) -- cycle ;
%Shape: Circle [id:dp31013131049455733] 
\draw  [fill={rgb, 255:red, 0; green, 0; blue, 0 }  ,fill opacity=1 ] (209.5,129.25) .. controls (209.5,127.73) and (208.27,126.5) .. (206.75,126.5) .. controls (205.23,126.5) and (204,127.73) .. (204,129.25) .. controls (204,130.77) and (205.23,132) .. (206.75,132) .. controls (208.27,132) and (209.5,130.77) .. (209.5,129.25) -- cycle ;
%Shape: Circle [id:dp6974921212464122] 
\draw  [fill={rgb, 255:red, 0; green, 0; blue, 0 }  ,fill opacity=1 ] (226.5,153.25) .. controls (226.5,151.73) and (225.27,150.5) .. (223.75,150.5) .. controls (222.23,150.5) and (221,151.73) .. (221,153.25) .. controls (221,154.77) and (222.23,156) .. (223.75,156) .. controls (225.27,156) and (226.5,154.77) .. (226.5,153.25) -- cycle ;
%Shape: Circle [id:dp9008736551399286] 
\draw  [fill={rgb, 255:red, 0; green, 0; blue, 0 }  ,fill opacity=1 ] (261.5,156.25) .. controls (261.5,154.73) and (260.27,153.5) .. (258.75,153.5) .. controls (257.23,153.5) and (256,154.73) .. (256,156.25) .. controls (256,157.77) and (257.23,159) .. (258.75,159) .. controls (260.27,159) and (261.5,157.77) .. (261.5,156.25) -- cycle ;
%Shape: Circle [id:dp6228182626556256] 
\draw  [fill={rgb, 255:red, 0; green, 0; blue, 0 }  ,fill opacity=1 ] (269.5,127.25) .. controls (269.5,125.73) and (268.27,124.5) .. (266.75,124.5) .. controls (265.23,124.5) and (264,125.73) .. (264,127.25) .. controls (264,128.77) and (265.23,130) .. (266.75,130) .. controls (268.27,130) and (269.5,128.77) .. (269.5,127.25) -- cycle ;
%Shape: Circle [id:dp5479890294434486] 
\draw  [fill={rgb, 255:red, 0; green, 0; blue, 0 }  ,fill opacity=1 ] (299.5,137.25) .. controls (299.5,135.73) and (298.27,134.5) .. (296.75,134.5) .. controls (295.23,134.5) and (294,135.73) .. (294,137.25) .. controls (294,138.77) and (295.23,140) .. (296.75,140) .. controls (298.27,140) and (299.5,138.77) .. (299.5,137.25) -- cycle ;
%Shape: Circle [id:dp3089327492109597] 
\draw  [fill={rgb, 255:red, 0; green, 0; blue, 0 }  ,fill opacity=1 ] (288.5,159.25) .. controls (288.5,157.73) and (287.27,156.5) .. (285.75,156.5) .. controls (284.23,156.5) and (283,157.73) .. (283,159.25) .. controls (283,160.77) and (284.23,162) .. (285.75,162) .. controls (287.27,162) and (288.5,160.77) .. (288.5,159.25) -- cycle ;
%Shape: Circle [id:dp9886276213844347] 
\draw  [fill={rgb, 255:red, 0; green, 0; blue, 0 }  ,fill opacity=1 ] (327.5,147.25) .. controls (327.5,145.73) and (326.27,144.5) .. (324.75,144.5) .. controls (323.23,144.5) and (322,145.73) .. (322,147.25) .. controls (322,148.77) and (323.23,150) .. (324.75,150) .. controls (326.27,150) and (327.5,148.77) .. (327.5,147.25) -- cycle ;
%Shape: Circle [id:dp9143497170188142] 
\draw  [fill={rgb, 255:red, 0; green, 0; blue, 0 }  ,fill opacity=1 ] (348.5,171.25) .. controls (348.5,169.73) and (347.27,168.5) .. (345.75,168.5) .. controls (344.23,168.5) and (343,169.73) .. (343,171.25) .. controls (343,172.77) and (344.23,174) .. (345.75,174) .. controls (347.27,174) and (348.5,172.77) .. (348.5,171.25) -- cycle ;
%Shape: Circle [id:dp7638081018323446] 
\draw  [fill={rgb, 255:red, 0; green, 0; blue, 0 }  ,fill opacity=1 ] (327.5,163.25) .. controls (327.5,161.73) and (326.27,160.5) .. (324.75,160.5) .. controls (323.23,160.5) and (322,161.73) .. (322,163.25) .. controls (322,164.77) and (323.23,166) .. (324.75,166) .. controls (326.27,166) and (327.5,164.77) .. (327.5,163.25) -- cycle ;
%Shape: Circle [id:dp6339529974614517] 
\draw  [fill={rgb, 255:red, 0; green, 0; blue, 0 }  ,fill opacity=1 ] (336.5,185.25) .. controls (336.5,183.73) and (335.27,182.5) .. (333.75,182.5) .. controls (332.23,182.5) and (331,183.73) .. (331,185.25) .. controls (331,186.77) and (332.23,188) .. (333.75,188) .. controls (335.27,188) and (336.5,186.77) .. (336.5,185.25) -- cycle ;
%Shape: Circle [id:dp6268513918650771] 
\draw  [fill={rgb, 255:red, 0; green, 0; blue, 0 }  ,fill opacity=1 ] (306.5,184.25) .. controls (306.5,182.73) and (305.27,181.5) .. (303.75,181.5) .. controls (302.23,181.5) and (301,182.73) .. (301,184.25) .. controls (301,185.77) and (302.23,187) .. (303.75,187) .. controls (305.27,187) and (306.5,185.77) .. (306.5,184.25) -- cycle ;
%Shape: Circle [id:dp34278707202126] 
\draw  [fill={rgb, 255:red, 0; green, 0; blue, 0 }  ,fill opacity=1 ] (265.5,190.25) .. controls (265.5,188.73) and (264.27,187.5) .. (262.75,187.5) .. controls (261.23,187.5) and (260,188.73) .. (260,190.25) .. controls (260,191.77) and (261.23,193) .. (262.75,193) .. controls (264.27,193) and (265.5,191.77) .. (265.5,190.25) -- cycle ;
%Shape: Circle [id:dp892285008405513] 
\draw  [fill={rgb, 255:red, 0; green, 0; blue, 0 }  ,fill opacity=1 ] (173.5,214.25) .. controls (173.5,212.73) and (172.27,211.5) .. (170.75,211.5) .. controls (169.23,211.5) and (168,212.73) .. (168,214.25) .. controls (168,215.77) and (169.23,217) .. (170.75,217) .. controls (172.27,217) and (173.5,215.77) .. (173.5,214.25) -- cycle ;
%Shape: Circle [id:dp30952600116655726] 
\draw  [fill={rgb, 255:red, 0; green, 0; blue, 0 }  ,fill opacity=1 ] (198.5,202.25) .. controls (198.5,200.73) and (197.27,199.5) .. (195.75,199.5) .. controls (194.23,199.5) and (193,200.73) .. (193,202.25) .. controls (193,203.77) and (194.23,205) .. (195.75,205) .. controls (197.27,205) and (198.5,203.77) .. (198.5,202.25) -- cycle ;
%Shape: Circle [id:dp6241009278241503] 
\draw  [fill={rgb, 255:red, 0; green, 0; blue, 0 }  ,fill opacity=1 ] (219.5,227.25) .. controls (219.5,225.73) and (218.27,224.5) .. (216.75,224.5) .. controls (215.23,224.5) and (214,225.73) .. (214,227.25) .. controls (214,228.77) and (215.23,230) .. (216.75,230) .. controls (218.27,230) and (219.5,228.77) .. (219.5,227.25) -- cycle ;
%Shape: Circle [id:dp8765289934541025] 
\draw  [fill={rgb, 255:red, 0; green, 0; blue, 0 }  ,fill opacity=1 ] (287.5,222.25) .. controls (287.5,220.73) and (286.27,219.5) .. (284.75,219.5) .. controls (283.23,219.5) and (282,220.73) .. (282,222.25) .. controls (282,223.77) and (283.23,225) .. (284.75,225) .. controls (286.27,225) and (287.5,223.77) .. (287.5,222.25) -- cycle ;
%Shape: Circle [id:dp02068179190017272] 
\draw  [fill={rgb, 255:red, 0; green, 0; blue, 0 }  ,fill opacity=1 ] (242.5,208.25) .. controls (242.5,206.73) and (241.27,205.5) .. (239.75,205.5) .. controls (238.23,205.5) and (237,206.73) .. (237,208.25) .. controls (237,209.77) and (238.23,211) .. (239.75,211) .. controls (241.27,211) and (242.5,209.77) .. (242.5,208.25) -- cycle ;
%Shape: Circle [id:dp4267328105363608] 
\draw  [fill={rgb, 255:red, 0; green, 0; blue, 0 }  ,fill opacity=1 ] (192.5,163.25) .. controls (192.5,161.73) and (191.27,160.5) .. (189.75,160.5) .. controls (188.23,160.5) and (187,161.73) .. (187,163.25) .. controls (187,164.77) and (188.23,166) .. (189.75,166) .. controls (191.27,166) and (192.5,164.77) .. (192.5,163.25) -- cycle ;
%Shape: Circle [id:dp18282058458076222] 
\draw  [fill={rgb, 255:red, 0; green, 0; blue, 0 }  ,fill opacity=1 ] (176.5,144.25) .. controls (176.5,142.73) and (175.27,141.5) .. (173.75,141.5) .. controls (172.23,141.5) and (171,142.73) .. (171,144.25) .. controls (171,145.77) and (172.23,147) .. (173.75,147) .. controls (175.27,147) and (176.5,145.77) .. (176.5,144.25) -- cycle ;
%Shape: Circle [id:dp18737035904913113] 
\draw  [fill={rgb, 255:red, 0; green, 0; blue, 0 }  ,fill opacity=1 ] (142.5,199.25) .. controls (142.5,197.73) and (141.27,196.5) .. (139.75,196.5) .. controls (138.23,196.5) and (137,197.73) .. (137,199.25) .. controls (137,200.77) and (138.23,202) .. (139.75,202) .. controls (141.27,202) and (142.5,200.77) .. (142.5,199.25) -- cycle ;
%Shape: Circle [id:dp8218871791240683] 
\draw  [fill={rgb, 255:red, 0; green, 0; blue, 0 }  ,fill opacity=1 ] (165.5,180.25) .. controls (165.5,178.73) and (164.27,177.5) .. (162.75,177.5) .. controls (161.23,177.5) and (160,178.73) .. (160,180.25) .. controls (160,181.77) and (161.23,183) .. (162.75,183) .. controls (164.27,183) and (165.5,181.77) .. (165.5,180.25) -- cycle ;
%Shape: Circle [id:dp44071802146706296] 
\draw  [fill={rgb, 255:red, 0; green, 0; blue, 0 }  ,fill opacity=1 ] (148.5,146.25) .. controls (148.5,144.73) and (147.27,143.5) .. (145.75,143.5) .. controls (144.23,143.5) and (143,144.73) .. (143,146.25) .. controls (143,147.77) and (144.23,149) .. (145.75,149) .. controls (147.27,149) and (148.5,147.77) .. (148.5,146.25) -- cycle ;
%Shape: Circle [id:dp5789899152118303] 
\draw  [fill={rgb, 255:red, 0; green, 0; blue, 0 }  ,fill opacity=1 ] (124.5,174.25) .. controls (124.5,172.73) and (123.27,171.5) .. (121.75,171.5) .. controls (120.23,171.5) and (119,172.73) .. (119,174.25) .. controls (119,175.77) and (120.23,177) .. (121.75,177) .. controls (123.27,177) and (124.5,175.77) .. (124.5,174.25) -- cycle ;
%Shape: Circle [id:dp6730096934891217] 
\draw  [fill={rgb, 255:red, 0; green, 0; blue, 0 }  ,fill opacity=1 ] (144.5,165.25) .. controls (144.5,163.73) and (143.27,162.5) .. (141.75,162.5) .. controls (140.23,162.5) and (139,163.73) .. (139,165.25) .. controls (139,166.77) and (140.23,168) .. (141.75,168) .. controls (143.27,168) and (144.5,166.77) .. (144.5,165.25) -- cycle ;

% Text Node
\draw (231,145) node [anchor=north west][inner sep=0.75pt]   [align=left] {\textbf{X}};
% Text Node
\draw (237,118) node [anchor=north west][inner sep=0.75pt]  [color={rgb, 255:red, 255; green, 255; blue, 255 }  ,opacity=1 ] [align=left] {\textbf{P}};

\end{tikzpicture}
\caption{Equivalence classes \cite{eq_class}}
\label{eq_classes_fig}
\end{figure}

Fig.~\ref{eq_classes_fig} represents the set of data (parameter values in our case) that the user wants to use to produce the regressions. The data is split up into subsets, each subset containing some of the original data. It can be depicted from the figure that the whole set is being included for the partitioning ensuring notion of completeness and the disjointness between subsets ensures non-redundancy \cite{eq_class}. The subsets created are derived from the type of Equivalence relations used. Since an Equivalence relation is used to split the data, a safe assumption can be made that all data within a subset behave similarly. This means the result of a test case using point \textbf{P} will be the same as all the points \textbf{X\textsubscript{1}} to \textbf{X\textsubscript{n}} within the subset \cite{eq_class}. This property of Equivalence classes helps to reduce the redundancy and decrease the number of regressions significantly.

Equivalence class verification can be divided into four major types:

\begin{figure}[h]
  \centering
  \begin{minipage}[b]{0.4\textwidth}
    \begin{tikzpicture}[x=0.75pt,y=0.75pt,yscale=-1,xscale=1]
%uncomment if require: \path (0,478); %set diagram left start at 0, and has height of 478

%Shape: Rectangle [id:dp3195416905530526] 
\draw  [color={rgb, 255:red, 255; green, 255; blue, 255 }  ,draw opacity=1 ][fill={rgb, 255:red, 74; green, 144; blue, 226 }  ,fill opacity=0.2 ] (260.5,179) -- (421.5,179) -- (421.5,261) -- (260.5,261) -- cycle ;
%Shape: Axis 2D [id:dp5864684608603756] 
\draw  (203,292) -- (470.5,292)(229.75,130) -- (229.75,310) (463.5,287) -- (470.5,292) -- (463.5,297) (224.75,137) -- (229.75,130) -- (234.75,137)  ;
%Straight Lines [id:da9731062473951577] 
\draw  [dash pattern={on 4.5pt off 4.5pt}]  (261,160) -- (261,300) ;
%Straight Lines [id:da33822733683394923] 
\draw  [dash pattern={on 4.5pt off 4.5pt}]  (301,160) -- (301,300) ;
%Straight Lines [id:da1448268711646643] 
\draw  [dash pattern={on 4.5pt off 4.5pt}]  (381,160) -- (381,300) ;
%Straight Lines [id:da09272657255037209] 
\draw  [dash pattern={on 4.5pt off 4.5pt}]  (421,160) -- (421,300) ;
%Straight Lines [id:da8541650216578274] 
\draw  [dash pattern={on 4.5pt off 4.5pt}]  (221,260) -- (440.5,260) ;
%Straight Lines [id:da2202348799891225] 
\draw  [dash pattern={on 4.5pt off 4.5pt}]  (221,220) -- (440.5,220) ;
%Straight Lines [id:da042695582029246504] 
\draw  [dash pattern={on 4.5pt off 4.5pt}]  (221,180) -- (440.5,180) ;
%Shape: Circle [id:dp15380583974328843] 
\draw  [fill={rgb, 255:red, 0; green, 0; blue, 0 }  ,fill opacity=1 ] (337,201) .. controls (337,198.79) and (338.79,197) .. (341,197) .. controls (343.21,197) and (345,198.79) .. (345,201) .. controls (345,203.21) and (343.21,205) .. (341,205) .. controls (338.79,205) and (337,203.21) .. (337,201) -- cycle ;
%Shape: Circle [id:dp7413461815699502] 
\draw  [fill={rgb, 255:red, 0; green, 0; blue, 0 }  ,fill opacity=1 ] (277,240) .. controls (277,237.79) and (278.79,236) .. (281,236) .. controls (283.21,236) and (285,237.79) .. (285,240) .. controls (285,242.21) and (283.21,244) .. (281,244) .. controls (278.79,244) and (277,242.21) .. (277,240) -- cycle ;
%Shape: Circle [id:dp9894269508256563] 
\draw  [fill={rgb, 255:red, 0; green, 0; blue, 0 }  ,fill opacity=1 ] (397,240) .. controls (397,237.79) and (398.79,236) .. (401,236) .. controls (403.21,236) and (405,237.79) .. (405,240) .. controls (405,242.21) and (403.21,244) .. (401,244) .. controls (398.79,244) and (397,242.21) .. (397,240) -- cycle ;

% Text Node
\draw (256,303) node [anchor=north west][inner sep=0.75pt]   [align=left] {a};
% Text Node
\draw (296,301) node [anchor=north west][inner sep=0.75pt]   [align=left] {b};
% Text Node
\draw (376,303) node [anchor=north west][inner sep=0.75pt]   [align=left] {c};
% Text Node
\draw (415,301) node [anchor=north west][inner sep=0.75pt]   [align=left] {d};
% Text Node
\draw (207,175) node [anchor=north west][inner sep=0.75pt]   [align=left] {e};
% Text Node
\draw (209,212) node [anchor=north west][inner sep=0.75pt]   [align=left] {f};
% Text Node
\draw (207,253) node [anchor=north west][inner sep=0.75pt]   [align=left] {g};
% Text Node
\draw (459,302) node [anchor=north west][inner sep=0.75pt]   [align=left] {X};
% Text Node
\draw (208,129) node [anchor=north west][inner sep=0.75pt]   [align=left] {Y};

\end{tikzpicture}
\caption{Weak normal Equivalence class verification \cite{eq_class}}
\label{weak_normal}
  \end{minipage}
  \hspace{2cm}
  \begin{minipage}[b]{0.4\textwidth}
    \begin{tikzpicture}[x=0.75pt,y=0.75pt,yscale=-1,xscale=1]
%uncomment if require: \path (0,478); %set diagram left start at 0, and has height of 478

%Shape: Rectangle [id:dp3195416905530526] 
\draw  [color={rgb, 255:red, 255; green, 255; blue, 255 }  ,draw opacity=1 ][fill={rgb, 255:red, 74; green, 144; blue, 226 }  ,fill opacity=0.2 ] (260.5,179) -- (421.5,179) -- (421.5,261) -- (260.5,261) -- cycle ;
%Shape: Axis 2D [id:dp5864684608603756] 
\draw  (203,292) -- (470.5,292)(229.75,130) -- (229.75,310) (463.5,287) -- (470.5,292) -- (463.5,297) (224.75,137) -- (229.75,130) -- (234.75,137)  ;
%Straight Lines [id:da9731062473951577] 
\draw  [dash pattern={on 4.5pt off 4.5pt}]  (261,160) -- (261,300) ;
%Straight Lines [id:da33822733683394923] 
\draw  [dash pattern={on 4.5pt off 4.5pt}]  (301,160) -- (301,300) ;
%Straight Lines [id:da1448268711646643] 
\draw  [dash pattern={on 4.5pt off 4.5pt}]  (381,160) -- (381,300) ;
%Straight Lines [id:da09272657255037209] 
\draw  [dash pattern={on 4.5pt off 4.5pt}]  (421,160) -- (421,300) ;
%Straight Lines [id:da8541650216578274] 
\draw  [dash pattern={on 4.5pt off 4.5pt}]  (221,260) -- (440.5,260) ;
%Straight Lines [id:da2202348799891225] 
\draw  [dash pattern={on 4.5pt off 4.5pt}]  (221,220) -- (440.5,220) ;
%Straight Lines [id:da042695582029246504] 
\draw  [dash pattern={on 4.5pt off 4.5pt}]  (221,180) -- (440.5,180) ;
%Shape: Circle [id:dp15380583974328843] 
\draw  [fill={rgb, 255:red, 0; green, 0; blue, 0 }  ,fill opacity=1 ] (337,201) .. controls (337,198.79) and (338.79,197) .. (341,197) .. controls (343.21,197) and (345,198.79) .. (345,201) .. controls (345,203.21) and (343.21,205) .. (341,205) .. controls (338.79,205) and (337,203.21) .. (337,201) -- cycle ;
%Shape: Circle [id:dp7413461815699502] 
\draw  [fill={rgb, 255:red, 0; green, 0; blue, 0 }  ,fill opacity=1 ] (277,240) .. controls (277,237.79) and (278.79,236) .. (281,236) .. controls (283.21,236) and (285,237.79) .. (285,240) .. controls (285,242.21) and (283.21,244) .. (281,244) .. controls (278.79,244) and (277,242.21) .. (277,240) -- cycle ;
%Shape: Circle [id:dp9894269508256563] 
\draw  [fill={rgb, 255:red, 0; green, 0; blue, 0 }  ,fill opacity=1 ] (397,240) .. controls (397,237.79) and (398.79,236) .. (401,236) .. controls (403.21,236) and (405,237.79) .. (405,240) .. controls (405,242.21) and (403.21,244) .. (401,244) .. controls (398.79,244) and (397,242.21) .. (397,240) -- cycle ;
%Shape: Circle [id:dp47052494926194277] 
\draw  [fill={rgb, 255:red, 0; green, 0; blue, 0 }  ,fill opacity=1 ] (337,240) .. controls (337,237.79) and (338.79,236) .. (341,236) .. controls (343.21,236) and (345,237.79) .. (345,240) .. controls (345,242.21) and (343.21,244) .. (341,244) .. controls (338.79,244) and (337,242.21) .. (337,240) -- cycle ;
%Shape: Circle [id:dp6308723847944224] 
\draw  [fill={rgb, 255:red, 0; green, 0; blue, 0 }  ,fill opacity=1 ] (397,201) .. controls (397,198.79) and (398.79,197) .. (401,197) .. controls (403.21,197) and (405,198.79) .. (405,201) .. controls (405,203.21) and (403.21,205) .. (401,205) .. controls (398.79,205) and (397,203.21) .. (397,201) -- cycle ;
%Shape: Circle [id:dp7930249785945129] 
\draw  [fill={rgb, 255:red, 0; green, 0; blue, 0 }  ,fill opacity=1 ] (277,201) .. controls (277,198.79) and (278.79,197) .. (281,197) .. controls (283.21,197) and (285,198.79) .. (285,201) .. controls (285,203.21) and (283.21,205) .. (281,205) .. controls (278.79,205) and (277,203.21) .. (277,201) -- cycle ;

% Text Node
\draw (256,303) node [anchor=north west][inner sep=0.75pt]   [align=left] {a};
% Text Node
\draw (296,301) node [anchor=north west][inner sep=0.75pt]   [align=left] {b};
% Text Node
\draw (376,303) node [anchor=north west][inner sep=0.75pt]   [align=left] {c};
% Text Node
\draw (415,301) node [anchor=north west][inner sep=0.75pt]   [align=left] {d};
% Text Node
\draw (207,175) node [anchor=north west][inner sep=0.75pt]   [align=left] {e};
% Text Node
\draw (209,212) node [anchor=north west][inner sep=0.75pt]   [align=left] {f};
% Text Node
\draw (207,253) node [anchor=north west][inner sep=0.75pt]   [align=left] {g};
% Text Node
\draw (459,302) node [anchor=north west][inner sep=0.75pt]   [align=left] {X};
% Text Node
\draw (208,129) node [anchor=north west][inner sep=0.75pt]   [align=left] {Y};

\end{tikzpicture}
\caption{Strong normal Equivalence class verification \cite{eq_class}}
\label{strong_normal}
  \end{minipage}
\end{figure}

\begin{figure}[h]
  \centering
  \begin{minipage}[b]{0.4\textwidth}
    \begin{tikzpicture}[x=0.75pt,y=0.75pt,yscale=-1,xscale=1]
%uncomment if require: \path (0,478); %set diagram left start at 0, and has height of 478

%Shape: Rectangle [id:dp3195416905530526] 
\draw  [color={rgb, 255:red, 255; green, 255; blue, 255 }  ,draw opacity=1 ][fill={rgb, 255:red, 74; green, 144; blue, 226 }  ,fill opacity=0.2 ] (260.5,179) -- (421.5,179) -- (421.5,261) -- (260.5,261) -- cycle ;
%Shape: Axis 2D [id:dp5864684608603756] 
\draw  (203,292) -- (470.5,292)(229.75,130) -- (229.75,310) (463.5,287) -- (470.5,292) -- (463.5,297) (224.75,137) -- (229.75,130) -- (234.75,137)  ;
%Straight Lines [id:da9731062473951577] 
\draw  [dash pattern={on 4.5pt off 4.5pt}]  (261,160) -- (261,300) ;
%Straight Lines [id:da33822733683394923] 
\draw  [dash pattern={on 4.5pt off 4.5pt}]  (301,160) -- (301,300) ;
%Straight Lines [id:da1448268711646643] 
\draw  [dash pattern={on 4.5pt off 4.5pt}]  (381,160) -- (381,300) ;
%Straight Lines [id:da09272657255037209] 
\draw  [dash pattern={on 4.5pt off 4.5pt}]  (421,160) -- (421,300) ;
%Straight Lines [id:da8541650216578274] 
\draw  [dash pattern={on 4.5pt off 4.5pt}]  (221,260) -- (440.5,260) ;
%Straight Lines [id:da2202348799891225] 
\draw  [dash pattern={on 4.5pt off 4.5pt}]  (221,220) -- (440.5,220) ;
%Straight Lines [id:da042695582029246504] 
\draw  [dash pattern={on 4.5pt off 4.5pt}]  (221,180) -- (440.5,180) ;
%Shape: Circle [id:dp15380583974328843] 
\draw  [fill={rgb, 255:red, 0; green, 0; blue, 0 }  ,fill opacity=1 ] (337,201) .. controls (337,198.79) and (338.79,197) .. (341,197) .. controls (343.21,197) and (345,198.79) .. (345,201) .. controls (345,203.21) and (343.21,205) .. (341,205) .. controls (338.79,205) and (337,203.21) .. (337,201) -- cycle ;
%Shape: Circle [id:dp7413461815699502] 
\draw  [fill={rgb, 255:red, 0; green, 0; blue, 0 }  ,fill opacity=1 ] (277,240) .. controls (277,237.79) and (278.79,236) .. (281,236) .. controls (283.21,236) and (285,237.79) .. (285,240) .. controls (285,242.21) and (283.21,244) .. (281,244) .. controls (278.79,244) and (277,242.21) .. (277,240) -- cycle ;
%Shape: Circle [id:dp9894269508256563] 
\draw  [fill={rgb, 255:red, 0; green, 0; blue, 0 }  ,fill opacity=1 ] (397,240) .. controls (397,237.79) and (398.79,236) .. (401,236) .. controls (403.21,236) and (405,237.79) .. (405,240) .. controls (405,242.21) and (403.21,244) .. (401,244) .. controls (398.79,244) and (397,242.21) .. (397,240) -- cycle ;
%Shape: Circle [id:dp26349771339901285] 
\draw  [fill={rgb, 255:red, 0; green, 0; blue, 0 }  ,fill opacity=1 ] (243,241) .. controls (243,238.79) and (244.79,237) .. (247,237) .. controls (249.21,237) and (251,238.79) .. (251,241) .. controls (251,243.21) and (249.21,245) .. (247,245) .. controls (244.79,245) and (243,243.21) .. (243,241) -- cycle ;
%Shape: Circle [id:dp692367624858315] 
\draw  [fill={rgb, 255:red, 0; green, 0; blue, 0 }  ,fill opacity=1 ] (397,166) .. controls (397,163.79) and (398.79,162) .. (401,162) .. controls (403.21,162) and (405,163.79) .. (405,166) .. controls (405,168.21) and (403.21,170) .. (401,170) .. controls (398.79,170) and (397,168.21) .. (397,166) -- cycle ;

% Text Node
\draw (256,303) node [anchor=north west][inner sep=0.75pt]   [align=left] {a};
% Text Node
\draw (296,301) node [anchor=north west][inner sep=0.75pt]   [align=left] {b};
% Text Node
\draw (376,303) node [anchor=north west][inner sep=0.75pt]   [align=left] {c};
% Text Node
\draw (415,301) node [anchor=north west][inner sep=0.75pt]   [align=left] {d};
% Text Node
\draw (207,175) node [anchor=north west][inner sep=0.75pt]   [align=left] {e};
% Text Node
\draw (209,212) node [anchor=north west][inner sep=0.75pt]   [align=left] {f};
% Text Node
\draw (207,253) node [anchor=north west][inner sep=0.75pt]   [align=left] {g};
% Text Node
\draw (459,302) node [anchor=north west][inner sep=0.75pt]   [align=left] {X};
% Text Node
\draw (208,129) node [anchor=north west][inner sep=0.75pt]   [align=left] {Y};

\end{tikzpicture}
\caption{Weak robust Equivalence class verification \cite{eq_class}}
\label{weak_robust}
  \end{minipage}
  \hspace{2cm}
  \begin{minipage}[b]{0.4\textwidth}
    \begin{tikzpicture}[x=0.75pt,y=0.75pt,yscale=-1,xscale=1]
%uncomment if require: \path (0,478); %set diagram left start at 0, and has height of 478

%Shape: Rectangle [id:dp3195416905530526] 
\draw  [color={rgb, 255:red, 255; green, 255; blue, 255 }  ,draw opacity=1 ][fill={rgb, 255:red, 74; green, 144; blue, 226 }  ,fill opacity=0.2 ] (260.5,179) -- (421.5,179) -- (421.5,261) -- (260.5,261) -- cycle ;
%Shape: Axis 2D [id:dp5864684608603756] 
\draw  (203,292) -- (470.5,292)(229.75,130) -- (229.75,310) (463.5,287) -- (470.5,292) -- (463.5,297) (224.75,137) -- (229.75,130) -- (234.75,137)  ;
%Straight Lines [id:da9731062473951577] 
\draw  [dash pattern={on 4.5pt off 4.5pt}]  (261,160) -- (261,300) ;
%Straight Lines [id:da33822733683394923] 
\draw  [dash pattern={on 4.5pt off 4.5pt}]  (301,160) -- (301,300) ;
%Straight Lines [id:da1448268711646643] 
\draw  [dash pattern={on 4.5pt off 4.5pt}]  (381,160) -- (381,300) ;
%Straight Lines [id:da09272657255037209] 
\draw  [dash pattern={on 4.5pt off 4.5pt}]  (421,160) -- (421,300) ;
%Straight Lines [id:da8541650216578274] 
\draw  [dash pattern={on 4.5pt off 4.5pt}]  (221,260) -- (440.5,260) ;
%Straight Lines [id:da2202348799891225] 
\draw  [dash pattern={on 4.5pt off 4.5pt}]  (221,220) -- (440.5,220) ;
%Straight Lines [id:da042695582029246504] 
\draw  [dash pattern={on 4.5pt off 4.5pt}]  (221,180) -- (440.5,180) ;
%Shape: Circle [id:dp15380583974328843] 
\draw  [fill={rgb, 255:red, 0; green, 0; blue, 0 }  ,fill opacity=1 ] (337,201) .. controls (337,198.79) and (338.79,197) .. (341,197) .. controls (343.21,197) and (345,198.79) .. (345,201) .. controls (345,203.21) and (343.21,205) .. (341,205) .. controls (338.79,205) and (337,203.21) .. (337,201) -- cycle ;
%Shape: Circle [id:dp7413461815699502] 
\draw  [fill={rgb, 255:red, 0; green, 0; blue, 0 }  ,fill opacity=1 ] (277,240) .. controls (277,237.79) and (278.79,236) .. (281,236) .. controls (283.21,236) and (285,237.79) .. (285,240) .. controls (285,242.21) and (283.21,244) .. (281,244) .. controls (278.79,244) and (277,242.21) .. (277,240) -- cycle ;
%Shape: Circle [id:dp9894269508256563] 
\draw  [fill={rgb, 255:red, 0; green, 0; blue, 0 }  ,fill opacity=1 ] (397,240) .. controls (397,237.79) and (398.79,236) .. (401,236) .. controls (403.21,236) and (405,237.79) .. (405,240) .. controls (405,242.21) and (403.21,244) .. (401,244) .. controls (398.79,244) and (397,242.21) .. (397,240) -- cycle ;
%Shape: Circle [id:dp47052494926194277] 
\draw  [fill={rgb, 255:red, 0; green, 0; blue, 0 }  ,fill opacity=1 ] (337,240) .. controls (337,237.79) and (338.79,236) .. (341,236) .. controls (343.21,236) and (345,237.79) .. (345,240) .. controls (345,242.21) and (343.21,244) .. (341,244) .. controls (338.79,244) and (337,242.21) .. (337,240) -- cycle ;
%Shape: Circle [id:dp6308723847944224] 
\draw  [fill={rgb, 255:red, 0; green, 0; blue, 0 }  ,fill opacity=1 ] (397,201) .. controls (397,198.79) and (398.79,197) .. (401,197) .. controls (403.21,197) and (405,198.79) .. (405,201) .. controls (405,203.21) and (403.21,205) .. (401,205) .. controls (398.79,205) and (397,203.21) .. (397,201) -- cycle ;
%Shape: Circle [id:dp7930249785945129] 
\draw  [fill={rgb, 255:red, 0; green, 0; blue, 0 }  ,fill opacity=1 ] (277,201) .. controls (277,198.79) and (278.79,197) .. (281,197) .. controls (283.21,197) and (285,198.79) .. (285,201) .. controls (285,203.21) and (283.21,205) .. (281,205) .. controls (278.79,205) and (277,203.21) .. (277,201) -- cycle ;
%Shape: Circle [id:dp5224160035470657] 
\draw  [fill={rgb, 255:red, 0; green, 0; blue, 0 }  ,fill opacity=1 ] (430,240) .. controls (430,237.79) and (431.79,236) .. (434,236) .. controls (436.21,236) and (438,237.79) .. (438,240) .. controls (438,242.21) and (436.21,244) .. (434,244) .. controls (431.79,244) and (430,242.21) .. (430,240) -- cycle ;
%Shape: Circle [id:dp7303868175364496] 
\draw  [fill={rgb, 255:red, 0; green, 0; blue, 0 }  ,fill opacity=1 ] (430,201) .. controls (430,198.79) and (431.79,197) .. (434,197) .. controls (436.21,197) and (438,198.79) .. (438,201) .. controls (438,203.21) and (436.21,205) .. (434,205) .. controls (431.79,205) and (430,203.21) .. (430,201) -- cycle ;
%Shape: Circle [id:dp746419026644076] 
\draw  [fill={rgb, 255:red, 0; green, 0; blue, 0 }  ,fill opacity=1 ] (397,273) .. controls (397,270.79) and (398.79,269) .. (401,269) .. controls (403.21,269) and (405,270.79) .. (405,273) .. controls (405,275.21) and (403.21,277) .. (401,277) .. controls (398.79,277) and (397,275.21) .. (397,273) -- cycle ;
%Shape: Circle [id:dp8295736560202394] 
\draw  [fill={rgb, 255:red, 0; green, 0; blue, 0 }  ,fill opacity=1 ] (337,273) .. controls (337,270.79) and (338.79,269) .. (341,269) .. controls (343.21,269) and (345,270.79) .. (345,273) .. controls (345,275.21) and (343.21,277) .. (341,277) .. controls (338.79,277) and (337,275.21) .. (337,273) -- cycle ;
%Shape: Circle [id:dp6104449251900974] 
\draw  [fill={rgb, 255:red, 0; green, 0; blue, 0 }  ,fill opacity=1 ] (277,273) .. controls (277,270.79) and (278.79,269) .. (281,269) .. controls (283.21,269) and (285,270.79) .. (285,273) .. controls (285,275.21) and (283.21,277) .. (281,277) .. controls (278.79,277) and (277,275.21) .. (277,273) -- cycle ;
%Shape: Circle [id:dp2193988533378226] 
\draw  [fill={rgb, 255:red, 0; green, 0; blue, 0 }  ,fill opacity=1 ] (243,273) .. controls (243,270.79) and (244.79,269) .. (247,269) .. controls (249.21,269) and (251,270.79) .. (251,273) .. controls (251,275.21) and (249.21,277) .. (247,277) .. controls (244.79,277) and (243,275.21) .. (243,273) -- cycle ;
%Shape: Circle [id:dp8877953879156839] 
\draw  [fill={rgb, 255:red, 0; green, 0; blue, 0 }  ,fill opacity=1 ] (430,273) .. controls (430,270.79) and (431.79,269) .. (434,269) .. controls (436.21,269) and (438,270.79) .. (438,273) .. controls (438,275.21) and (436.21,277) .. (434,277) .. controls (431.79,277) and (430,275.21) .. (430,273) -- cycle ;
%Shape: Circle [id:dp26349771339901285] 
\draw  [fill={rgb, 255:red, 0; green, 0; blue, 0 }  ,fill opacity=1 ] (243,241) .. controls (243,238.79) and (244.79,237) .. (247,237) .. controls (249.21,237) and (251,238.79) .. (251,241) .. controls (251,243.21) and (249.21,245) .. (247,245) .. controls (244.79,245) and (243,243.21) .. (243,241) -- cycle ;
%Shape: Circle [id:dp9062034714020735] 
\draw  [fill={rgb, 255:red, 0; green, 0; blue, 0 }  ,fill opacity=1 ] (243,201) .. controls (243,198.79) and (244.79,197) .. (247,197) .. controls (249.21,197) and (251,198.79) .. (251,201) .. controls (251,203.21) and (249.21,205) .. (247,205) .. controls (244.79,205) and (243,203.21) .. (243,201) -- cycle ;
%Shape: Circle [id:dp5662014569105551] 
\draw  [fill={rgb, 255:red, 0; green, 0; blue, 0 }  ,fill opacity=1 ] (243,166) .. controls (243,163.79) and (244.79,162) .. (247,162) .. controls (249.21,162) and (251,163.79) .. (251,166) .. controls (251,168.21) and (249.21,170) .. (247,170) .. controls (244.79,170) and (243,168.21) .. (243,166) -- cycle ;
%Shape: Circle [id:dp6556154268379637] 
\draw  [fill={rgb, 255:red, 0; green, 0; blue, 0 }  ,fill opacity=1 ] (277,166) .. controls (277,163.79) and (278.79,162) .. (281,162) .. controls (283.21,162) and (285,163.79) .. (285,166) .. controls (285,168.21) and (283.21,170) .. (281,170) .. controls (278.79,170) and (277,168.21) .. (277,166) -- cycle ;
%Shape: Circle [id:dp7203999786548791] 
\draw  [fill={rgb, 255:red, 0; green, 0; blue, 0 }  ,fill opacity=1 ] (337,166) .. controls (337,163.79) and (338.79,162) .. (341,162) .. controls (343.21,162) and (345,163.79) .. (345,166) .. controls (345,168.21) and (343.21,170) .. (341,170) .. controls (338.79,170) and (337,168.21) .. (337,166) -- cycle ;
%Shape: Circle [id:dp692367624858315] 
\draw  [fill={rgb, 255:red, 0; green, 0; blue, 0 }  ,fill opacity=1 ] (397,166) .. controls (397,163.79) and (398.79,162) .. (401,162) .. controls (403.21,162) and (405,163.79) .. (405,166) .. controls (405,168.21) and (403.21,170) .. (401,170) .. controls (398.79,170) and (397,168.21) .. (397,166) -- cycle ;
%Shape: Circle [id:dp4032200258680627] 
\draw  [fill={rgb, 255:red, 0; green, 0; blue, 0 }  ,fill opacity=1 ] (430,166) .. controls (430,163.79) and (431.79,162) .. (434,162) .. controls (436.21,162) and (438,163.79) .. (438,166) .. controls (438,168.21) and (436.21,170) .. (434,170) .. controls (431.79,170) and (430,168.21) .. (430,166) -- cycle ;

% Text Node
\draw (256,303) node [anchor=north west][inner sep=0.75pt]   [align=left] {a};
% Text Node
\draw (296,301) node [anchor=north west][inner sep=0.75pt]   [align=left] {b};
% Text Node
\draw (376,303) node [anchor=north west][inner sep=0.75pt]   [align=left] {c};
% Text Node
\draw (415,301) node [anchor=north west][inner sep=0.75pt]   [align=left] {d};
% Text Node
\draw (207,175) node [anchor=north west][inner sep=0.75pt]   [align=left] {e};
% Text Node
\draw (209,212) node [anchor=north west][inner sep=0.75pt]   [align=left] {f};
% Text Node
\draw (207,253) node [anchor=north west][inner sep=0.75pt]   [align=left] {g};
% Text Node
\draw (459,302) node [anchor=north west][inner sep=0.75pt]   [align=left] {X};
% Text Node
\draw (208,129) node [anchor=north west][inner sep=0.75pt]   [align=left] {Y};

\end{tikzpicture}
\caption{Strong robust Equivalence class verification \cite{eq_class}}
\label{strong_robust}
  \end{minipage}
\end{figure}

\subsubsection{Weak Normal Equivalence Class Verification}
It is based on the assumption that an error is not often caused as a result of two or more faults occurring simultaneously. Therefore, weak Equivalence class verification takes one variable from each Equivalence class \cite{eq_class}.

\subsubsection{Strong Normal Equivalence Class Verification}
It is based on the assumption that the errors generally result in a combination of faults. Therefore, strong Equivalence class verification takes every combination of elements formed as a result of the Cartesian product of the Equivalence relation \cite{eq_class}.

\subsubsection{Weak Robust Equivalence Class Verification}
In the weak normal Equivalence class verification, variable from each Equivalence class is usually verified. However, verification for invalid values as well is done in this case. Therefore, weak robust Equivalence class verification takes one invalid value and the remaining values will all be valid \cite{eq_class}.

\subsubsection{Strong Robust Equivalence Class Verification}
Strong robust Equivalence class verification takes all valid and invalid elements of the Cartesian product of all the Equivalence classes \cite{eq_class}.

The choice of the type of Equivalence class verification for the methodology depends on certain factors. Some of the types use invalid elements as well for constructing the regressions. However, it is not necessary to verify values from the invalid classes if we use strongly typed languages such as \acrshort{VHDL} for design. Nevertheless, negative testing in hardware verification is out of scope since we would eventually verify functionality that would never exist. It is also found that most design bugs often exist at the boundary values of an Equivalence class \cite{eq_partitioning}. In addition to this, the ISO 26262 standards also recommend using boundary values with Equivalence classes \cite{iso}. Therefore, considering the facts mentioned above, we should use strong normal Equivalence class verification with a combination of boundary values for our methodology. This means we will not just use any value from each of the Equivalence classes, instead, take one value from the maxima and one value from the minima of each of the Equivalence classes for the verification.

The Equivalence classes, as mentioned before, are a part of the top-down approach and are driven by the design specifications. These Equivalence classes are defined using Equivalence relation. Nevertheless, it is also important to verify the Equivalence classes in order to make sure that the classes defined are correct. This can be done using formal verification and writing properties, as shown in Listing \ref{eq_class_fv} as an example. This approach will also make sure that the design specifications and the designed implementation are tightly coupled.
\vspace{0.25cm}
\begin{lstlisting}[language=Verilog, float=h, caption=Example of \acrshort{SVA} property for verifying Equivalence classes, label={eq_class_fv}]
property Eq_Class;
    (Address_Width <= THRESHOLD)
  |->
    (Memory_Access == DIRECT);
endproperty
AP_Eq_Class : assert property(Eq_Class);
\end{lstlisting}

The property mentioned in Listing \ref{eq_class_fv} states that if the Address\_Width is less than or equal to the defined threshold, the memory access is always direct. Proving this property using formal verification makes sure that there are two Equivalence classes for the parameter Address\_Width, one having values less than the threshold and the other one having values greater than the threshold.

As already mentioned, the process of including the Equivalence classes and subsequently, modifying the Pairwise template is automated. First, the Equivalence classes are identified and mentioned on the Requirements tool as a requirement for verification. Once the requirement is mentioned on the Requirements tool for the Equivalence class, an XML file is generated containing all the information and the Python script extracts the Equivalence classes and then modifies the Pairwise template.

The bottom-up approach is primarily influenced by design implementation. In this sub-step, formal structural checking is performed to identify the design blocks inside the top-level suitable for block-level formal verification. The parameters used by the block is then neglected in the Pairwise generation by modifying the Pairwise template. The formal structural checking is explained in the subsequent subsection.

\subsection{Formal Structural Checking}
Formal structural checking is used in the methodology as a bottom-up approach to reduce the number of regressions. It is based on the concept of \acrfull{COI} in formal verification. The \acrshort{COI}, as shown in Fig.~\ref{coi}, can be defined intuitively as the cone that contains all the logic, which is required to prove or disprove the property under evaluation \cite{fvbook}. In a similar concept, the \acrshort{COI} for a signal is the cone that contains all the signals that affect or are affected by the signal. This essentially means all other signals in fan-in and fan-out of the signal under evaluation are the \acrshort{COI}. This concept is used to identify the parameters that are used by a single block inside the top-level. Once the block is identified, the signals using this parameter are identified, and the fan-out of this signal is traced to check if the signal is connected to the output of the top-level. Once this is confirmed, the block is said to be suitable for block-level formal verification. The formal verification of that block is carried out, and the completeness metrics (structural coverage) are analyzed. The flow for identifying the blocks suitable for block-level formal verification using formal structural checking is shown in Fig.~\ref{formal_flow}. The flow is fully automated using any formal verification tool and scripts written in Python.

\tikzset{every picture/.style={line width=1pt}} %set default line width to 0.75pt        
\begin{figure}[h]
\centering
%\begin{adjustbox}{min width=11cm}
\begin{tikzpicture}[x=0.75pt,y=0.75pt,yscale=-1,xscale=1]
%uncomment if require: \path (0,478); %set diagram left start at 0, and has height of 478

%Rounded Rect [id:dp8324210336052416] 
\draw   (100,120) .. controls (100,108.95) and (108.95,100) .. (120,100) -- (250.5,100) .. controls (261.55,100) and (270.5,108.95) .. (270.5,120) -- (270.5,180) .. controls (270.5,191.05) and (261.55,200) .. (250.5,200) -- (120,200) .. controls (108.95,200) and (100,191.05) .. (100,180) -- cycle ;
%Shape: Triangle [id:dp7085663028834985] 
\draw  [fill={rgb, 255:red, 173; green, 212; blue, 131 }  ,fill opacity=0.3 ] (270.89,148.85) -- (100.03,179.99) -- (100.03,122.29) -- cycle ;
%Straight Lines [id:da20885532708515497] 
\draw    (270.89,148.85) -- (289.5,148.99) ;
\draw [shift={(291.5,149)}, rotate = 180.42] [color={rgb, 255:red, 0; green, 0; blue, 0 }  ][line width=0.75]    (10.93,-3.29) .. controls (6.95,-1.4) and (3.31,-0.3) .. (0,0) .. controls (3.31,0.3) and (6.95,1.4) .. (10.93,3.29)   ;

% Text Node
\draw (175,105) node [anchor=north west][inner sep=0.75pt]  [font=\footnotesize] [align=left] {Irrelevant logic};
% Text Node
\draw (173,118) node [anchor=north west][inner sep=0.75pt]  [font=\footnotesize] [align=left] {for the property};
% Text Node
\draw (175,169) node [anchor=north west][inner sep=0.75pt]  [font=\footnotesize] [align=left] {Irrelevant logic};
% Text Node
\draw (173,182) node [anchor=north west][inner sep=0.75pt]  [font=\footnotesize] [align=left] {for the property};
% Text Node
\draw (106,137) node [anchor=north west][inner sep=0.75pt]  [font=\footnotesize] [align=left] {Relevant};
% Text Node
\draw (115,153) node [anchor=north west][inner sep=0.75pt]  [font=\footnotesize] [align=left] {logic};
% Text Node
\draw (185,144) node [anchor=north west][inner sep=0.75pt]  [font=\footnotesize] [align=left] {COI};
% Text Node
\draw (74,133) node [anchor=north west][inner sep=0.75pt]  [font=\fontsize{4.33em}{5.2em}\selectfont] [align=left] {\{};
% Text Node
\draw (296,143) node [anchor=north west][inner sep=0.75pt]  [font=\footnotesize] [align=left] {Property \textbf{P}};
% Text Node
\draw (39,145) node [anchor=north west][inner sep=0.75pt]  [font=\footnotesize] [align=left] {Inputs};

\end{tikzpicture}
\caption{\acrfull{COI}}
\label{coi}
%\end{adjustbox}
\end{figure}

As an example, Fig.~\ref{block_level_fig} shows a top-level \acrshort{SoC} that contains three design blocks. Out of these design blocks, \say{Block 3} uses parameter \say{P1}, and the output of the signals being affected by this parameter is the top-level output of the \acrshort{SoC}. This means the output of signals affected by \say{P1} in \say{Block 3} are not affecting any other logic inside the top-level \acrshort{SoC}. Therefore, \say{Block 3} is suitable for block-level formal verification. Design parameters used only by a single design block inside the top-level design, and the signals that are affected by the parameter and are the output of the top-level are identified for block-level formal verification. The parameter used by the identified block is then neglected while generating the Pairwise regressions by modifying the Pairwise template. Since formal verification is an exhaustive verification method, the properties written to capture the intent of the design block are verified for all possible values of the parameter. Therefore, the parameter can be safely excluded from the Pairwise regressions without compromising the quality of verification.

\tikzset{every picture/.style={line width=1pt}} %set default line width to 0.75pt        
\begin{figure}[h]
\centering
\begin{tikzpicture}[x=0.75pt,y=0.75pt,yscale=-1,xscale=1]
%uncomment if require: \path (0,427); %set diagram left start at 0, and has height of 427

%Rounded Rect [id:dp853971072853128] 
\draw   (151,82) .. controls (151,64.33) and (165.33,50) .. (183,50) -- (378.5,50) .. controls (396.17,50) and (410.5,64.33) .. (410.5,82) -- (410.5,178) .. controls (410.5,195.67) and (396.17,210) .. (378.5,210) -- (183,210) .. controls (165.33,210) and (151,195.67) .. (151,178) -- cycle ;
%Rounded Rect [id:dp221774649098895] 
\draw   (173,87) .. controls (173,82.58) and (176.58,79) .. (181,79) -- (235,79) .. controls (239.42,79) and (243,82.58) .. (243,87) -- (243,111) .. controls (243,115.42) and (239.42,119) .. (235,119) -- (181,119) .. controls (176.58,119) and (173,115.42) .. (173,111) -- cycle ;
%Rounded Rect [id:dp5437764965215084] 
\draw   (281,148) .. controls (281,143.58) and (284.58,140) .. (289,140) -- (343,140) .. controls (347.42,140) and (351,143.58) .. (351,148) -- (351,172) .. controls (351,176.42) and (347.42,180) .. (343,180) -- (289,180) .. controls (284.58,180) and (281,176.42) .. (281,172) -- cycle ;
%Straight Lines [id:da009433324848433733] 
\draw    (410.5,82) -- (445,82) ;
\draw [shift={(447,82)}, rotate = 180] [color={rgb, 255:red, 0; green, 0; blue, 0 }  ][line width=0.75]    (10.93,-3.29) .. controls (6.95,-1.4) and (3.31,-0.3) .. (0,0) .. controls (3.31,0.3) and (6.95,1.4) .. (10.93,3.29)   ;
%Straight Lines [id:da47955159154980853] 
\draw    (410.5,178) -- (445,178) ;
\draw [shift={(447,178)}, rotate = 180] [color={rgb, 255:red, 0; green, 0; blue, 0 }  ][line width=0.75]    (10.93,-3.29) .. controls (6.95,-1.4) and (3.31,-0.3) .. (0,0) .. controls (3.31,0.3) and (6.95,1.4) .. (10.93,3.29)   ;
%Straight Lines [id:da25352358632315664] 
\draw    (410.5,114) -- (445,114) ;
\draw [shift={(447,114)}, rotate = 180] [color={rgb, 255:red, 0; green, 0; blue, 0 }  ][line width=0.75]    (10.93,-3.29) .. controls (6.95,-1.4) and (3.31,-0.3) .. (0,0) .. controls (3.31,0.3) and (6.95,1.4) .. (10.93,3.29)   ;
%Straight Lines [id:da8982686252390961] 
\draw    (410.5,146) -- (445,146) ;
\draw [shift={(447,146)}, rotate = 180] [color={rgb, 255:red, 0; green, 0; blue, 0 }  ][line width=0.75]    (10.93,-3.29) .. controls (6.95,-1.4) and (3.31,-0.3) .. (0,0) .. controls (3.31,0.3) and (6.95,1.4) .. (10.93,3.29)   ;
%Straight Lines [id:da5163832254355214] 
\draw    (114.5,82) -- (149,82) ;
\draw [shift={(151,82)}, rotate = 180] [color={rgb, 255:red, 0; green, 0; blue, 0 }  ][line width=0.75]    (10.93,-3.29) .. controls (6.95,-1.4) and (3.31,-0.3) .. (0,0) .. controls (3.31,0.3) and (6.95,1.4) .. (10.93,3.29)   ;
%Straight Lines [id:da19641565235507885] 
\draw    (114.5,178) -- (149,178) ;
\draw [shift={(151,178)}, rotate = 180] [color={rgb, 255:red, 0; green, 0; blue, 0 }  ][line width=0.75]    (10.93,-3.29) .. controls (6.95,-1.4) and (3.31,-0.3) .. (0,0) .. controls (3.31,0.3) and (6.95,1.4) .. (10.93,3.29)   ;
%Straight Lines [id:da9877556909999512] 
\draw    (114.5,114) -- (149,114) ;
\draw [shift={(151,114)}, rotate = 180] [color={rgb, 255:red, 0; green, 0; blue, 0 }  ][line width=0.75]    (10.93,-3.29) .. controls (6.95,-1.4) and (3.31,-0.3) .. (0,0) .. controls (3.31,0.3) and (6.95,1.4) .. (10.93,3.29)   ;
%Straight Lines [id:da5162123789557187] 
\draw    (114.5,146) -- (149,146) ;
\draw [shift={(151,146)}, rotate = 180] [color={rgb, 255:red, 0; green, 0; blue, 0 }  ][line width=0.75]    (10.93,-3.29) .. controls (6.95,-1.4) and (3.31,-0.3) .. (0,0) .. controls (3.31,0.3) and (6.95,1.4) .. (10.93,3.29)   ;
%Straight Lines [id:da013282600436109071] 
\draw    (351,159) -- (385.5,159) ;
\draw [shift={(387.5,159)}, rotate = 180] [color={rgb, 255:red, 0; green, 0; blue, 0 }  ][line width=0.75]    (10.93,-3.29) .. controls (6.95,-1.4) and (3.31,-0.3) .. (0,0) .. controls (3.31,0.3) and (6.95,1.4) .. (10.93,3.29)   ;
%Straight Lines [id:da6470035023068417] 
\draw [color={rgb, 255:red, 65; green, 117; blue, 5 }  ,draw opacity=1 ]   (387.5,159) -- (410.5,146) ;
%Shape: Rectangle [id:dp533995040620985] 
\draw  [color={rgb, 255:red, 208; green, 2; blue, 27 }  ,draw opacity=1 ] (148,158) -- (154.5,158) -- (154.5,165) -- (148,165) -- cycle ;
%Straight Lines [id:da1168833340574158] 
\draw    (244.5,159) -- (279,159) ;
\draw [shift={(281,159)}, rotate = 180] [color={rgb, 255:red, 0; green, 0; blue, 0 }  ][line width=0.75]    (10.93,-3.29) .. controls (6.95,-1.4) and (3.31,-0.3) .. (0,0) .. controls (3.31,0.3) and (6.95,1.4) .. (10.93,3.29)   ;
%Straight Lines [id:da048610677500721566] 
\draw [color={rgb, 255:red, 208; green, 2; blue, 27 }  ,draw opacity=1 ]   (151.25,161.5) -- (244.5,159) ;
%Straight Lines [id:da13126806266413227] 
\draw    (243,97) -- (267.5,97) ;
\draw [shift={(269.5,97)}, rotate = 180] [color={rgb, 255:red, 0; green, 0; blue, 0 }  ][line width=0.75]    (10.93,-3.29) .. controls (6.95,-1.4) and (3.31,-0.3) .. (0,0) .. controls (3.31,0.3) and (6.95,1.4) .. (10.93,3.29)   ;
%Rounded Rect [id:dp566728914305946] 
\draw   (316,95) .. controls (316,90.58) and (319.58,87) .. (324,87) -- (378,87) .. controls (382.42,87) and (386,90.58) .. (386,95) -- (386,119) .. controls (386,123.42) and (382.42,127) .. (378,127) -- (324,127) .. controls (319.58,127) and (316,123.42) .. (316,119) -- cycle ;
%Straight Lines [id:da6183161907708121] 
\draw    (289.5,106) -- (314,106) ;
\draw [shift={(316,106)}, rotate = 180] [color={rgb, 255:red, 0; green, 0; blue, 0 }  ][line width=0.75]    (10.93,-3.29) .. controls (6.95,-1.4) and (3.31,-0.3) .. (0,0) .. controls (3.31,0.3) and (6.95,1.4) .. (10.93,3.29)   ;
%Straight Lines [id:da022580440313028305] 
\draw    (269.5,97) -- (289.5,106) ;
%Straight Lines [id:da8330109286691199] 
\draw    (386,106) -- (400.5,106) ;
\draw [shift={(402.5,106)}, rotate = 180] [color={rgb, 255:red, 0; green, 0; blue, 0 }  ][line width=0.75]    (10.93,-3.29) .. controls (6.95,-1.4) and (3.31,-0.3) .. (0,0) .. controls (3.31,0.3) and (6.95,1.4) .. (10.93,3.29)   ;
%Straight Lines [id:da5950139744535683] 
\draw    (402.5,106) -- (410.5,114) ;

% Text Node
\draw (288,60) node [anchor=north west][inner sep=0.75pt]   [align=left] {SoC Top-Level};
% Text Node
\draw (292,153) node [anchor=north west][inner sep=0.75pt]   [align=left] {Block 3};
% Text Node
\draw (184,93) node [anchor=north west][inner sep=0.75pt]   [align=left] {Block 1};
% Text Node
\draw (50,154) node [anchor=north west][inner sep=0.75pt]  [color={rgb, 255:red, 208; green, 2; blue, 27 }  ,opacity=1 ] [align=left] {Parameter P1};
% Text Node
\draw (66,95) node [anchor=north west][inner sep=0.75pt]   [align=left] {Inputs};
% Text Node
\draw (453,122) node [anchor=north west][inner sep=0.75pt]   [align=left] {Outputs};
% Text Node
\draw (327,100) node [anchor=north west][inner sep=0.75pt]   [align=left] {Block 2};

\end{tikzpicture}
\caption{Identifying block suitable for block-level verification using formal structural checking}
\label{block_level_fig}
\end{figure}
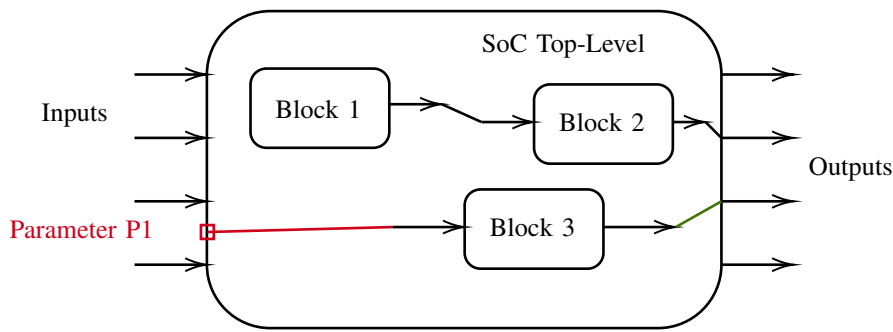

\tikzstyle{startstop} = [rectangle, rounded corners, minimum width=3cm, minimum height=1cm,text centered, draw=black]
\tikzstyle{io} = [trapezium, trapezium left angle=70, trapezium right angle=110, minimum width=3cm, minimum height=1cm, text centered, draw=black]
\tikzstyle{process} = [rectangle, minimum width=3cm, minimum height=1cm, text centered, draw=black]
\tikzstyle{decision} = [diamond, minimum width=3cm, minimum height=1cm, text width=4cm, aspect=2, text centered, draw=black]
\tikzstyle{arrow} = [thick,->,>=stealth]
\begin{figure}[h]
\centering
\resizebox{0.4\textwidth}{!}{%
%\begin{adjustbox}{max width=0.4\textwidth}
\begin{tikzpicture}[node distance=1.5cm]
\node (start) [startstop] {START};
\node (pro1) [process, below of=start] {Extract Parameters};
\node (pro2) [process, below of=pro1] {Identify blocks using these Parameters};
\node (dec1) [decision, below of=pro2, yshift=-1cm] {Only one block uses the Parameter?};
\node (pro3a) [process, below of=dec1, yshift=-1.5cm] {Check signals that uses the Parameter};
\node (pro3b) [process, right of=dec1, xshift=5cm] {No block found suitable};
\node (pro4) [process, below of=pro3a] {Check fanout of signals};
\node (dec2) [decision, below of=pro4, yshift=-1.5cm] {Signal is connected to the output of Top-level?};
\node (pro5a) [process, below of=dec2, yshift=-1.5cm] {Block suitable for Block-level verification};
\node (pro5b) [process, right of=dec2, xshift=5.5cm] {No block found suitable};
\draw [arrow] (start) -- (pro1);
\draw [arrow] (pro1) -- (pro2);
\draw [arrow] (pro2) -- (dec1);
\draw [arrow] (dec1) -- node[anchor=east] {Yes} (pro3a);
\draw [arrow] (dec1) -- node[anchor=south] {No} (pro3b);
\draw [arrow] (pro3a) -- (pro4);
\draw [arrow] (pro4) -- (dec2);
\draw [arrow] (dec2) -- node[anchor=east] {Yes} (pro5a);
\draw [arrow] (dec2) -- node[anchor=south] {No} (pro5b);
\end{tikzpicture}}
\caption{Formal structural checking flow (automated using formal tool and Python scripts)}
\label{formal_flow}
%\end{adjustbox}
\end{figure}
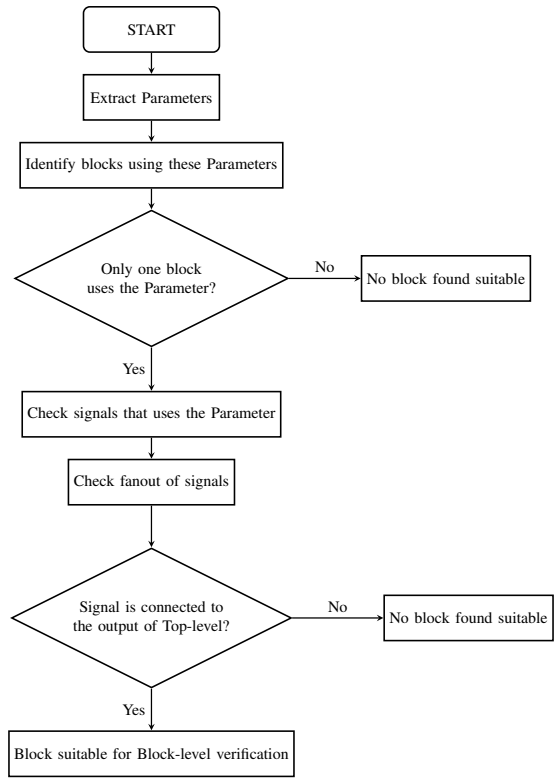

\section{ConfigVerMet: Execute}
After executing the prepare and extract steps, the execute step is implemented. In this step, Pairwise regressions are generated considering the Pairwise template using the Microsoft \acrshort{PICT} tool. Since the regressions generated are optimized using several regression reduction techniques, it is referred to as \say{Optimized Pairwise Regressions}. After generating the optimized regressions, a SystemVerilog configuration coverage file is generated automatically that contains covergroups, coverpoints and cross-coverpoints for the SystemVerilog-\acrshort{UVM} setup. This is important to verify in the end to check if the configuration coverage goals are met. Additionally, a \acrfull{VSIF} is also automatically generated that contains the regression information for the Cadence vManager regression tool. In principle, setup files for other \acrshort{EDA} vendors can also be generated.

\section{Application and Results}

\subsection{The Design Under Verification}
The design implementation, considered for implementing the proposed methodology is a microprocessor \acrshort{IP}, as shown in Fig.~\ref{duv}. The microprocessor \acrshort{IP} should fulfil the standards of ISO 26262 and is graded as \acrshort{ASIL}-D category product. Since the developed methodology addresses the \acrshort{SEooC} designs concerning the verification of all relevant configurations, the microprocessor \acrshort{IP} which also happens to be a \acrshort{SEooC}, is suitable as a \acrshort{DUV} for applying the developed methodology.

\tikzset{every picture/.style={line width=1pt}} %set default line width to 0.75pt        
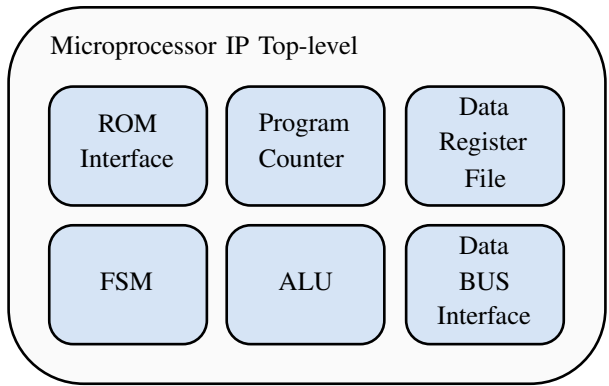
\begin{figure}[h]
\centering
\begin{tikzpicture}[x=0.75pt,y=0.75pt,yscale=-1,xscale=1]
%uncomment if require: \path (0,478); %set diagram left start at 0, and has height of 478

%Rounded Rect [id:dp979047197359711] 
\draw  [fill={rgb, 255:red, 155; green, 155; blue, 155 }  ,fill opacity=0.05 ] (80.5,118) .. controls (80.5,97.01) and (97.51,80) .. (118.5,80) -- (343.5,80) .. controls (364.49,80) and (381.5,97.01) .. (381.5,118) -- (381.5,232) .. controls (381.5,252.99) and (364.49,270) .. (343.5,270) -- (118.5,270) .. controls (97.51,270) and (80.5,252.99) .. (80.5,232) -- cycle ;
%Rounded Rect [id:dp3969545289173253] 
\draw  [fill={rgb, 255:red, 74; green, 144; blue, 226 }  ,fill opacity=0.2 ] (101,132) .. controls (101,125.37) and (106.37,120) .. (113,120) -- (168.5,120) .. controls (175.13,120) and (180.5,125.37) .. (180.5,132) -- (180.5,168) .. controls (180.5,174.63) and (175.13,180) .. (168.5,180) -- (113,180) .. controls (106.37,180) and (101,174.63) .. (101,168) -- cycle ;
%Rounded Rect [id:dp25408146674618237] 
\draw  [fill={rgb, 255:red, 74; green, 144; blue, 226 }  ,fill opacity=0.2 ] (190.5,132) .. controls (190.5,125.37) and (195.87,120) .. (202.5,120) -- (258,120) .. controls (264.63,120) and (270,125.37) .. (270,132) -- (270,168) .. controls (270,174.63) and (264.63,180) .. (258,180) -- (202.5,180) .. controls (195.87,180) and (190.5,174.63) .. (190.5,168) -- cycle ;
%Rounded Rect [id:dp19869711743339624] 
\draw  [fill={rgb, 255:red, 74; green, 144; blue, 226 }  ,fill opacity=0.2 ] (281,132) .. controls (281,125.37) and (286.37,120) .. (293,120) -- (348.5,120) .. controls (355.13,120) and (360.5,125.37) .. (360.5,132) -- (360.5,168) .. controls (360.5,174.63) and (355.13,180) .. (348.5,180) -- (293,180) .. controls (286.37,180) and (281,174.63) .. (281,168) -- cycle ;
%Rounded Rect [id:dp2523433237573154] 
\draw  [fill={rgb, 255:red, 74; green, 144; blue, 226 }  ,fill opacity=0.2 ] (101,202) .. controls (101,195.37) and (106.37,190) .. (113,190) -- (168.5,190) .. controls (175.13,190) and (180.5,195.37) .. (180.5,202) -- (180.5,238) .. controls (180.5,244.63) and (175.13,250) .. (168.5,250) -- (113,250) .. controls (106.37,250) and (101,244.63) .. (101,238) -- cycle ;
%Rounded Rect [id:dp7348089330993934] 
\draw  [fill={rgb, 255:red, 74; green, 144; blue, 226 }  ,fill opacity=0.2 ] (190.5,202) .. controls (190.5,195.37) and (195.87,190) .. (202.5,190) -- (258,190) .. controls (264.63,190) and (270,195.37) .. (270,202) -- (270,238) .. controls (270,244.63) and (264.63,250) .. (258,250) -- (202.5,250) .. controls (195.87,250) and (190.5,244.63) .. (190.5,238) -- cycle ;
%Rounded Rect [id:dp9875722197434789] 
\draw  [fill={rgb, 255:red, 74; green, 144; blue, 226 }  ,fill opacity=0.2 ] (281,202) .. controls (281,195.37) and (286.37,190) .. (293,190) -- (348.5,190) .. controls (355.13,190) and (360.5,195.37) .. (360.5,202) -- (360.5,238) .. controls (360.5,244.63) and (355.13,250) .. (348.5,250) -- (293,250) .. controls (286.37,250) and (281,244.63) .. (281,238) -- cycle ;

% Text Node
\draw (295,229) node [anchor=north west][inner sep=0.75pt]   [align=left] {Interface};
% Text Node
\draw (100,93) node [anchor=north west][inner sep=0.75pt]   [align=left] {Microprocessor IP Top-level};
% Text Node
\draw (306,124) node [anchor=north west][inner sep=0.75pt]   [align=left] {Data};
% Text Node
\draw (296,141) node [anchor=north west][inner sep=0.75pt]   [align=left] {Register};
% Text Node
\draw (309,160) node [anchor=north west][inner sep=0.75pt]   [align=left] {File};
% Text Node
\draw (124,132) node [anchor=north west][inner sep=0.75pt]   [align=left] {ROM};
% Text Node
\draw (115,150) node [anchor=north west][inner sep=0.75pt]   [align=left] {Interface};
% Text Node
\draw (205,132) node [anchor=north west][inner sep=0.75pt]   [align=left] {Program};
% Text Node
\draw (205,150) node [anchor=north west][inner sep=0.75pt]   [align=left] {Counter};
% Text Node
\draw (125,212) node [anchor=north west][inner sep=0.75pt]   [align=left] {FSM};
% Text Node
\draw (215,212) node [anchor=north west][inner sep=0.75pt]   [align=left] {ALU};
% Text Node
\draw (306,194) node [anchor=north west][inner sep=0.75pt]   [align=left] {Data};
% Text Node
\draw (306,212) node [anchor=north west][inner sep=0.75pt]   [align=left] {BUS};

\end{tikzpicture}
\caption{Block diagram of the \acrshort{DUV}}
\label{duv}
\end{figure}

The microprocessor \acrshort{IP} is based on the Harvard architecture and therefore, has separated control and data path. The \acrshort{RTL} is written in \acrshort{VHDL} and has six design parameters. The parameters are either feature enabling/disabling parameters or bus-width parameterizations. The parameters are:
\begin{itemize}
    \setlength\itemsep{0.5em}
    \item \textbf{Feature\_1:} values 0 to 1
    \item \textbf{Feature\_2:} values 0 to 1
    \item \textbf{Address\_Range\_1:} values 1 to 32767
    \item \textbf{Address\_Width\_1:} values 6 to 16
    \item \textbf{Address\_Width\_2:} values 1 to 14
    \item \textbf{Address\_Width\_3:} values 8 to 15
\end{itemize}

In order to perform an exhaustive verification considering all possible combinations of the parameters mentioned above, we would end up having billions of regressions. It is clear that the number of exhaustive regression is not reasonable, and therefore, the developed methodology is suitable for identifying the relevant configurations for the verification. To exercise the methodology on the microprocessor \acrshort{IP}, we used an existing SystemVerilog-\acrshort{UVM} testbench setup that takes 24 hours or one day for one regression (meaning one configuration or combination of parameter values) to complete. This means, for a brute-force approach, it would take more than 10 million years to complete the regressions. To overcome the aforementioned challenges, the developed methodology is applied to the \acrshort{DUV}.

\tikzset{every picture/.style={line width=0.75pt}} %set default line width to 0.75pt
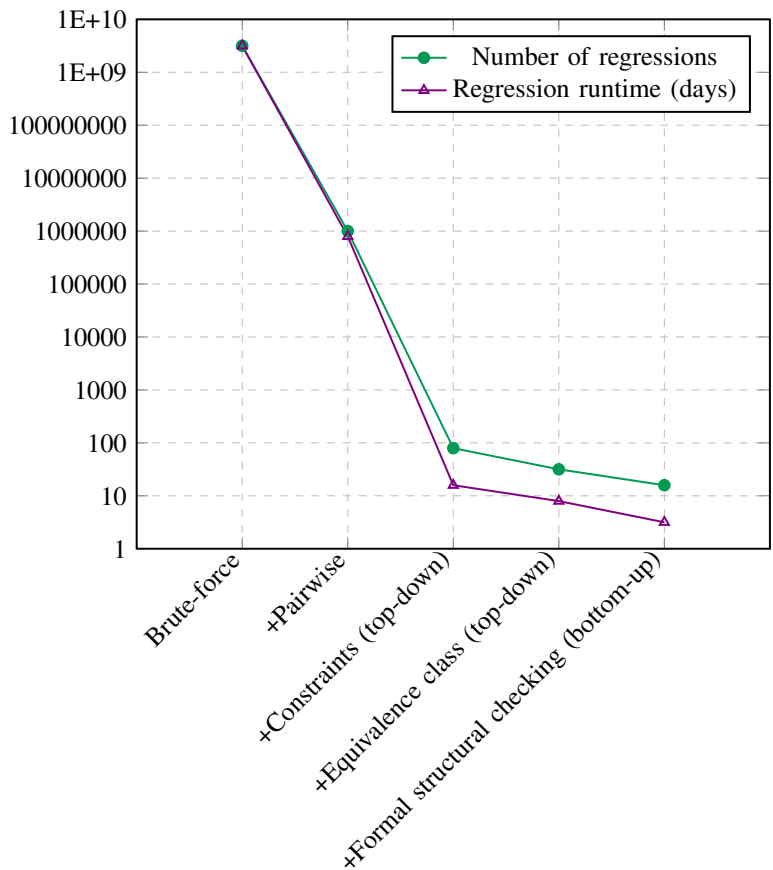
\begin{figure}[h]
\centering
    \begin{tikzpicture}
    \begin{axis}[
        xmin=0, xmax=6,
        ymin=1, ymax=11,
        xtick={1,2,3,4,5},
        ytick={1,2,3,4,5,6,7,8,9,10,11},
        yticklabels={1,10,100,1000,10000,100000,1000000,10000000,100000000,1E+09,1E+10},
        xticklabels={Brute-force, +Pairwise, +Constraints (top-down), +Equivalence class (top-down), +Formal structural checking (bottom-up)},
        legend pos=north east,
        ymajorgrids=true,
        xmajorgrids=true,
        grid style=dashed,
        xlabel style={yshift=-1cm},
        x tick label style={rotate=45, anchor=east},
    ]
    
    \addplot[
        color=ForestGreen,
        mark=*,
        ]
        coordinates {
        (1,10.5)(2,7)(3,2.9)(4,2.5)(5,2.2)
        };
        \addlegendentry{Number of regressions}
    
    \addplot[
        color=violet,
        mark=triangle,
        ]
        coordinates {
        (1,10.5)(2,6.9)(3,2.2)(4,1.9)(5,1.5)
        };
        \addlegendentry{Regression runtime (days)}
    \end{axis}
    \end{tikzpicture}
\caption{Decrement in regressions and regression runtime with added techniques}
\label{result_plot}
\end{figure}

Fig.~\ref{result_plot} shows that with added techniques, the number of regressions and the regression runtime decreased significantly into a more realistic figure. We started with billions of regressions in the brute-force approach. The regressions (and the regression runtime) went down to millions after the addition of Pairwise technique. After adding the constraints, the regressions were 89 and the regression runtime was 20 days. As soon as the Equivalence classes were introduced, the regressions went down to 33 with a runtime of around a week. Finally, after introducing the formal structural checking flow, the regressions (or optimized pairwise regressions) were 17 and the runtime was 3.5 days.

\subsection{Formal Verification of Blocks}
The block suitable for block-level formal verification using the formal structural checking flow was found to be the Data BUS Interface. This block is a combinatorial logic that mainly executes case statements. Since the block is combinatorial, it is even more suitable for formal verification \cite{formal_friendly}.

\begin{figure}[h]
    \centering
    \includegraphics[width=\textwidth]{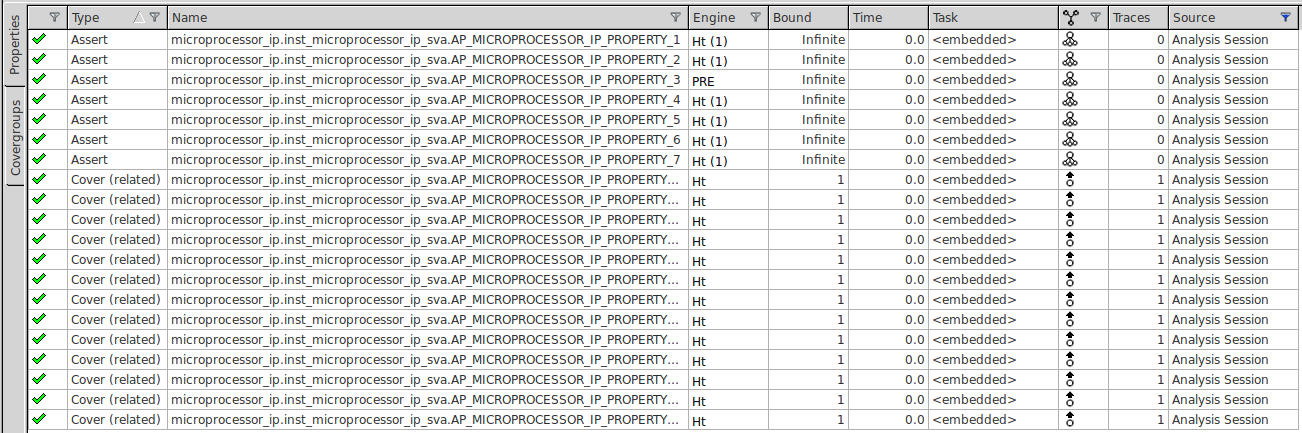}
    \caption{Properties proved in formal tool}
    \label{fv_proof}
\end{figure}

\begin{figure}[h]
    \centering
     \includegraphics[width=\textwidth]{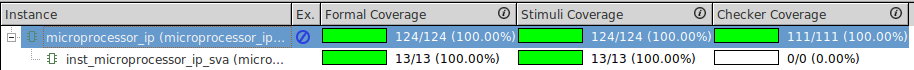}
    \caption{Coverage results in formal tool}
    \label{fv_cov}
\end{figure}

There were 21 properties written to verify the block out of which seven were assert properties and 14 were cover properties. All the properties hold, as shown in Fig.~\ref{fv_proof} and the coverage was collected for completeness metrics analysis, as shown in Fig.~\ref{fv_cov}. The properties were verified for all the 11 values of the parameter \say{Address\_Width\_1}.

The properties proved in the formal tool during verification are necessary to prove the functional correctness of the design. However, a more significant question, if these properties are necessary and sufficient to prove the functional correctness of the design in every possible valid scenario, remains open. Formal coverage plays an important role to answer this question. The formal completeness analysis obtained from the formal coverage provides a very low likelihood that significant errors have been overlooked in verification.

The stimuli coverage, as shown in Fig.~\ref{fv_cov} is the reachability of the formal environment. This includes toggle coverage and branch coverage. The checker coverage is for the assertion completeness and essentially checks for the \acrshort{RTL} code coverage. The formal coverage is the consolidated stimuli and checker coverage.

%\renewcommand{\arraystretch}{1.5}
%\begin{table}[h]
%\caption{Effort and savings with Formal structural checking for the Microprocessor \acrshort{IP}}
\begin{table}[h]
\caption{Effort and savings with formal structural checking for the microprocessor \acrshort{IP}}
\begin{center}
\resizebox{\textwidth}{!}{%
\centering
%\begin{adjustbox}{max width=\textwidth}
%\setlength{\tabcolsep}{10pt}
 \begin{tabular}{|p{0.25\textwidth}|p{0.25\textwidth}|p{0.25\textwidth}|p{0.25\textwidth}|}
 \hline
 \textbf{} & \textbf{Without Formal Structural Check} & \textbf{With Formal Structural Check} & \textbf{Savings} \\
 \hline
 \textbf{No. of regressions} & 33 & 17 & \SI{50}{\percent} \\
 \hline
 \textbf{Regression runtime for one regression} & 12 hours & 8 hours + 8 mins of \acrshort{FV} & $\sim$\SI{50}{\percent} \\
 \hline
 \textbf{Effort in writing properties} & No effort & $\sim$0.5 day & Extra effort in verification \\
 \hline
 \textbf{Probability of finding bug} & High & Very high & Can potentially find more bugs \\
 \hline
\end{tabular}}
%\end{adjustbox}
\label{fv_effort}
\end{center}
\end{table}

To access the benefits of using formal structural checking, Table \ref{fv_effort} is created. It is clear from the table that although some extra effort was required to implement the formal verification (since there is an existing SV-UVM testbench to verify the complete \acrshort{SoC}), the savings in terms of the number of regressions and regression runtime are significant.

\section{Empirical Observations}\label{observations}
There were some empirical observations made during the course of the methodology development and implementation. Some of them substantiate the techniques used in the methodology while others hint in the future direction of possibilities to reduce the number of regressions and consequently, the regression runtime even further.

\subsection{Scalability of Formal Structural Checking}\label{scalability}
Formal structural checking is applied in the extract step of the regression plan generation and is a part of the bottom-up approach. It is essentially used as a regression reduction technique where candidates suitable for block-level verification are identified using the \acrshort{COI} concept. Several other techniques, such as Pairwise and Equivalence classes are also used in the methodology. Pairwise and Equivalence class techniques are already used in software testing to reduce the number of testcases. They have proved to be feasible and scalable for this purpose. Formal structural checking, on the other hand, is a new concept, and the scalability of this technique should be studied.

\begin{figure}[h]
\centering
\begin{tikzpicture}
\begin{axis}[
    ybar=6pt,
    bar width=0.5cm,
    %width=11cm, %remove if RISC-V is removed
    enlargelimits=0.15,
    legend style={at={(0.5,-0.15)},
      anchor=north,legend columns=-1},
    ylabel={Regressions},
    symbolic x coords={Microprocessor \acrshort{IP},HiCoVec,Arm Cortex M0+},
    xtick=data,
    nodes near coords,
    nodes near coords align={vertical},
    ]
\addplot[fill=Peach] coordinates {(Microprocessor \acrshort{IP},33) (HiCoVec,135) (Arm Cortex M0+,160)};
\addplot[fill=YellowGreen] coordinates {(Microprocessor \acrshort{IP},17) (HiCoVec,60) (Arm Cortex M0+,160)};
\legend{Without Block-level verification,With Block-level verification}
\end{axis}
\end{tikzpicture}
\caption{Number of regressions with and without using formal structural checking}
\label{fv_scalability}
\end{figure}
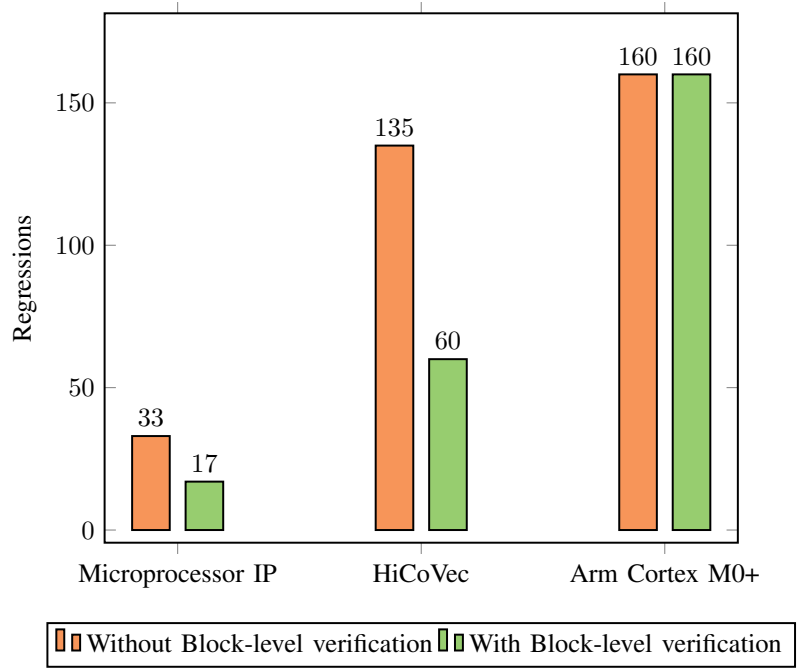

To address the question of scalability of formal structural checking, this technique was applied on some open-source and other processor \acrshort{IP}s. Fig.~\ref{fv_scalability} shows the number of regressions with and without using this technique for processors like Arm Cortex M0+ and Highly Configurable Vector Processor (HiCoVec). It can be depicted from the graph that a decrement of \SI{48.48}{\percent} for the microprocessor \acrshort{IP} and \SI{55.55}{\percent} for HiCoVec processor is achieved using formal structural checking. However, since no block was found suitable for block-level verification in case of Arm Cortex M0+ processor, there was no decrement in the number of regressions. However, since Arm Cortex M0+ has mostly feature enabling/disabling parameters (therefore either 0 or 1 as possible values for them), just using the Pairwise technique would suffice the goal of reducing the number of regressions. Overall, the graph shows that the technique can be applied to any design \acrshort{IP} with positive outcomes. This fact makes formal structural checking a scalable technique.

The regressions were run based on the optimized configurations identified from the regression plan generation. Formal verification for blocks was also carried out with positive results. The configuration coverage was collected in the regression management tool after the regressions finished and can be seen in Fig.~\ref{config_cov_result}.

\tikzset{every picture/.style={line width=1pt}} %set default line width to 0.75pt        
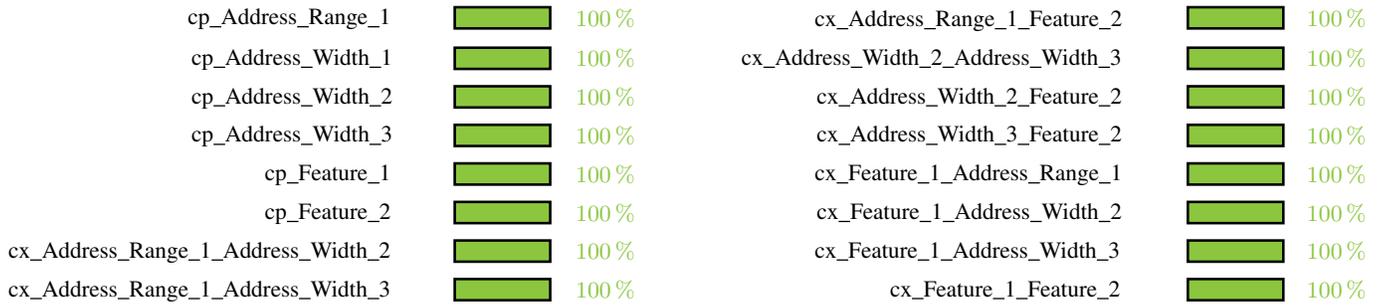
\begin{figure}[h]
\centering
\resizebox{\textwidth}{!}{%
%\begin{adjustbox}{width=\textwidth}
\begin{tikzpicture}[x=0.75pt,y=0.75pt,yscale=-1,xscale=1]
%uncomment if require: \path (0,696); %set diagram left start at 0, and has height of 696

%Shape: Rectangle [id:dp7286918552863675] 
\draw  [fill={LimeGreen}] (211,10) -- (260.5,10) -- (260.5,21) -- (211,21) -- cycle ;
%Shape: Rectangle [id:dp11459947718681862] 
\draw  [fill={LimeGreen}] (211,31) -- (260.5,31) -- (260.5,42) -- (211,42) -- cycle ;
%Shape: Rectangle [id:dp14034051600047204] 
\draw  [fill={LimeGreen}] (211,51) -- (260.5,51) -- (260.5,62) -- (211,62) -- cycle ;
%Shape: Rectangle [id:dp7496067521727252] 
\draw  [fill={LimeGreen}] (211,71) -- (260.5,71) -- (260.5,82) -- (211,82) -- cycle ;
%Shape: Rectangle [id:dp23339040510590836] 
\draw  [fill={LimeGreen}] (211,91) -- (260.5,91) -- (260.5,102) -- (211,102) -- cycle ;
%Shape: Rectangle [id:dp48790987581941114] 
\draw  [fill={LimeGreen}] (211,111) -- (260.5,111) -- (260.5,122) -- (211,122) -- cycle ;
%Shape: Rectangle [id:dp6900293979526715] 
\draw  [fill={LimeGreen}] (211,131) -- (260.5,131) -- (260.5,142) -- (211,142) -- cycle ;
%Shape: Rectangle [id:dp06019852748692989] 
\draw  [fill={LimeGreen}] (211,151) -- (260.5,151) -- (260.5,162) -- (211,162) -- cycle ;
%Shape: Rectangle [id:dp2408355407386138] 
\draw  [fill={LimeGreen}] (591,10) -- (640.5,10) -- (640.5,21) -- (591,21) -- cycle ;
%Shape: Rectangle [id:dp2432458283164125] 
\draw  [fill={LimeGreen}] (591,31) -- (640.5,31) -- (640.5,42) -- (591,42) -- cycle ;
%Shape: Rectangle [id:dp15846981641383806] 
\draw  [fill={LimeGreen}] (591,51) -- (640.5,51) -- (640.5,62) -- (591,62) -- cycle ;
%Shape: Rectangle [id:dp9798706677045681] 
\draw  [fill={LimeGreen}] (591,71) -- (640.5,71) -- (640.5,82) -- (591,82) -- cycle ;
%Shape: Rectangle [id:dp5140381240378118] 
\draw  [fill={LimeGreen}] (591,91) -- (640.5,91) -- (640.5,102) -- (591,102) -- cycle ;
%Shape: Rectangle [id:dp37641085363526217] 
\draw  [fill={LimeGreen}] (591,111) -- (640.5,111) -- (640.5,122) -- (591,122) -- cycle ;
%Shape: Rectangle [id:dp7341276817820546] 
\draw  [fill={LimeGreen}] (591,131) -- (640.5,131) -- (640.5,142) -- (591,142) -- cycle ;
%Shape: Rectangle [id:dp4250793118884395] 
\draw  [fill={LimeGreen}] (591,151) -- (640.5,151) -- (640.5,162) -- (591,162) -- cycle ;

% Text Node
\draw (71,9) node [anchor=north west][inner sep=0.75pt]  [font=\small] [align=left] {cp\_Address\_Range\_1};
% Text Node
\draw (73,30) node [anchor=north west][inner sep=0.75pt]  [font=\small] [align=left] {cp\_Address\_Width\_1};
% Text Node
\draw (73,50) node [anchor=north west][inner sep=0.75pt]  [font=\small] [align=left] {cp\_Address\_Width\_2};
% Text Node
\draw (73,70) node [anchor=north west][inner sep=0.75pt]  [font=\small] [align=left] {cp\_Address\_Width\_3};
% Text Node
\draw (111,90) node [anchor=north west][inner sep=0.75pt]  [font=\small] [align=left] {cp\_Feature\_1};
% Text Node
\draw (111,110) node [anchor=north west][inner sep=0.75pt]  [font=\small] [align=left] {cp\_Feature\_2};
% Text Node
\draw (-22,130) node [anchor=north west][inner sep=0.75pt]  [font=\small] [align=left] {cx\_Address\_Range\_1\_Address\_Width\_2};
% Text Node
\draw (-22,150) node [anchor=north west][inner sep=0.75pt]  [font=\small] [align=left] {cx\_Address\_Range\_1\_Address\_Width\_3};
% Text Node
\draw (396,10) node [anchor=north west][inner sep=0.75pt]  [font=\small] [align=left] {cx\_Address\_Range\_1\_Feature\_2};
% Text Node
\draw (358,30) node [anchor=north west][inner sep=0.75pt]  [font=\small] [align=left] {cx\_Address\_Width\_2\_Address\_Width\_3};
% Text Node
\draw (397,50) node [anchor=north west][inner sep=0.75pt]  [font=\small] [align=left] {cx\_Address\_Width\_2\_Feature\_2};
% Text Node
\draw (397,70) node [anchor=north west][inner sep=0.75pt]  [font=\small] [align=left] {cx\_Address\_Width\_3\_Feature\_2};
% Text Node
\draw (396,90) node [anchor=north west][inner sep=0.75pt]  [font=\small] [align=left] {cx\_Feature\_1\_Address\_Range\_1};
% Text Node
\draw (397,110) node [anchor=north west][inner sep=0.75pt]  [font=\small] [align=left] {cx\_Feature\_1\_Address\_Width\_2};
% Text Node
\draw (396,130) node [anchor=north west][inner sep=0.75pt]  [font=\small] [align=left] {cx\_Feature\_1\_Address\_Width\_3};
% Text Node
\draw (435,150) node [anchor=north west][inner sep=0.75pt]  [font=\small] [align=left] {cx\_Feature\_1\_Feature\_2};
% Text Node
\draw (272,9) node [anchor=north west][inner sep=0.75pt]  [font=\small] [align=left] {\textcolor{LimeGreen}{\textbf{\SI{100}{\percent}}}};
% Text Node
\draw (272,30) node [anchor=north west][inner sep=0.75pt]  [font=\small] [align=left] {\textcolor{LimeGreen}{\textbf{\SI{100}{\percent}}}};
% Text Node
\draw (272,50) node [anchor=north west][inner sep=0.75pt]  [font=\small] [align=left] {\textcolor{LimeGreen}{\textbf{\SI{100}{\percent}}}};
% Text Node
\draw (272,70) node [anchor=north west][inner sep=0.75pt]  [font=\small] [align=left] {\textcolor{LimeGreen}{\textbf{\SI{100}{\percent}}}};
% Text Node
\draw (272,90) node [anchor=north west][inner sep=0.75pt]  [font=\small] [align=left] {\textcolor{LimeGreen}{\textbf{\SI{100}{\percent}}}};
% Text Node
\draw (272,110) node [anchor=north west][inner sep=0.75pt]  [font=\small] [align=left] {\textcolor{LimeGreen}{\textbf{\SI{100}{\percent}}}};
% Text Node
\draw (272,130) node [anchor=north west][inner sep=0.75pt]  [font=\small] [align=left] {\textcolor{LimeGreen}{\textbf{\SI{100}{\percent}}}};
% Text Node
\draw (272,150) node [anchor=north west][inner sep=0.75pt]  [font=\small] [align=left] {\textcolor{LimeGreen}{\textbf{\SI{100}{\percent}}}};
% Text Node
\draw (651,9) node [anchor=north west][inner sep=0.75pt]  [font=\small] [align=left] {\textcolor{LimeGreen}{\textbf{\SI{100}{\percent}}}};
% Text Node
\draw (651,30) node [anchor=north west][inner sep=0.75pt]  [font=\small] [align=left] {\textcolor{LimeGreen}{\textbf{\SI{100}{\percent}}}};
% Text Node
\draw (651,50) node [anchor=north west][inner sep=0.75pt]  [font=\small] [align=left] {\textcolor{LimeGreen}{\textbf{\SI{100}{\percent}}}};
% Text Node
\draw (651,70) node [anchor=north west][inner sep=0.75pt]  [font=\small] [align=left] {\textcolor{LimeGreen}{\textbf{\SI{100}{\percent}}}};
% Text Node
\draw (651,90) node [anchor=north west][inner sep=0.75pt]  [font=\small] [align=left] {\textcolor{LimeGreen}{\textbf{\SI{100}{\percent}}}};
% Text Node
\draw (651,110) node [anchor=north west][inner sep=0.75pt]  [font=\small] [align=left] {\textcolor{LimeGreen}{\textbf{\SI{100}{\percent}}}};
% Text Node
\draw (651,130) node [anchor=north west][inner sep=0.75pt]  [font=\small] [align=left] {\textcolor{LimeGreen}{\textbf{\SI{100}{\percent}}}};
% Text Node
\draw (651,150) node [anchor=north west][inner sep=0.75pt]  [font=\small] [align=left] {\textcolor{LimeGreen}{\textbf{\SI{100}{\percent}}}};

\end{tikzpicture}}
\caption{Configuration coverage results for the microprocessor IP}
\label{config_cov_result}
%\end{adjustbox}
\end{figure}

\subsection{Dependency and Independency of Parameters}
During the course of the development of the methodology, efforts were made to find the dependency and independency between the parameters. To realize this, a dependency table was created using the \acrshort{CBAW} technique \cite{param_ver_cadence} and performing some structural checks in the formal tool.

%\renewcommand{\arraystretch}{1.5}
%\begin{table}[h]
%\caption{Dependency table for the Microprocessor \acrshort{IP}}
%\centering
%\begin{adjustbox}{max width=\textwidth}
%\setlength{\tabcolsep}{10pt}
% \begin{tabular}{c c c c c c c}
% \hline
% \textbf{Parameters} & \textbf{Feature\_1} & \textbf{Address\_Range\_1} & \textbf{Address\_Width\_1} %& \textbf{Address\_Width\_2} & \textbf{Address\_Width\_3} & \textbf{Feature\_2} \\
% \hline\hline
% \textbf{Feature\_1} & N/A & \checkmark & \checkmark & \checkmark & \checkmark & x \\
% \hline
% \textbf{Address\_Range\_1} & \checkmark & N/A & \checkmark & \checkmark & \checkmark & \checkmark \\
% \hline
% \textbf{Address\_Width\_1} & \checkmark & \checkmark & N/A & \checkmark & \checkmark & \checkmark \\
% \hline
% \textbf{Address\_Width\_2} & \checkmark & \checkmark & \checkmark & N/A & \checkmark & \checkmark \\
% \hline
% \textbf{Address\_Width\_3} & \checkmark & \checkmark & \checkmark & \checkmark & N/A & \checkmark \\
% \hline
% \textbf{Feature\_2} & x & \checkmark & \checkmark & \checkmark & \checkmark & N/A \\
% \hline
%\end{tabular}
%\end{adjustbox}
%\label{dep_table}
%\end{table}
%
%The following formulas were used to find the percentage of dependency and independency where %\textit{N} is the number of design parameters:
%
%\begin{equation}
%\textrm{Dependency} = \frac{\textrm{\#\checkmark}}{N \cdot (N-1)} \cdot \SI{100}{\percent}
%\end{equation}
%
%\begin{equation}
%\textrm{Independency} = \frac{\textrm{\#x}}{N \cdot (N-1)} \cdot \SI{100}{\percent}
%\end{equation}
%
The dependency for the microprocessor \acrshort{IP} was found to be \SI{93.33}{\percent}, whereas the independency was found to be \SI{6.66}{\percent}. It is clear that most of the parameters are dependent and a similar result was obtained for other open-source processors as well. When performed the dependency check in open-source RISC-V processor PicoRV32 \cite{riscv}, \SI{46.66}{\percent} parameters were found to be dependent whereas, \SI{53.33}{\percent} parameters were independent. This is due to the fact that many of the parameters were feature enabling/disabling parameters, and it was observed that generally the feature enabling/disabling parameters are independent of each other. On the other hand, bus width parameterizations are dependent.

\tikzset{every picture/.style={line width=1pt}} %set default line width to 0.75pt        
\begin{figure}[h]
\centering
\begin{tikzpicture}[x=0.75pt,y=0.75pt,yscale=-1,xscale=1]

%Flowchart: Manual Operation [id:dp9107331951922693] 
\draw  [fill={rgb, 255:red, 74; green, 144; blue, 226 }  ,fill opacity=0.2 ] (288.43,133.48) -- (230.25,153.2) -- (230.25,49.12) -- (288.43,68.43) -- cycle ;
%Straight Lines [id:da9015648606668383] 
\draw    (288.43,101.48) -- (310.43,101.48) ;
%Straight Lines [id:da7444354178414021] 
\draw    (201,76) -- (230,76) ;
%Straight Lines [id:da7113258971076806] 
\draw    (201,126) -- (230,126) ;
%Straight Lines [id:da8828803747755887] 
\draw    (259,144) -- (259,166) ;

% Text Node
\draw (131,94) node [anchor=north west][inner sep=0.75pt]   [align=left] {Data path};
% Text Node
\draw (217,169) node [anchor=north west][inner sep=0.75pt]   [align=left] {Control path};
% Text Node
\draw (241,94) node [anchor=north west][inner sep=0.75pt]   [align=left] {MUX};
% Text Node
\draw (315,94) node [anchor=north west][inner sep=0.75pt]   [align=left] {out};

\end{tikzpicture}
\caption{Control and data path}
\label{fv_path}
\end{figure}
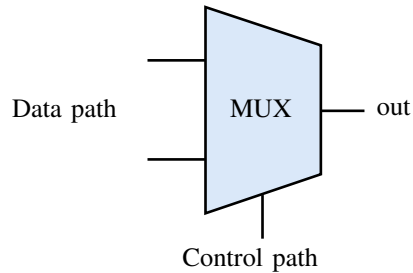

When investigated more into the results concluded from the structural checks, it was observed that most of the parameters are dependent via control path (e.g., through MUX select signals) and not via data path, as shown in Fig.~\ref{fv_path}. This observation was made using some structural checks performed in in the formal tool.

\section{Conclusion and Future Work}
We developed a semi-formal verification methodology to achieve efficient configuration coverage of highly configurable digital designs. To demonstrate its added values, we applied the methodology on a microprocessor \acrshort{IP}. The results are positive as we identified 17 configurations for verification with regression runtime of 3.5 days. We achieved \SI{100}{\percent} optimized configuration coverage identified using the developed methodology and unvieled nine bugs out of which eight were testbench bugs (SV-UVM) and one was a \acrshort{RTL} bug. In the end, our methodology is automated, efficient, complied with the ISO 26262 guidelines and is scalable.

As a future work, we would focus on developing a concrete definition of the in-/dependency of parameters based on the findings in section \ref{observations} so that the number of regressions and consequently, the regression runtime can be reduced even further. Another improvement could have been made by increasing the number of parallel runs in the simulator being used. This would have definitely reduced the regression runtime even more than the one observed in the microprocessor \acrshort{IP}, giving an opportunity for running even more amount of regressions. However, using multiple simulation runs would require check-out licenses for the tool and would cost money at the end. Therefore, a trade-off needs to be set to address this possibility and find an optimum balance between time and cost.

%\section*{References}
%\bibliographystyle{./bibliography/IEEEtran}
%\bibliography{./bibliography/IEEEabrv,./bibliography/IEEEexample}
\printbibliography

\end{document}